\documentclass[12pt, draftclsnofoot, onecolumn]{IEEEtran}

\usepackage{graphicx,psfrag,epsfig,epsf,latexsym,hhline,amsmath,amssymb,multirow}
\usepackage[usenames,dvipsnames]{pstricks}
\usepackage{pst-plot}
\usepackage{pstricks-add}
\usepackage{pifont}
\usepackage{graphicx}
\usepackage{amsthm}
\usepackage{algorithm}
\usepackage{algpseudocode}
\usepackage[noadjust]{cite}
\usepackage{array}
\usepackage{blindtext}
\usepackage{etoolbox}
\usepackage{comment}
\usepackage{epstopdf}
\usepackage{makecell}
\usepackage{aligned-overset}
\usepackage{hyperref}
\usepackage{longtable,tabularx,ltxtable,ragged2e}
\usepackage{filecontents}
\newcolumntype{M}[1]{>{\centering\arraybackslash}m{#1}}
\newcolumntype{P}[1]{>{\centering\arraybackslash}p{#1}}

\newcolumntype{Y}{>{\RaggedRight\arraybackslash}X}

\newcommand{\mbb}[1]{\mathbb{#1}}

\newcommand{\msf}[1]{\mathsf{#1}}

\newcommand{\SNR}{\msf{SNR}}
\newcommand{\INR}{\msf{INR}}

\newcommand{\E}{\mathbb{E}}

\newcommand{\iid}{i.\@i.\@d.\ }

\theoremstyle{definition}
\newtheorem{define}{Definition}
\newtheorem{example}{Example}
\newtheorem{theorem}{Theorem}
\newtheorem{lemma}{Lemma}
\newtheorem{proposition}{Proposition}
\newtheorem{remark}{Remark}
\newtheorem{corollary}{Corollary}

\begin{document}
\title{Discrete Signaling and Treating Interference as Noise for the Gaussian Interference Channel}
\author{Min Qiu, Yu-Chih Huang, and Jinhong Yuan


\thanks{This work was presented in part at the 2019 IEEE Global Communications Conference (GLOBECOM) \cite{Qiu19Globecom} and will be presented in part at the 2020 IEEE Internal Symposium on Information Theory (ISIT) \cite{Qiu2020ISIT}.

M. Qiu, and J. Yuan are with the School of Electrical Engineering and Telecommunications, University of New South Wales, Sydney, NSW, 2052 Australia (e-mail: min.qiu@unsw.edu.au; j.yuan@unsw.edu.au).

Y.-C. Huang is with the Institute of Communications Engineering, National Yang Ming Chiao Tung University, Hsinchu City 30010, Taiwan (e-mail: jerryhuang@nctu.edu.tw).
}%
}

\maketitle

\begin{abstract}
The two-user Gaussian interference channel (G-IC) is revisited, with a particular focus on practically amenable discrete input signalling and treating interference as noise (TIN) receivers. The corresponding deterministic interference channel (D-IC) is first investigated and coding schemes that can achieve the entire capacity region of D-IC under TIN are proposed. These schemes are then \emph{systematically} translate into multi-layer superposition coding schemes based on purely discrete inputs for the real-valued G-IC. Our analysis shows that the proposed scheme is able to achieve the \emph{entire} capacity region to within a constant gap for all channel parameters. To the best of our knowledge, this is the first constant-gap result under purely discrete signalling and TIN for the entire capacity region and all the interference regimes. Furthermore, the approach is extended to obtain coding schemes based on discrete inputs for the complex-valued G-IC. For such a scenario, the minimum distance and the achievable rate of the proposed scheme under TIN are analyzed, which takes into account the effects of random phase rotations introduced by the channels. Simulation results show that our scheme is capable of approaching the capacity region of the complex-valued G-IC and significantly outperforms Gaussian signalling with TIN in various interference regimes.
\end{abstract}


\section{Introduction}\label{sec:intro}
Interference is one of the key challenges in wireless networks where multiple transmissions share and compete for the same medium resource \cite{Network_IT}. To study this problem, it is essential to start with one of the most fundamental channel models: the two-user Gaussian interference channel (G-IC), which is described by the following input-output relationship
\begin{align}
\msf{Y}_1 = h_{11}\msf{X}_1+h_{12}\msf{X}_2+\msf{Z}_1, \label{eq:gic_1}\\
\msf{Y}_2 = h_{21}\msf{X}_1+h_{22}\msf{X}_2+\msf{Z}_2, \label{eq:gic_2}
\end{align}
where $\forall k,\bar{k} \in \{1,2\}$, $\msf{X}_k$ is user $k$'s signal intended for receiver $k$ and is subject to a unit power constraint $\E[\|\msf{X}_k \|^2]\leq1$, $\msf{Z}_k$ is the additive white Gaussian noise with zero mean and unit variance, and $h_{k\bar{k}}$ is the channel between transmitter $k$ and receiver $\bar{k}$, which is fixed and known to all transmitters and receivers. For notation simplicity, define user $k$'s signal-to-noise ratio (SNR) and interference-to-noise ratio (INR) as $\SNR_k \triangleq |h_{kk}|^2$ and $\INR_k \triangleq |h_{k\bar{k}}|^2$ for $k\neq \bar{k}$, respectively. This channel is referred to as the complex G-IC when all the variables are complex-valued and is referred to as the real G-IC when all the variables are real-valued. The interference regimes of the channel can be characterized into: very weak when $\INR_{\bar{k}}(1+\INR_k) \leq \SNR_k$ \cite{7051266}, weak when $\INR_{\bar{k}} \leq \SNR_k$ \cite{4675741}, strong when $\INR_{\bar{k}} \geq \SNR_k$ \cite{4675741}, very strong when $\SNR^2_k(1+\SNR^2_{\bar{k}}) \leq \INR^2_{\bar{k}}$ \cite{1056416}, and mixed when $\SNR_k \geq \INR_{\bar{k}},\SNR_{\bar{k}}\leq  \INR_k$ or $\SNR_k \leq \INR_{\bar{k}},\SNR_{\bar{k}} \geq  \INR_k$ \cite{4675741}.

For this channel, after a long pursuit \cite{Sato1977,1055812}, the capacity region can now be characterized to within $1/2$ bits per channel use \cite{4675741}. For some special cases where the interference are either strong or very strong \cite{10.1109/TIT.1975.1055432,1056416,1057340} or very weak (and symmetric) \cite{5075903}, the exact characterizations of the capacity regions are also available. The key ingredients for deriving these results are a tight converse bound \cite{4675741} and the use of Han-Kobayashi (HK) scheme \cite{1056307,4544957} along with Gaussian signaling. The main idea of the HK scheme is to split the message at each transmitter into a common message and a private message, while the common message needs to be successfully decoded and subtracted out first at both intended and unintended receivers. However, such a successive interference cancellation (SIC) procedure would introduce extra decoding latency and complexity and may compromise the security of the transmissions.

\subsection{Motivation}
Compared to SIC, treating interference as noise (TIN) is much appealing in practice as it simply involves single-user decoding. Due to its low decoding complexity and latency, there has been a growing interest in characterizing the behavior of TIN decoding in various channel models, e.g., \cite{7051266,7106540,7322253,7422814,7486985,7496850,7742969,8113535,8700247,9112224}. For the class of interference channels, it is well-known that when the interference is sufficiently weak in the sense that each user's desired signal strength is no less than the sum of the strongest interference strengths from and to this user \cite{7051266}, Gaussian signaling with TIN is constant-gap optimal for the two-user G-IC \cite{4777648,4777617,5075903} and it is optimal in the $K$-user G-IC from a generalized degrees of freedom perspective \cite{7051266}. However, for other interference regimes, adopting Gaussian signaling with TIN usually achieves significantly suboptimal results due to excessive interference. On the other hand, encouraging results can be found in \cite{Dobrushin1963,Ahlswede1973} where the capacity region of the interference channel is shown to be achievable with each receiver performing single-user decoding, i.e., TIN. However, due to the multi-letter nature of the results in \cite{Dobrushin1963,Ahlswede1973}, the capacity region is hard to compute and the capacity-achieving input distributions are difficult to find. Nonetheless, these results reveal that the suboptimality of TIN is {\it not fundamental} to the problem itself; but merely the limitation of the existing schemes.

Although most of the achievements with TIN adopt Gaussian input distributions, one may still suspect that the Gaussian input distributions are the main source of the suboptimality of TIN. This can be seen by noting that Gaussian is the best input distribution for the power constrained point-to-point Gaussian channel, but also the worst noise (or interference when it is treated as noise) for such the channel \cite{Cover:2006:EIT:1146355}. In addition, it is still very difficult (if not impossible) to implement Gaussian signaling in the current communication systems. On the other hand, discrete signaling can behave differently from Gaussian signaling when being treated as noise as several works have reported larger rate regions obtained by discrete signaling with TIN over Gaussian signaling with TIN, which we will discuss shortly. Further, in the current communication systems, e.g., 4G \cite{TS136212} and 5G \cite{TS138212}, channel coded discrete modulations are the sole approach to carry and transmit messages. In light of the above considerations, to unleash the full potential of interference channels in practical communication systems, the study of employing discrete inputs and TIN in the G-IC is of utmost both practical and theoretical importance.

\subsection{Relevant Works}
Recently, a number of research has been conducted on discrete inputs and TIN for the interference channel \cite{6939682,7282748,7451210,ShuoLithesis}. In \cite{6939682}, it was shown that it is possible to achieve higher rate when one user adopts discrete inputs while the other user adopts Gaussian inputs. Furthermore, Dytso \textit{et al.} \cite{7451210} showed that employing mixed pulse amplitude modulation (PAM) and Gaussian inputs at each user can achieve the capacity region of the real-valued G-IC within a gap of at most $O(\log_2(\log_2(\min(\SNR,\INR))/\eta))$ \footnote{The asymmetric very strong interference regime and some subregimes of the symmetric weak interference regime can be achieved by purely discrete inputs to within a constant gap \cite{7451210}.} up to a Lebesgue measure $\eta \in (0,1]$. The rationale behind this success is that under TIN, the structure of discrete interference can be harnessed by carefully designing the power allocation for the discrete and continuous parts of the mixed inputs. Despite the huge step made in \cite{7451210} towards eliminating the need of SIC, a mixed discrete and Gaussian input still involves Gaussian distributions and hence is far from being practical. On the other hand, the author in \cite{ShuoLithesis} (also its conference version \cite{7282748}) constructed schemes with TIN for the symmetric deterministic interference channel (D-IC) \cite{doi:10.1002/ett.1287} (i.e., the linear deterministic approximation of the G-IC \cite{Avestimehr11}) and translated the schemes into \emph{purely} discrete PAM signalings with TIN for the symmetric real-valued G-IC. After the translation, the author showed that the translated scheme can achieve the symmetric capacity of the symmetric real-valued G-IC to within a constant gap \cite{ShuoLithesis}, under the assumption that the channel gains are powers of 2. All the above works have demonstrated that discrete signalings are promising for handling interference when they are treated as noise. That being said, it remains unclear whether it is possible for purely discrete inputs with TIN to achieve the \emph{entire} capacity region of the two-user \textit{general} G-IC (which can be symmetric or asymmetric) to within a constant gap for all interference regimes and all channel parameters. Further, the above works only consider the real G-IC. For the complex G-IC, the performance of discrete signalings with TIN could be severely affected by the phase rotation introduced in each communication link. Consequently, it is difficult to directly apply the results from the real G-IC to complex G-IC, although this is not a problem for circularly symmetric Gaussian input signaling.

%


\subsection{Contributions}
In this work, we continue the quest of designing (asymptotically) optimal input distributions that can achieve the capacity region to within a constant gap for the general G-IC. In particular, for practical relevance, we focus solely on {\it purely discrete} input distributions at the encoders and {\it TIN} at the decoders. Our goal here is not to obtain sharpened bounds on the achievable rate of discrete inputs, but rather to further push the frontiers of discrete signaling with TIN in other interference regimes and show its (constant-gap) optimality. Specifically, we focus on the not very weak, not very strong and mixed interference regimes. As for other regimes, the constant-gap optimality of discrete signaling with TIN has been shown. The main contributions of the papers are as follows:
\begin{itemize}
\item We use a three-step approach to prove the constant-gap optimality of purely discrete signaling with TIN for the general G-IC for all interference regimes: Step 1) We first look into the general D-IC model \cite{doi:10.1002/ett.1287} and systematically construct novel coding schemes with TIN that are proven to achieve the \emph{entire capacity region} for \emph{all interference regimes}; Step 2) We translate the scheme for the D-IC into a multi-layer superposition coding scheme based on the commonly used PAM for the general G-IC; Step 3) By using the connection between the D-IC and G-IC, we prove that the translated schemes are capable of achieving any rate pair inside the capacity region of the real-valued G-IC to within a constant gap for all interference regimes. To the best of our knowledge, this is the first time that discrete signaling with TIN is proven to be \emph{constant-gap optimal} in the general G-IC for all the interference regimes.

\begin{itemize}
\item In Step 1), we propose novel coding scheme to achieve the whole capacity region of the general D-IC with TIN. This allows us to obtain purely discrete input distributions for the G-IC from the proposed scheme in Step 2). Our scheme is different from \cite{doi:10.1002/ett.1287} which achieves the capacity region with HK schemes; and it is also a highly non-trivial generalization of \cite{ShuoLithesis} which only considers achieving the symmetric capacity of the symmetric D-IC and thus no longer suffices for our purpose. Specifically, we propose two types of schemes to achieve every corner point of the capacity region for all interference regimes. The capacity region of the general D-IC can then be achieved by the proposed schemes together with time-sharing. With the ``achieving corner point'' approach, the design of achievable schemes and the achievability proof are greatly simplified compared to \cite{ShuoLithesis} (See Remark \ref{remark2}).

\item In Step 2), we translate each sub-matrix in the D-IC model into an independent discrete modulation. These discrete modulations are then scaled and superimposed together to form the composite discrete constellation. Different from \cite{7451210} which directly construct schemes for the G-IC, we translate schemes from the D-IC and take advantage of the fact that there exists a universal constant gap between the D-IC and the G-IC \cite{doi:10.1002/ett.1287}. With this, we only need to focus on achieving the rate pair of the D-IC in the G-IC setting. Moreover, we consider general channel parameters that can be any real values, rather than restricting to
    powers of 2 as in \cite{ShuoLithesis}. Under this setting, the power and the size of each independent discrete signal are carefully designed such that the proposed scheme is robust to the real channel gains that are not necessarily powers of 2.

\item In Step 3), we establish some useful tools to lower bound the minimum distance of a multi-layer superimposed signals in the proposed scheme (Lemmas \ref{lem:normalization}-\ref{fact:1}) as we show that the gap to capacity is a function of the minimum distance by using an Ozarow-type bound \cite{ozarow90}. Unlike \cite{ShuoLithesis} directly citing a minimum distance bound as a fact\footnote{It can be easily shown that when the channel parameters are not powers of 2, Facts 1-4 hinged by the proof in \cite{ShuoLithesis} no longer hold.}, we rigorously prove that the minimum distance under the proposed scheme is lower bounded by a constant independent of all channel parameters and interference regimes. This allows us to prove that for every interference regime, the proposed scheme is capable of achieving any rate pair inside the capacity region of the real-valued G-IC to within a constant gap regardless of channel parameters. It is also worth noting that a two-layer scheme based on PAM inputs was mentioned in \cite[Sec. VIII-C]{7451210} that may be good for the moderately weak interference regime. Our results can be deemed as a significant extension of the two layer scheme to multi-layer and to cover all the interference regimes. Moreover, our analysis offers a complete understanding and new insights for the multi-layer inputs schemes.

\end{itemize}

\item For the complex-valued G-IC, we translate the proposed scheme from the D-IC into a multi-layer superposition coding scheme based on the commonly used quadratic amplitude modulation (QAM). We then establish a useful tool for lower bounding the achievable rate of a discrete input drawn from an irregular two-dimensional constellation. With this tool, we obtain a lower bound on the achievable rate of the proposed schemes in the complex-valued G-IC, where the gap to the capacity is a function of the minimum Euclidean distance of the superimposed constellation. Although obtaining a closed form expression of the minimum distance of the superimposed constellation is difficult due to random phase distortions experienced by different links, we still manage to show that the phase rotations that result in zero minimum distance constitute a set of Lebesgue measure zero. Simulation results are provided to show that the proposed scheme can operate close to the capacity outer bound \cite{4675741} and the achievable rate region of the Gaussian HK scheme \cite{4675741} and it significantly outperforms Gaussian inputs with TIN.

\end{itemize}

\subsection{Paper Organization}
The rest of the paper is organized as follows. Section \ref{sec:proposed} introduces the D-IC model and the proposed achievable schemes with TIN that are proven to be capacity achieving. In Section \ref{sec:RGIC}, the scheme proposed for the D-IC is translated into a multi-layer superposition coding scheme based on PAM with TIN for the real-valued G-IC. The constant-gap optimality of the proposed scheme is also rigorously proven step-by-step in this section. Section \ref{sec:CGIC} presents the proposed schemes for the complex G-IC, followed by the achievable rate analysis and simulations. Finally, the paper concludes in Section \ref{sec:conclude}.

\subsection{Notations}
This paper uses the following notations. $\mathbb{Z},\mathbb{N},\mathbb{R}$ and $\mathbb{C}$ represent the sets of integers, natural numbers, real numbers and complex numbers, respectively. Random variables are written in uppercase Sans Serif font, e.g., $\msf{X}$. For $x\in\mathbb{R}$, $\left\lfloor x\right\rfloor = y \in \mathbb{Z}$ gives the nearest integer $y\leq x$. For a set $\mathcal{S}$, $|\mathcal{S}|$ outputs the cardinality of $\mathcal{S}$. For integers $a,b$ and $b>a$, $[a:b]$ denotes the set $\{a, a+1,\ldots, b\}$. For $z \in \mathbb{C}$, $\Re(z)$ and $\Im(z)$ represent the real and imaginary part of $z$, respectively. The binary field and the collection of binary matrices of size $m\times n$ are denoted by $\mathbb{F}_2$ and $\mathbb{F}_2^{m,n}$, respectively. PAM$(|\Lambda|,d_{\min}(\Lambda))$ represents the uniform distribution over a conventional PAM constellation $\Lambda$ with mean $\mathbb{E}[\Lambda] = 0$, cardinality $|\Lambda|$, minimum distance $d_{\min}(\Lambda)$ and with average energy $\E[\| \Lambda\|^2] = d_{\min}^2(\Lambda)\frac{|\Lambda|^2-1}{12}$. Similarly, QAM$(|\Lambda|,d_{\min}(\Lambda))$ represents the uniform distribution over a conventional QAM constellation $\Lambda$ with mean $\mathbb{E}[\Lambda] = 0$, cardinality $|\Lambda|$, minimum distance $d_{\min}(\Lambda)$ and with average energy $\E[\| \Lambda\|^2] = d_{\min}^2(\Lambda)\frac{|\Lambda|-1}{6}$. The symbol $\exists!$ denotes unique existence. We reserve $k,\bar{k} \in \{1,2\}$ to be the user indexes such that $\bar{k}=1$ if $k=2$ and $\bar{k}=2$ if $k=1$.

\section{The Linear Deterministic Interference Channel}\label{sec:proposed}
In this section, we first look into the linear D-IC as an approximation to the G-IC model and propose a family of capacity achieving schemes. The schemes obtained here will be systematically translated into coding schemes for real and complex G-IC in Section \ref{sec:RGIC} and Section \ref{sec:CGIC}, respectively.

\subsection{Channel Model}\label{sec:DIC_model}
The channel model for the two-user D-IC is defined as \cite{doi:10.1002/ett.1287}
\begin{align}\label{eq:DIC_model}
\msf{Y}_k = \boldsymbol{S}^{q-n_{kk}}\msf{X}_k \oplus \boldsymbol{S}^{q-n_{k\bar{k}}}\msf{X}_{\bar{k}},
\end{align}
where the multiplication and summation are over $\mathbb{F}_2$, $n_{kk} \triangleq \lfloor\log_2\SNR_k\rfloor$ and $n_{k\bar{k}} \triangleq \lfloor\log_2\INR_1\rfloor$ are for approximating the complex G-IC; while $n_{kk} \triangleq \lfloor\frac{1}{2}\log_2\SNR_k\rfloor$ and $n_{k\bar{k}} \triangleq \lfloor\frac{1}{2}\log_2\INR_k\rfloor$ are adopted for approximating the real G-IC, $q=\max\{n_{11},n_{12},n_{21},n_{22}\}$, $\boldsymbol{S}$ is a $q \times q$ shift matrix,
\begin{align}
\boldsymbol{S}=\begin{bmatrix}
0 & 0 & 0 & \ldots & 0 \\
1 & 0 & 0 & \ldots & 0 \\
0 & 1 & 0 & \ldots & 0 \\
\vdots & & & \ddots & \vdots\\
0 & \cdots & 0 & 1 & 0\\
\end{bmatrix},
\end{align}
and $\msf{X}_k,\msf{Y}_k \in \mathbb{F}_2^q$ are binary column vectors representing the discrete channel inputs and outputs, respectively, for user $k$. Each entry of the input column vector represents a \emph{power level}. The highest element of $\msf{X}_k$ is called the highest power level while the lowest element is called the lowest power level. The channel operation $\boldsymbol{S}^n$ for some natural number $n\leq q$ is modeled as a set of noiseless bit pipes such that with $\boldsymbol{S}^n\msf{X}_k$, only the highest $q-n$ bits of $\msf{X}_k$ would be received losslessly while the lowest $n$ bits of $\msf{X}_k$ are below the noise level and get truncated.

The capacity region of this channel is characterized in \cite{doi:10.1002/ett.1287} and is summarized in Theorem \ref{the:1} for completeness.
\begin{theorem} \label{the:1}
The capacity region of the D-IC is the set of non-negative rate pair $(r_1,r_2)$ satisfying:
\begin{align}
r_k\leq& n_{kk} \label{eq:dmodcap1} \\
r_k+r_{\bar{k}} \leq &\max\{n_{kk}-n_{k\bar{k}},0\}+\max\{n_{\bar{k}\bar{k}},n_{k\bar{k}}\} \label{eq:dmodcap2}\\
r_k+r_{\bar{k}} \leq& \max\{n_{\bar{k}k},\max\{n_{kk}-n_{k\bar{k}},0\}\} 
+\max\{n_{k\bar{k}},\max\{n_{\bar{k}\bar{k}}-n_{\bar{k}k},0\}\} \label{eq:dmodcap4}\\
2r_k+r_{\bar{k}} \leq &\max\{n_{kk},n_{\bar{k}k}\}+\max\{n_{kk}-n_{k\bar{k}},0\} 
+\max\{n_{k\bar{k}},\max\{n_{\bar{k}\bar{k}}-n_{\bar{k}k},0\}\} \label{eq:dmodcap5}.
\end{align}
Moreover, the above capacity region is within \emph{a constant gap} to the capacity region of the (real and complex) G-IC.
\end{theorem}

It should be noted that the above capacity region of the D-IC was shown to be achievable by Han-Kobayashi scheme together with time-sharing in \cite{doi:10.1002/ett.1287}. In contrast, we show in the following that by carefully designing the transmission scheme, the capacity region can be achieved with TIN together with time-sharing.

\subsection{Main Result}
We state the main result of this section in the following.
\begin{theorem}\label{the:main}
For the two-user G-IC in \eqref{eq:DIC_model}, there exist a pair of input distributions $(\msf{X}_1,\msf{X}_2)$ such that any rate pair inside the capacity region can be achieved by using TIN together with time-sharing.
\end{theorem}
\begin{IEEEproof}
The proof is provided in the next subsection, where the detailed achievable schemes are given in Appendix \ref{app:proof}.
\end{IEEEproof}

\subsection{Proof of Theorem 2}\label{sec:2c}
Let $\msf{U}_k$ be user $k$'s message vector of length $r_k$ with i.i.d. entries drawn independently and uniformly distributed over $\mathbb{F}_2$. And let $\msf{X}_k = \boldsymbol{G}_k \msf{U}_k$ be the channel input for user $k$, where $\boldsymbol{G}_k \in \mathbb{F}_2^{q,r_k}$ is a generator matrix. We also let $\boldsymbol{A}_k \triangleq \boldsymbol{S}^{q-n_{kk}}$ and $\boldsymbol{B}_k \triangleq \boldsymbol{S}^{q-n_{k\bar{k}}}$ represent the channels of the D-IC.

The achievable rate of user $k$ with single-user decoding (i.e., TIN) can be derived as
\begin{align}
I(\msf{X}_k;\msf{Y}_k) &= H(\msf{Y}_k) - H(\msf{Y}_k|\msf{X}_k) \nonumber \\
& = H(\boldsymbol{S}^{q-n_{kk}}\boldsymbol{G}_k \msf{U}_k\oplus \boldsymbol{S}^{q-n_{k\bar{k}}}\boldsymbol{G}_{\bar{k}} \msf{U}_{\bar{k}}) - H(\boldsymbol{S}^{q-n_{k\bar{k}}}\boldsymbol{G}_{\bar{k}} \msf{U}_{\bar{k}}) \nonumber  \\
& = \text{rank}([\boldsymbol{S}^{q-n_{kk}}\boldsymbol{G}_k \; \boldsymbol{S}^{q-n_{k\bar{k}}}\boldsymbol{G}_{\bar{k}}]) - \text{rank}(\boldsymbol{S}^{q-n_{k\bar{k}}}\boldsymbol{G}_{\bar{k}}) \nonumber  \\
& = \text{rank}([\boldsymbol{A}_k\boldsymbol{G}_k \; \boldsymbol{B}_k\boldsymbol{G}_{\bar{k}}]) - \text{rank}(\boldsymbol{B}_k\boldsymbol{G}_{\bar{k}}), \label{eq:I1}
\end{align}
where the multiplication and addition are over $\mathbb{F}_2$.

From this point onwards, the problem becomes designing $\boldsymbol{G}_1$ and $\boldsymbol{G}_2$ such that \\ $(I(\msf{X}_1;\msf{Y}_1),I(\msf{X}_2;\msf{Y}_2))=(r_1,r_2)$ for any integer rate pair $(r_1,r_2)$ inside the capacity region defined in Theorem \ref{the:1}.

In our proposed scheme, we decompose the generator matrix into $M_k$ submatrices, $\boldsymbol{G}_k = [\boldsymbol{E}^T_{k,1},\ldots,\boldsymbol{E}^T_{k,M_k}]^T$. Define $\mathcal{J}_k \triangleq \{j_k:j_k \in [1:M_k],\boldsymbol{E}_{k,j_k} \neq \boldsymbol{0}\}$. There are $|\mathcal{J}_k|$ binary submatrices and $M_k-|\mathcal{J}_k|$ all-zero submatrices. Specifically,
\begin{align}\label{eq:EK}
\boldsymbol{E}_{k,j_k} = \left\{\begin{array}{ll} \boldsymbol{F}_{k,i_k},& j_k \in \mathcal{J}_k,f(j_k)=i_k,\\
\mathbf{0} ,& j_k \in [1:M_k] \setminus \mathcal{J}_k,
\end{array}\right.
\end{align}
where $\boldsymbol{F}_{k,i_k} \in \mathbb{F}_2^{m_{i_k},r_k}$, $i_k \in [1:L_k]$, $f:\mathcal{J}_k \rightarrow [1:L_k]$ is a surjective mapping function and thus $L_k \leq |\mathcal{J}_k|$. The rows of $\boldsymbol{F}_{k,i_k}$ are \emph{linearly independent} and satisfy $m_{i_k} \leq r_k$, i.e., $\text{rank}(\boldsymbol{F}_{k,i_k})=m_{i_k}$. Note that $m_{i_k}$ is a crucial design parameter for achieving the target rate $r_k$. The two binary matrices $\boldsymbol{F}_{a,b}$ and $\boldsymbol{F}_{c,d}$ are linearly independent as long as $(a,b) \neq (c,d)$. Moreover, the position of submatrix $\boldsymbol{E}_{k,j_k}$ in $\boldsymbol{G}_k$ determines the power level of the bits generated by $\boldsymbol{E}_{k,j_k}$. In order to achieve the capacity region, we propose two types of schemes and whether to use a type I or type II scheme depends on the interference regime. The schemes are defined as follows.

\begin{define}\label{def_type12}
An achievable scheme is referred to as a type I scheme if $\forall j_k,j'_k\in \mathcal{J}_k,j_k \neq j'_k$ such that $\boldsymbol{E}_{k,j_k}\neq \boldsymbol{E}_{k,j'_k}$ and $|\mathcal{J}_k|=L_k$ for both users. An achievable scheme is referred to as a type II scheme if $ \exists j_k \in \mathcal{J}_k, \exists ! j'_k \in \mathcal{J}_k,j_k\neq j'_k$ such that $\boldsymbol{E}_{k,j_k}=\boldsymbol{E}_{k,j'_k}=\boldsymbol{F}_{k,i_k}$ and $|\mathcal{J}_k|=L_k+1$ for either one user or both users.
\end{define}

\begin{remark}
In a type I scheme, every binary submatrix $\boldsymbol{E}_{k,j_k}$ is linear independent to any other binary submatrix $\boldsymbol{E}_{k',j_{k'}}$ as in Definition \ref{def_type12}. Thus, we say that each submatrix only occupy one power level. In a type II scheme, there is a binary submatrix that appears in two different sets of rows in the generator matrix, and we say that this submatrix occupies two different power levels. When a type I scheme is used, we design the scheme in such a way that all the desired bits are in the power levels that are not occupied by the interfered bits at both users. This allows the receiver to easily distinguish its intended bits from the interfered bits and thus facilitates the use of TIN to achieve a target rate pair. A type II scheme contains all the features of a type I scheme, except that each of the bits in the block occupying two different power levels will have only one interference-free replica and the other one interfered by the other user's bit. The purpose of introducing these bits is to maximize the achievable rate of their intended user without interfering the other user's bits that occupy the same power levels. In what follows, we use an example to demonstrate the above ideas. 
\end{remark}

\begin{figure}[t!]
	\centering
\includegraphics[width=2.0in,clip,keepaspectratio]{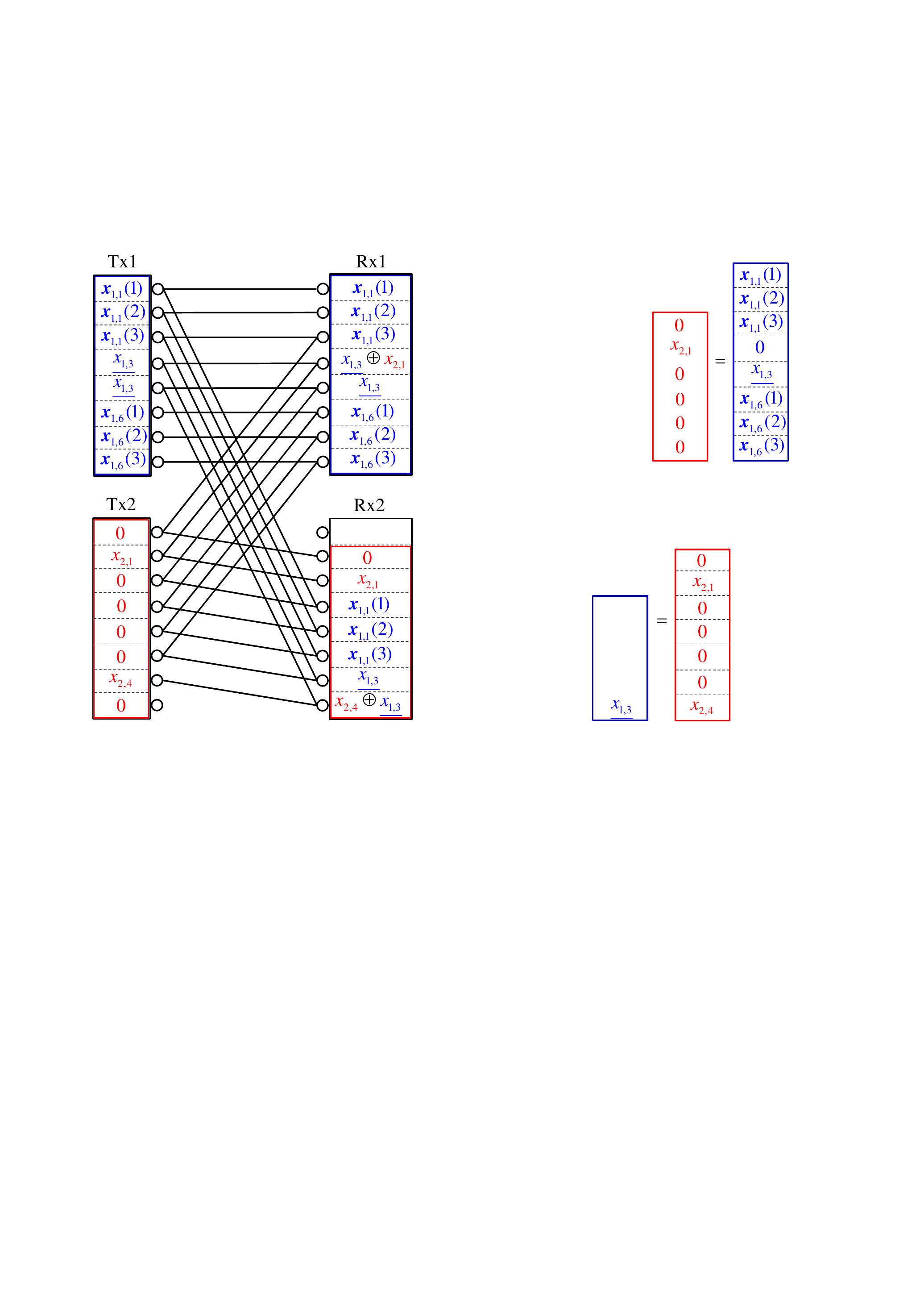}
\caption{Coding scheme for the example channel to achieve $(r_1,r_2) = (7,2)$.}
\label{fig:example1}
\end{figure}

\begin{example}\label{example_new}
Consider $n_{11}=8,n_{22}=7,n_{12}=6,n_{21} = 5$. This set of parameters belong to the weak interference regime, for which our proposed scheme is a type II scheme given in \eqref{eq:weak1_2a_E} in Appendix \ref{sec:weak_12_example}(a). Assume that we want to achieve the target rate pair $(r_1,r_2) = (7,2)$ on the capacity region of the D-IC. The generator matrices are
\begin{align}
\boldsymbol{G}_1=\begin{bmatrix}
\boldsymbol{E}_{1,1}\\
\boldsymbol{E}_{1,2}\\
\boldsymbol{E}_{1,3}\\
\boldsymbol{E}_{1,4}
\end{bmatrix}=
\begin{bmatrix}
\boldsymbol{F}_{1,1}\\
\underline{\boldsymbol{F}_{1,3}}\\
\underline{\boldsymbol{F}_{1,3}}\\
\boldsymbol{F}_{1,6}\\
\end{bmatrix},
\boldsymbol{G}_2=\begin{bmatrix}
\boldsymbol{E}_{2,1}\\
\boldsymbol{E}_{2,2}\\
\boldsymbol{E}_{2,3}\\
\boldsymbol{E}_{2,4}\\
\boldsymbol{E}_{2,5}
\end{bmatrix}= \begin{bmatrix}
\boldsymbol{0}^{1,2} \\
\boldsymbol{F}_{2,1}\\
\boldsymbol{0}^{4,2}\\
\boldsymbol{F}_{2,4}\\
\boldsymbol{0}^{1,2}
\end{bmatrix},
\end{align}
where $\boldsymbol{F}_{1,1},\boldsymbol{F}_{1,6}\in\mathbb{F}_2^{3,7}$, $\boldsymbol{F}_{1,3}\in\mathbb{F}_2^{1,7}$, and $\boldsymbol{F}_{2,1},\boldsymbol{F}_{2,4}\in\mathbb{F}_2^{1,2}$. We underline submatrix $\boldsymbol{F}_{1,3}$ to stress that it is associated with two power levels. The message vectors for users 1 and 2 are $\boldsymbol{x}_1=[\boldsymbol{x}_{1,1},\underline{x_{1,3}},\underline{x_{1,3}},\boldsymbol{x}_{1,6}]^T=[\boldsymbol{x}_{1,1}(1),\boldsymbol{x}_{1,1}(2),\boldsymbol{x}_{1,1}(3),\underline{x_{1,3}},\underline{x_{1,3}},\boldsymbol{x}_{1,6}(1),\boldsymbol{x}_{1,6}(2),\boldsymbol{x}_{1,6}(3)]^T$ and $\boldsymbol{x}_2=[0,x_{2,1},\boldsymbol{0},x_{2,4},0]^T=[0,x_{2,1},0,0,0,0,x_{2,4},0]^T$, respectively, where each element inside the message vector is generated by the corresponding submatrix in the respective generator matrix and $\boldsymbol{x}_{1,1}(1)$ denotes the first bit of subvector $\boldsymbol{x}_{1,1}$ and so forth.

An illustration for the above type II scheme is depicted in Fig. \ref{fig:example1}, where each circle represents a bit. A bit is said to be above the noise level at a receiver if it is emitting from its circle at a transmitter and passed noiselessly through the edge to the circle at that receiver. A bit is said to be below the noise level at a receiver if there is no edge between its circle at a transmitter and any circle at that receiver. Each bit occupies a power level and the power levels are separated by dash lines. There are 8 power levels in total and we number them from the highest power level to the lowest power level as power levels $8,\ldots,1$. In this example, transmitters 1 and 2 are intended to transmit 7 and 2 bits of information, $\boldsymbol{x}_1$ and $\boldsymbol{x}_2$ to users 1 and 2, respectively. The received bits get shifted down due to the channel effects on the incoming communication links.

At receiver 1, user 1's bit in a power level (i.e., power levels 1-4 and 6-8) that is not interfered by any of user 2's bits, can be successfully received. Note that every bit in a type I scheme has this feature. Although $\underline{x_{1,3}}$ and $x_{2,1}$ are aligned in power level 5, user 1 can still retrieve $\underline{x_{1,3}}$ in power level 4.\footnote{For the ease of presentation, we refer to ``two signals are aligned'' as that two signals occupy the same power level in the D-IC. The reader should not confuse this with any form of interference alignment \cite{4567443,6846359}.} As a result, user 1 successfully receives 7 bits. At receiver 2, the bit in power level 6 can be received successfully. Although $x_{2,4}$ and $\underline{x_{1,3}}$ are aligned in power level 1, user 2 can use the knowledge of interfering bit $\underline{x_{1,3}}$ received in power level 2 to obtain its intended bit $x_{2,4}$.\footnote{This process corresponds to Gaussian elimination when computing the ranks in \eqref{eq:case5ba1_user2} and \eqref{eq:case5ba1_user2a} and it should not be confused with any form of SIC.} As a result, user 2 receivers 2 bits. \hspace{104mm}$\blacksquare$
\end{example}

Since TIN has already been proved to be constant-gap optimal in the very weak interference regime \cite{7051266} and in the very strong interference regime \cite{7451210}, hence, we only focus on the weak, strong and mixed interference regimes. The characterizations of the interference regimes for the D-IC follow that of the G-IC in Section \ref{sec:intro}: weak interference regime when $n_{11}\geq n_{21},n_{22}\geq n_{12}$; the strong interference regime when $n_{11}\leq n_{21},n_{22}\leq n_{12}$; and the mixed interference regime when $n_{12}\leq n_{22},n_{21}\geq n_{11}$ or $n_{12}\geq n_{22},n_{21}\leq n_{11}$. We consider $n_{11}\geq n_{22}$ without loss of generality. Since we do not consider the very weak \cite{7051266} and very strong interference regimes \cite{1056416}, $n_{11},n_{22},n_{12}$, and $n_{21}$ should not satisfy either $\min\{n_{11},n_{22}\}\geq n_{12}+n_{21}$ or $\min\{n_{12},n_{21}\}\geq n_{11}+n_{22}$.

Note that the capacity region in Theorem \ref{the:1} is also defined by the convex hull of all corner points (including $(0,0)$). That is, the boundary of the region is formed by all corner points and the line segment between each pair of neighboring corner points. Hence, for the achievability proof, we are targeting on achieving every corner point of the capacity region of the D-IC. Once all the corner points are achieved, the capacity region can then be achieved by the proposed schemes together with time-sharing. For the rest of the proof, we show the detailed design of the generator matrices in Appendix \ref{app:proof}. We also provide the details of all interference subregimes and a pointer to the achievability proof of each subregime in Table \ref{table:result}. Although we focus on achieving corner points, we also showcase that the proposed scheme is capable of achieving all the integer rate pairs inside the capacity region without time-sharing for some subregimes in Appendix \ref{sec:weak1}.

\begin{table}[ht!]
  \centering
 \caption{Interference regimes and schemes}\label{table:result}
\begin{tabular}{|c|c|c|c|c|c|}
\hline
 Interference regimes &  Subregimes   & Scheme  \\
 \hline
  	&	Weak 1: $n_{11}>n_{22}>n_{12}>n_{21}$	  	     & Appendix \ref{sec:weak1} and Table \ref{table:weak1} \\
  Weak: $n_{12}\leq n_{22},n_{21}\leq n_{11}$	&	Weak 2: $n_{11}>n_{22}>n_{21}>n_{12}$  	     & Table \ref{table:weak2} \\
          &		Weak 3: $n_{11}>n_{21}>n_{22}>n_{12}$ & Table \ref{table:weak3}  \\
\hline
  	&	Strong 1: $n_{12}>n_{21}>n_{11}>n_{22}$	  	     & Table \ref{table:strong1} \\
  Strong: $n_{11}\leq n_{21},n_{22}\leq n_{12}$	&	Strong 2: $n_{21}>n_{12}>n_{11}>n_{22}$  	     & Table \ref{table:strong2} \\
          &		Strong 3: $n_{21}>n_{11}>n_{12}>n_{22}$ & Table \ref{table:strong3}  \\
\hline
  	&	Mixed 1: $n_{11}>n_{12}>n_{22}>n_{21}$	  	     & Table \ref{table:mix1} \\
  	&	Mixed 2: $n_{11}>n_{21}>n_{12}>n_{22}$  	     & Table \ref{table:mix2} \\
    Mixed: $n_{12}\leq n_{22},n_{21}\geq n_{11}$      &		Mixed 3: $n_{11}>n_{12}>n_{21}>n_{22}$ & Same as Mixed 2  \\
      or   $n_{12}\geq n_{22},n_{21}\leq n_{11}$        &		Mixed 4: $n_{12}>n_{11}>n_{21}>n_{22}$ & Table \ref{table:mix4}  \\
                          &		Mixed 5: $n_{12}>n_{11}>n_{22}>n_{21}$ & Table \ref{table:mix5}  \\
                                        &		Mixed 6: $n_{21}>n_{11}>n_{22}>n_{12}$ & Table \ref{table:mix6}  \\
\hline
\end{tabular}
\end{table}

\begin{remark}\label{remark2}
We emphasize that the ``achieving corner point'' approach has greatly simplified the achievable schemes and the achievability proof compared to the ``achieving symmetric capacity'' approach in \cite{ShuoLithesis}. This can be seen by looking into the weak and strong interference regimes of the symmetric D-IC, whose symmetric capacity is not a corner point. In order to achieve the symmetric capacity of these two regimes, the author in \cite{ShuoLithesis} had to divide the interference regime into an infinite number of subregimes. As a result, the generator matrices of the corresponding achievable scheme can have infinite number of submatrices (see Eq. $(2.18)$ and Eq. $(2.23)$ of \cite{ShuoLithesis}). In our case, we successfully circumvent this difficulty by targeting on corner points. Consequently, we only need to consider a finite number of sub-regimes and the generator matrices have a finite number of submatrices. In addition, the scheme proposed in \cite{ShuoLithesis} requires two time slots to achieve the symmetric capacity in the aforementioned two regimes (See Eq. $(2.33)$- Eq. $(2.36)$ of \cite{ShuoLithesis}) while all of our capacity-achieving schemes only use one time slot. We will see in Section \ref{sec:RGIC} that these successes in the D-IC also lead to much simpler achievable schemes for the G-IC.
\end{remark}

When characterizing each subregime of the considered above regimes, we only present the relationships between channel parameters as strict inequalities since any relationship with equalities automatically belongs to a special case of that subregime. When the relationship has equalities, the number of rows of some submatrices of $\boldsymbol{G}_k$ become zero. It is also possible that some corner points can become single-user achievable rate points, i.e., $(n_{11},0)$ and $(0,n_{22})$. When this happens, their associated achievable schemes become single-user achievable schemes. Hence, the case $n_{11}=n_{22}=n_{12}=n_{21}$ is excluded as the only corner points are the $(n_{11},0)$ and $(0,n_{22})$, which is off the interest here.


\section{The Real-Valued Gaussian Interference Channel}\label{sec:RGIC}
In this section, we propose purely discrete input distributions by systematically translating the schemes for the D-IC to the real G-IC. The constant-gap optimality of the proposed discrete input distributions is then shown. Unless specified otherwise, the definitions and their associated notations from Section \ref{sec:proposed} are continued to be used for the rest of the paper.

\subsection{Main Result}
We state the main result of this section in the following.
\begin{theorem}\label{the:main1}
For the two-user real G-IC in \eqref{eq:gic_1} and \eqref{eq:gic_2}, there exist a pair of purely discrete input distributions $(\msf{X}_1,\msf{X}_2)$ such that any rate pair inside the capacity region can be achieved to within a constant gap by using TIN together with time-sharing, where the gap is independent of all channel parameters and interference regimes.
\end{theorem}

In what follows, we describe the proposed scheme and the proof for Theorem \ref{the:main1}.

\subsection{Proposed Schemes}
First, we denote the difference between the actual channel value and the corresponding quantized value (in the D-IC) as $\beta_{kk} \triangleq \frac{1}{2}\log_2\SNR_k-n_{kk}$ and $\beta_{k\bar{k}} \triangleq \frac{1}{2}\log_2\INR_{k}-n_{k\bar{k}}$. Thus, $\beta_{kk},\beta_{k\bar{k}} \in [0,1)$.

The proposed distributions in the D-IC are systematically translated into a multi-layer superposition PAM signaling, where each PAM's power level and cardinality can be directly derived from $\boldsymbol{G}_1$ and $\boldsymbol{G}_2$ in our proposed scheme for the D-IC. Specifically, user $k$'s signal is given by
\begin{align}\label{eq:X1}
\msf{X}_k = 2^{-q}\sum_{i_k=1}^{L_k}2^{\sum_{i=j_k+1}^{M_k}\text{row}(\boldsymbol{E}_{k,i})}\rho_{k,i_k}\msf{F}_{k,i_k},
\end{align}
where $2^{-q}$ is the normalization factor for satisfying the power constraint $\mathbb{E}[\|\msf{X}_k\|^2] \leq 1$ as shown in Lemma \ref{lem:normalization} in Appendix \ref{appendix:lemma}, $\text{row}(.)$ outputs the number of rows, $\msf{F}_{k,i_k}$ is a discrete random variable and the cardinality of its support is associated with $\text{rank}(\boldsymbol{F}_{k,i_k})$, $\boldsymbol{E}_{k,j_k} = \boldsymbol{F}_{k,i_k}$ when $f(j_k) = i_k$, $j_k \in \mathcal{J}_k$, and $\boldsymbol{E}_{k,j_k} = \boldsymbol{0}$ when $j_k \in [1:M_k] \setminus \mathcal{J}_k$ following \eqref{eq:EK}, $2^{\sum_{i=j_k+1}^{M_k}\text{row}(\boldsymbol{E}_{k,i})}$ and $\rho_{k,i_k} \in [1,2)$ are the power scaling factor and power adjustment, respectively, for $\msf{F}_{k,i_k}$.

In a type I scheme, $\msf{F}_{k,i_k}\sim \text{PAM}(2^{\text{rank}(\boldsymbol{F}_{k,i_k})-1},1)$ and is uniquely associated with one power scaling factor. Moreover, $\rho_{k,i_k}=1$ (i.e., no power adjustment is required).

In a type II scheme, $\msf{F}_{k,i_k}\sim \text{PAM}(2^{\text{rank}(\boldsymbol{F}_{k,i_k})-2},1)$. If $\boldsymbol{E}_{k,j_k}=\boldsymbol{E}_{k,j'_k}=\boldsymbol{F}_{k,i_k}$ for $j_k \neq j'_k$ and $j_k,j'_k \in \mathcal{J}_k$, then the power scaling factor of $\msf{F}_{k,i_k}$ is $2^{\sum_{i=j_k+1}^{M_k}\text{row}(\boldsymbol{E}_{k,i})}+2^{\sum_{i=j'_k+1}^{M_k}\text{row}(\boldsymbol{E}_{k,i})}$. 
Consider $j_k<j'_k$ without loss of generality. Given user $\bar{k}$'s generator matrix $\boldsymbol{G}_{\bar{k}} = [\boldsymbol{E}^T_{\bar{k},1},\ldots,\boldsymbol{E}^T_{\bar{k},M_{\bar{k}}}]^T$ with binary submatrix $\boldsymbol{E}_{\bar{k},j_{\bar{k}}} = \boldsymbol{F}_{\bar{k},i_{\bar{k}}}$ and $j_{\bar{k}}\in \mathcal{J}_{\bar{k}}$, the power adjustments for any pair of $(\msf{F}_{k,i_k},\msf{F}_{\bar{k},i_{\bar{k}}})$ follow
\begin{align}\label{eq:rho}
&(\rho_{k,i_k},\rho_{\bar{k},i_{\bar{k}}}) =\left\{\begin{array}{ll} (2^{\max\{\beta_{kk},\beta_{k\bar{k}} \}-\beta_{kk}},2^{\max\{\beta_{kk},\beta_{k\bar{k}} \}-\beta_{kk}}),& \text{C1} ,\\
(2^{\max\{\beta_{\bar{k}\bar{k}},\beta_{\bar{k}k} \}-\beta_{\bar{k}k}},2^{\max\{\beta_{\bar{k}\bar{k}},\beta_{\bar{k}k} \}-\beta_{\bar{k}\bar{k}}}), &\text{C2},\\
(1,1), & \text{Otherwise}.
\end{array}\right.\\
\text{C1}:&n_{kk}+\sum\nolimits_{i=j_k+1}^{M_k}\text{row}(\boldsymbol{E}_{k,i}) \leq n_{k\bar{k}}+\sum\nolimits_{i=j_{\bar{k}}+1}^{M_{\bar{k}}}\text{row}(\boldsymbol{E}_{\bar{k},i}) < n_{kk}+\sum\nolimits_{i=j_k}^{M_k}\text{row}(\boldsymbol{E}_{k,i}), \nonumber\\
\text{C2}:&n_{\bar{k}\bar{k}}+\sum\nolimits_{i=j_{\bar{k}}+1}^{M_{\bar{k}}}\text{row}(\boldsymbol{E}_{{\bar{k}},i})
\leq n_{\bar{k}k}+\sum\nolimits_{i=j_k+1}^{M_k}\text{row}(\boldsymbol{E}_{k,i})< n_{\bar{k}\bar{k}}+\sum\nolimits_{i=j_{\bar{k}}}^{M_{\bar{k}}}\text{row}(\boldsymbol{E}_{{\bar{k}},i}). \nonumber
\end{align}
Conditions C1 and C2 in \eqref{eq:rho} correspond to scenarios in the D-IC, where for binary matrix $\boldsymbol{F}_{k,i_{k}}$ in $\boldsymbol{G}_k$, there uniquely exists a matrix consisting of $t+1$ consecutive binary submatrices $[\boldsymbol{F}^T_{\bar{k},i_{\bar{k}}-t},\ldots,\boldsymbol{F}^T_{\bar{k},i_{\bar{k}}}]^T$ in $\boldsymbol{G}_{\bar{k}}$ such that both matrices occupy the same rows which are in a higher position than that of the replica of $\boldsymbol{F}_{k,i_{k}}$ in matrix $[\boldsymbol{A}_k\boldsymbol{G}_k \; \boldsymbol{B}_k\boldsymbol{G}_{\bar{k}}]$ and matrix $[\boldsymbol{A}_{\bar{k}}\boldsymbol{G}_{\bar{k}} \; \boldsymbol{B}_{\bar{k}}\boldsymbol{G}_k]$ (i.e., at receiver $k$ and $\bar{k}$), respectively, and $t\in \{0,1\}$ by design. It should also be noted that either C1 or C2 is active for $\boldsymbol{F}_{k,i_{k}}$ in a type II scheme. Moreover, in our design for any pair of the submatrices that occupy the same rows, only one of them has a replica and occupies two different power levels in the D-IC. We emphasize that the power adjustment is a constant to enforce perfect alignment for $(\msf{F}_{k,i_k},\msf{F}_{\bar{k},i_{\bar{k}}})$ at the receiver when the channel gains are not powers of 2. The detailed explanation is given in Section \ref{sec:type2_dmin}.

For type I and type II schemes, the reduction on the order of the cardinality of a PAM signal is to reduce the extra interference between each PAM signals caused by the mismatch between the expected channel gains (powers of 2) and the actual channel gains. It is also used to avoid the carry over from the signal with two power levels to the signal above in a type II scheme. The details will soon become clear when we analyze the minimum distance. 

\begin{remark}\label{remark_bound_lk}
In our scheme, we translate each non-zero submatrix in $\boldsymbol{G}_k$ into a PAM constellation. Note that each submatrix can be further partitioned into a number of submatrices and can be translated into the superposition of multiple independent PAM constellations. Therefore, in its ultimate form, each non-zero row of $\boldsymbol{G}_k$ could be translated into a binary phase shift keying signal. However, this is completely unnecessary. Moreover, for each independent PAM modulation, we have to added a one-bit (two-bit for type II schemes) guard interval as mentioned above. Hence, in our translation, we tend to keep the number of independent PAM modulations, i.e., $L_k$, small. The largest number of $L_k$ among all of our schemes for the D-IC in Appendix \ref{app:proof} is 9, which occurs for the scheme in \eqref{eq:ex_two_align}.
\end{remark}

\subsection{Proof of Theorem \ref{the:main1}}\label{sec:proof_the2}

It has been shown in \cite{doi:10.1002/ett.1287} that the capacity region of the D-IC $\mathcal{C}_{\text{D-IC}}$ and that of the G-IC $\mathcal{C}_{\text{G-IC}}$ satisfy $\mathcal{C}_{\text{G-IC}} \subseteq \mathcal{C}_{\text{D-IC}} +c$ for some constant $c>0$ for the real G-IC. In what follows, we will show that the rate region $\mathcal{R}^{\text{TIN}}_{\text{G-IC}}$ achieved by our discrete input distribution given in \eqref{eq:X1} with TIN satisfies $\mathcal{C}_{\text{D-IC}} \subseteq \mathcal{R}^{\text{TIN}}_{\text{G-IC}} + c'  $ for some constant $c'>0$. Hence, any rate pair inside the capacity region of the real-valued G-IC can be achieved by our scheme to within a constant gap, i.e., $\mathcal{C}_{\text{G-IC}} \subseteq \mathcal{R}^{\text{TIN}}_{\text{G-IC}} + c''  $ for some constant $c''>0$. Next, we prove the constant gap result in six steps.

\subsubsection{Bounding User $k$'s Achievable Rate as a function of Minimum Distance}
With the channel model in \eqref{eq:gic_1} and \eqref{eq:gic_2}, user $k$'s mutual information is
\begin{align}\label{eq:user1_GC}
I(\msf{X}_k;\msf{Y}_k) =&h(\msf{Y}_k)-h(\msf{Y}_k|\msf{X}_k) \nonumber \\
=&h(h_{kk}\msf{X}_1+h_{k\bar{k}}\msf{X}_{\bar{k}}+\msf{Z}_k)-h(h_{k\bar{k}}\msf{X}_{\bar{k}}+\msf{Z}_k) \nonumber \\
=&h(h_{kk}\msf{X}_1+h_{k\bar{k}}\msf{X}_{\bar{k}}+\msf{Z}_k)-h(\msf{Z}_k)-(h(h_{k\bar{k}}\msf{X}_{\bar{k}}+\msf{Z}_k)-h(\msf{Z}_k)) \nonumber \\
=& I(h_{kk}\msf{X}_k+h_{k\bar{k}}\msf{X}_{\bar{k}};h_{kk}\msf{X}_k+h_{k\bar{k}}\msf{X}_{\bar{k}}+\msf{Z}_k)-I(h_{k\bar{k}}\msf{X}_{\bar{k}};h_{k\bar{k}}\msf{X}_{\bar{k}}+\msf{Z}_k).
\end{align}

To bound $I(h_{kk}\msf{X}_k+h_{k\bar{k}}\msf{X}_{\bar{k}};h_{kk}\msf{X}_k+h_{k\bar{k}}\msf{X}_{\bar{k}}+\msf{Z}_k)$, we first note that
\begin{subequations}\label{eq:decompose_X}
\begin{align}
&h_{kk}\msf{X}_k+h_{k\bar{k}}\msf{X}_{\bar{k}} =\sqrt{\SNR_k}\msf{X}_k+\sqrt{\INR_k}\msf{X}_{\bar{k}}  \label{eq:17}\\
 &= \underbrace{2^{n_{kk}+\beta_{kk}-q}\sum_{i_k=1}^{L_k}2^{\sum_{i=j_k+1}^{M_k}\text{row}(\boldsymbol{E}_{k,i})}\rho_{k,i_k}\msf{F}_{k,i_k}}_{\msf{X}_k^{+}+\msf{X}_k^{-}} 
+\underbrace{2^{n_{k\bar{k}}+\beta_{k\bar{k}}-q}\sum_{i_{\bar{k}}=1}^{L_{\bar{k}}}2^{\sum_{i=j_{\bar{k}}+1}^{M_{\bar{k}}}\text{row}(\boldsymbol{E}_{{\bar{k}},i})}\rho_{\bar{k},i_{\bar{k}}}\msf{F}_{\bar{k},i_{\bar{k}}}}_{\msf{X}_{\bar{k}}^{+}+\msf{X}_{\bar{k}}^{-}}  \label{eq:I1_AB}
\end{align}
\end{subequations}
where in \eqref{eq:I1_AB} we decompose the superimposed signals of users $k$ and $\bar{k}$ from \eqref{eq:17} into two parts, respectively, and the superscripts ``$+$'' and ``$-$'' mean that the signals are above and below the noise level, respectively, from receiver $k$'s perspective. Each decomposed signal has the same form as that of the signal before decomposition, except that the range of $i_k$ and $i_{\bar{k}}$ are from the following subsets of $[1:L_k]$ and $[1:L_{\bar{k}}]$, respectively,
\begin{align}
\mathcal{A}_k &\triangleq \left\{i_k: \sum\nolimits_{i=j_k+1}^{M_k}\text{row}(\boldsymbol{E}_{k,i})\geq q-n_{kk}\right\}, \;\text{for} \; \msf{X}_k^{+}, \label{eq:XA_set_k} \\
\mathcal{B}_k &\triangleq \left\{i_k: \sum\nolimits_{i=j_k+1}^{M_k}\text{row}(\boldsymbol{E}_{k,i})< q-n_{kk}\right\}, \;\text{for} \; \msf{X}_k^{-},  \label{eq:XB_set_k} \\
\mathcal{A}_{\bar{k}} &\triangleq \left\{i_{\bar{k}}: \sum\nolimits_{i=j_{\bar{k}}}^{M_{\bar{k}}}\text{row}(\boldsymbol{E}_{{\bar{k}},i})\geq q-n_{k\bar{k}}\right\},\;\text{for} \; \msf{X}_{\bar{k}}^{+}, \label{eq:XA_set_k1} \\
\mathcal{B}_{\bar{k}} &\triangleq \left\{i_{\bar{k}}: \sum\nolimits_{i=j_{\bar{k}}}^{M_{\bar{k}}}\text{row}(\boldsymbol{E}_{{\bar{k}},i})< q-n_{k\bar{k}}\right\}, \;\text{for} \; \msf{X}_{\bar{k}}^{-}. \label{eq:XB_set_k1}
\end{align}
Note that $|\mathcal{A}_k| \leq L_k$ and $|\mathcal{A}_{\bar{k}}| \leq L_{\bar{k}}$. It is also worth noting that $2^{q-n_{kk}}$ and $2^{q-n_{k\bar{k}}}$ are the noise levels for users $k$ and $\bar{k}$, respectively. This follows the D-IC model in Section \ref{sec:DIC_model}, where $(q-n_{kk})$ and $(q-n_{k\bar{k}})$ bits of $\msf{X}_k$ and $\msf{X}_{\bar{k}}$, respectively, are shifted down to below the noise level at receiver $k$. In fact, the signals above the noise level are those translated from the binary submatrices in $[\boldsymbol{A}_k\boldsymbol{G}_k \; \boldsymbol{B}_k\boldsymbol{G}_{\bar{k}}]$. Next, we give a detailed example to show the signals above and below the noise level for the type II scheme illustrated in Example \ref{example_new}.

\begin{example}\label{example:1}
Consider the scheme in Appendix \ref{sec:weak_12_example}(a). In this case, $q =n_{11}$. We assume $\beta_{11}>\beta_{12}$ and $\beta_{21}>\beta_{22}$. From user 1's perspective,
\begin{align}
\msf{X}^{+}_1 =&\msf{X}_1 = 2^{\beta_{11}}(\msf{F}_{1,6}+2^{n_{12}+n_{21}-n_{22}-t_1}\msf{F}_{1,5}+2^{n_{22}+n_{11}-n_{12}-n_{21}-t_2}\msf{F}_{1,4}
\nonumber \\
&+(2^{n_{11}-n_{21}}+2^{n_{22}+n_{11}-n_{12}-n_{21}})\msf{F}_{1,3}+2^{n_{21}-t_1}\msf{F}_{1,2}+2^{n_{21}}\msf{F}_{1,1}), \\
\msf{X}^{+}_2 =&2^{\beta_{12}}(2^{n_{12}+n_{21}-n_{22}}\msf{F}_{2,2}+2^{n_{22}+n_{11}-n_{12}-n_{21}}\cdot2^{\beta_{11}-\beta_{12}}\msf{F}_{2,1}),\\
\msf{X}^{-}_1 =&0,\\
\msf{X}^{-}_2 =&2^{\beta_{12}}(2^{n_{12}-n_{22}}\msf{F}_{2,4}+2^{2(n_{12}+n_{21}-n_{22})-n_{11}}\msf{F}_{2,3}),
\end{align}
where $(\rho_{1,3},\rho_{2,1})=(2^{\beta_{11}-\beta_{11}},2^{\beta_{11}-\beta_{12}})=(1,2^{\beta_{11}-\beta_{12}})$ and the rest of the power adjustments are 1. From user 2's perspective,
\begin{align}
\msf{X}^{+}_1  =& 2^{\beta_{21}}(2^{n_{12}+2n_{21}-n_{11}-n_{22}-t_1}\msf{F}_{1,5}+2^{n_{22}-n_{12}-t_2}\msf{F}_{1,4}
+(1+2^{n_{22}-n_{12}})\msf{F}_{1,3} \nonumber \\
&+2^{2n_{21}-n_{11}-t_1}\msf{F}_{1,2}+2^{2n_{21}-n_{11}}\msf{F}_{1,1}), \\
\msf{X}^{+}_2 =&\msf{X}_2 =2^{\beta_{22}}(\msf{F}_{2,4}+2^{2n_{21}-n_{11}}\msf{F}_{2,3}+2^{n_{21}}\msf{F}_{2,2}+2^{2n_{22}+n_{11}-2n_{12}-n_{21}}\cdot 2^{\beta_{11}-\beta_{12}}\msf{F}_{2,1}),\\
\msf{X}^{-}_1 =&2^{\beta_{21}-n_{11}}\msf{F}_{1,6},\\
\msf{X}^{-}_2 =&0.
\end{align}
\hspace{16.0cm}$\blacksquare$
\end{example}

With \eqref{eq:decompose_X}, we further bound the mutual information $I(h_{kk}\msf{X}_k+h_{k\bar{k}}\msf{X}_{\bar{k}};h_{kk}\msf{X}_k+h_{k\bar{k}}\msf{X}_{\bar{k}}+\msf{Z}_k)$
\begin{subequations}\label{eq:I1_part1}
\begin{align}
I(h_{kk}\msf{X}_k+h_{k\bar{k}}\msf{X}_{\bar{k}};h_{kk}\msf{X}_k&+h_{k\bar{k}}\msf{X}_{\bar{k}}+\msf{Z}_k)
= h(\msf{X}_k^{+}+\msf{X}_k^{-}+\msf{X}_{\bar{k}}^{+}+\msf{X}_{\bar{k}}^{-}+\msf{Z}_k)-h(\msf{Z}_k)  \\
&\geq h(\msf{X}_k^{+}+\msf{X}_{\bar{k}}^{+}+\msf{Z}_k)-h(\msf{Z}_k) \label{eq:I1_part1_b} \\
&\geq I(\msf{X}_k^{+}+\msf{X}_{\bar{k}}^{+};\msf{X}_k^{+}+\msf{X}_{\bar{k}}^{+}+\msf{Z}_k)  \\
&\geq H(\msf{X}_k^{+}+\msf{X}_{\bar{k}}^{+}) - \frac{1}{2}\log_2 2\pi e \left(\frac{1}{ d^2_{\min}(\msf{X}_k^{+}+\msf{X}_{\bar{k}}^{+})} + \frac{1}{12}\right), \label{eq:I1_part1_c}
\end{align}
\end{subequations}
where the lower bound in \eqref{eq:I1_part1_b} does not result in too much loss since $(\msf{X}_{k}^{-}+\msf{X}_{\bar{k}}^{-})$ is already below the noise level as will be shown shortly, and \eqref{eq:I1_part1_c} follows from an Ozarow-type bound \cite{ozarow90} for the achievable rate of a uniform input distribution over a one-dimensional constellation in \cite[Prop. 1]{7451210}.

\begin{remark}
Although there exist better bounds for the mutual information for discrete inputs (e.g., in \cite{7451210}), we opt to use a type of Ozarow-Wyner bound \cite{ozarow90} due to its simplicity for enabling closed-form analytical computation (see also \cite[Sec. II-A]{7451210}).
\end{remark}

With \eqref{eq:decompose_X}-\eqref{eq:XB_set_k1}, we further bound $I(h_{k\bar{k}}\msf{X}_{\bar{k}};h_{k{\bar{k}}}\msf{X}_{\bar{k}}+\msf{Z}_k)$ from \eqref{eq:user1_GC} as
\begin{subequations}\label{eq:I1_part2}
\begin{align}
&I(h_{k\bar{k}}\msf{X}_{\bar{k}};h_{k{\bar{k}}}\msf{X}_{\bar{k}}+\msf{Z}_k) = h(h_{k\bar{k}}\msf{X}_{\bar{k}}+\msf{Z}_k)-h(\msf{X}^{-}_{\bar{k}}+\msf{Z}_k) 
+h(\msf{X}^{-}_{\bar{k}}+\msf{Z}_k) - h(\msf{Z}_k) \nonumber\\
&=I(\msf{X}^{+}_{\bar{k}};h_{k\bar{k}}\msf{X}_{\bar{k}}+\msf{Z}_k)+I(\msf{X}^{-}_{\bar{k}};\msf{X}^{-}_{\bar{k}}+\msf{Z}_k) \nonumber\\
&\leq  H(\msf{X}^{+}_{\bar{k}})+\frac{1}{2}\log_2(1+\E[\|\msf{X}^{-}_{\bar{k}}\|^2]) \nonumber\\
&\overset{\eqref{eq:XB_set_k1}}{=}H(\msf{X}^{+}_{\bar{k}})+\frac{1}{2}\log_2\bigg(1+\E\bigg[\bigg\|2^{n_{k\bar{k}}+\beta_{k\bar{k}}-q}\sum_{i_{\bar{k}} \in \mathcal{B}_{\bar{k}}}2^{\sum_{i=j_{\bar{k}}+1}^{M_{\bar{k}}}\text{row}(\boldsymbol{E}_{{\bar{k}},i})}\rho_{\bar{k},i_{\bar{k}}}\msf{F}_{\bar{k},i_{\bar{k}}}\bigg\|^2\bigg]\bigg) \nonumber \\
&\leq H(\msf{X}^{+}_{\bar{k}})+\frac{1}{2}\log_2\left(1+\max\{2^{2\beta_{k\bar{k}}}\rho^2_{\bar{k},i_{\bar{k}}}\}2^{2(n_{k\bar{k}}-q)}\E[\|\text{PAM}(2^{q-n_{k\bar{k}}},1)\|^2]\right) \label{eq:I1_part2_a}\\
&\leq H(\msf{X}^{+}_{\bar{k}})+\frac{1}{2}\log_2\left(1+2^{2(n_{k\bar{k}}-q)+4}\cdot \frac{2^{2(q-n_{k\bar{k}})}-1}{12}\right)\label{eq:I1_part2_b} \\
&\leq H(\msf{X}^{+}_{\bar{k}})+\frac{1}{2}\log_2\left(\frac{7}{3}\right),
\end{align}
\end{subequations}
where \eqref{eq:I1_part2_a} follows that there are at most $q-n_{k\bar{k}}$ bits of $\msf{X}_{\bar{k}}$ below the noise level in the D-IC and their corresponding $q-n_{k\bar{k}}$ rows in $\boldsymbol{G}_{\bar{k}}$ can be translated into a $\text{PAM}(2^{q-n_{k\bar{k}}},1)$ which gives the largest possible constellation for $\msf{X}^{-}_{\bar{k}}$, and \eqref{eq:I1_part2_b} follows from \eqref{eq:rho} that $\max\{2^{\beta_{k\bar{k}}}\rho_{\bar{k},i_{\bar{k}}}\} \leq 4$.

Substituting \eqref{eq:I1_part1} and \eqref{eq:I1_part2} into \eqref{eq:user1_GC} gives
\begin{align}\label{eq:user1_rate}
I(\msf{X}_k;\msf{Y}_k) & \geq  H(\msf{X}^{+}_k+\msf{X}^{+}_{\bar{k}}) -  H(\msf{X}^{+}_{\bar{k}})-\frac{1}{2}\log_2 2\pi e \left(\frac{1}{ d^2_{\min}(\msf{X}^{+}_k+\msf{X}^{+}_{\bar{k}})} + \frac{1}{12}\right)-\frac{1}{2}\log_2\left(\frac{7}{3}\right).
\end{align}

In what follows, we analyze the cardinality and the minimum distance of $(\msf{X}^{+}_k+\msf{X}^{+}_{\bar{k}})$. We present the detailed analysis for our proposed two types of schemes separately. This together with \eqref{eq:user1_rate} and the converse bound in \cite{4675741} will allow us to complete the proof of Theorem \ref{the:main1}.

\subsubsection{Bounding the Minimum Distance Under Type I Scheme}\label{sec:tyep1_dmin}
When a type I scheme is used, $(\msf{X}^{+}_k+\msf{X}^{+}_{\bar{k}})$ can be written in the following form
\begin{align}\label{eq:sc_const_temp}
\msf{X}^{+}_k+\msf{X}^{+}_{\bar{k}}&= \sum_{l=1}^LP_l\msf{V}_l 
 \in \sum_{l=1}^L 2^{\sum_{i=1}^l(\alpha_i+m_{i-1})+\beta_l}\Lambda_l\triangleq \Lambda_{\Sigma}, 
\end{align}
where $\msf{V}_l\sim \text{PAM}(2^{m_l-1},1)$ with support $\Lambda_l$, representing \emph{either} $\msf{F}_{k,i_k}$ \emph{or} $\msf{F}_{\bar{k},i_{\bar{k}}}$ with $i_k \in \mathcal{A}_k$ and $i_{\bar{k}} \in \mathcal{A}_{\bar{k}}$, $m_l$ is either $\text{rank}(\boldsymbol{F}_{k,i_k})$ or $\text{rank}(\boldsymbol{F}_{\bar{k},i_{\bar{k}}})$ and $m_0 = 0$, $\rho_l=1$ for type I scheme and thus does not appear in \eqref{eq:sc_const_temp}, $\alpha_l \geq 0$ is the \emph{number of rows} of the \emph{all-zero submatrix} between the binary submatrices associated with $\msf{V}_l$ and $\msf{V}_{l-1}$ in $[\boldsymbol{A}_k\boldsymbol{G}_k \; \boldsymbol{B}_k\boldsymbol{G}_{\bar{k}}]$ in the D-IC, $\beta_l \in \{\beta_{kk},\beta_{k\bar{k}}\}$ is the channel difference associated with $\msf{V}_l$, $P_l \triangleq 2^{\sum_{i=1}^l\alpha_i+m_{i-1}+\beta_l}$ is the \emph{overall power coefficient} including the power level and the channel gain, i.e., $2^{\sum_{i=1}^l\alpha_i+m_{i-1}}$ is either $2^{n_{kk}-q+\sum_{i=j_k+1}^{M_k}\text{row}(\boldsymbol{E}_{k,i})}$ or $2^{n_{k\bar{k}}-q+\sum_{i=j_{\bar{k}}+1}^{M_{\bar{k}}}\text{row}(\boldsymbol{E}_{{\bar{k}},i})}$ and $P_l >P_{l-1}$, $L=|\mathcal{A}_k|+|\mathcal{A}_{\bar{k}}|$, $\Lambda_{\Sigma}$ is the overall constellation. We then have the following proposition for $(\msf{X}^{+}_k+\msf{X}^{+}_{\bar{k}})$.
\begin{proposition}\label{prop:dmin1}
$(\msf{X}^{+}_k+\msf{X}^{+}_{\bar{k}})$ is uniformly distributed over $\Lambda_{\Sigma}$ defined in \eqref{eq:sc_const_temp} satisfying
\begin{align}
&\text{i)}\; |\Lambda_{\Sigma}| = 2^{\sum_{l=1}^Lm_l-L}, \nonumber \\
&\text{ii)}\; d_{\min}(\Lambda_{\Sigma}) = 2^{\alpha_1+\beta_1}\geq 1, \nonumber \\
&\text{iii)}\; 1-2^{\sum_{l=1}^L(m_l+\alpha_l)-1}< \lambda < 2^{\sum_{l=1}^L(m_l+\alpha_l)-1}-1 ,\forall\lambda \in \Lambda_{\Sigma}. \nonumber
\end{align}
\end{proposition}
\begin{IEEEproof}
See Appendix \ref{proof:prop:dmin1}.
\end{IEEEproof}

\begin{figure}[t!]
	\centering
\includegraphics[width=3.43in,clip,keepaspectratio]{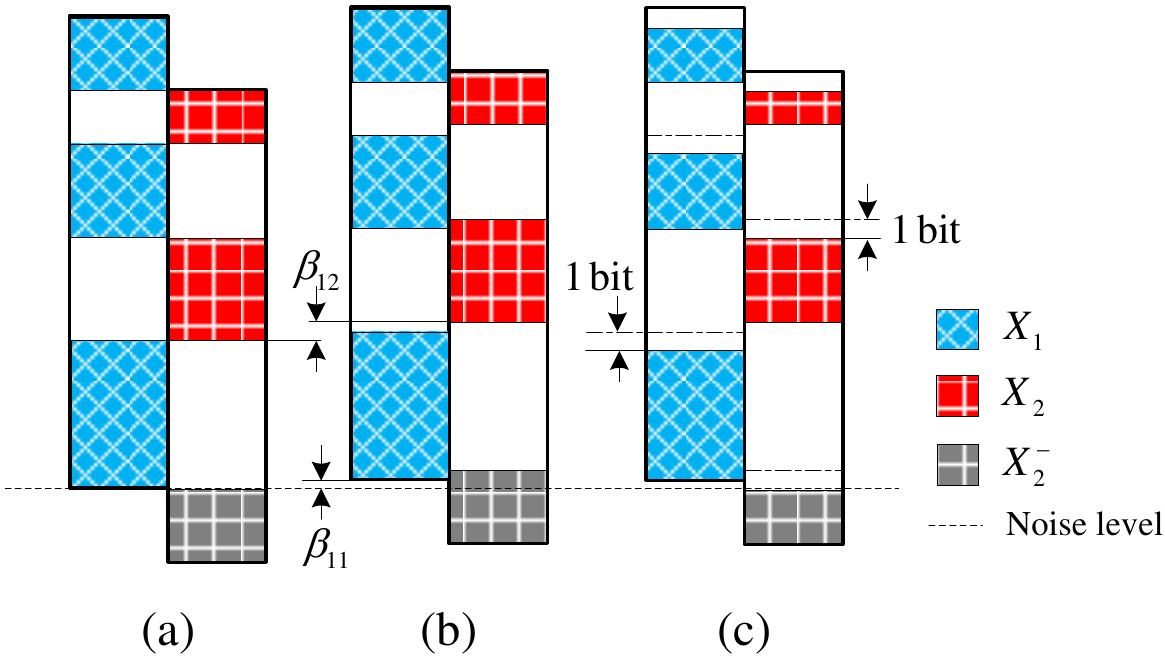}
\caption{Visual illustration of the desired and interference signals observed by receiver 1 when (a) the channel gain is a power of 2 and no reduction on the cardinality, (b) the channel gain is not a power of 2 and there is no `$-1$' to the order of the cardinality, (c) the channel gain is not a power of 2 and with `$-1$' reduction on the cardinality of each signal.}
\label{fig:type_I_rx}
\end{figure}

The intuition behind how the `$-1$' on the cardinality allows us to guarantee non-vanishing minimum distance is illustrated in Fig. \ref{fig:type_I_rx}, where we visualize the desired signals (user 1's signals) and interference (user 2's signals) as well as their power levels from receiver 1's perspective in the D-IC model. Specifically, each signal is represented by a grid, whose position is determined by its power level in the D-IC. As it can be seen in Fig. \ref{fig:type_I_rx}(a), when the channel gain of each signal is a power of 2 with $\beta_l=0$, the minimum distance is not vanished since the desired signal and the interference are disjoint. When the channel gains are not powers of 2 as shown in Fig. \ref{fig:type_I_rx}(b), the interference could be stronger than expected, which shrinks the minimum distance. In Fig. \ref{fig:type_I_rx}(c), the `$-1$' on the cardinality serves as a guard interval to maintain a large minimum distance. Since the power spacing between the least significant bit in the desired signal and most significant bit in the interference signals is determined by the difference $\beta_{11}-\beta_{12} \in (-1,1)$, adding a 1-bit guard interval suffices to guarantee a constant minimum distance even in the worst case.

\subsubsection{Bounding the Achievable Rate of User $k$ Under Type I Scheme}
With Proposition \ref{prop:dmin1} and \eqref{eq:user1_rate}, user $k$'s achievable rate is lower bounded by
\begin{subequations}\label{eq:type1_rate}
\begin{align}
I(\msf{X}_k;\msf{Y}_k) \geq & \bigg(\sum_{i_k \in \mathcal{A}_k} \text{rank}(\boldsymbol{F}_{k,i_k})-|\mathcal{A}_k|+\sum_{i_{\bar{k}} \in \mathcal{A}_{\bar{k}}} \text{rank}(\boldsymbol{F}_{\bar{k},i_{\bar{k}}})-|\mathcal{A}_{\bar{k}}|\bigg) \nonumber \\
&-\bigg(\sum_{i_{\bar{k}} \in \mathcal{A}_{\bar{k}}} \text{rank}(\boldsymbol{F}_{\bar{k},i_{\bar{k}}})-|\mathcal{A}_{\bar{k}}|\bigg) 
-\frac{1}{2}\log_2 2\pi e \left(\frac{1}{ d^2_{\min}(\msf{X}^{+}_k+\msf{X}^{+}_{\bar{k}})} + \frac{1}{12}\right)-\frac{1}{2}\log_2\left( \frac{7}{3}\right) \nonumber \\
=&\sum_{i_k \in \mathcal{A}_k} \text{rank}(\boldsymbol{F}_{k,i_k})- |\mathcal{A}_k|-\frac{1}{2}\log_2 2\pi e \left(\frac{1}{ d^2_{\min}(\msf{X}_A)} + \frac{1}{12}\right)-\frac{1}{2}\log_2\left( \frac{7}{3}\right) \nonumber \\
=& r_k-|\mathcal{A}_k|-\frac{1}{2}\log_2 2\pi e \left(\frac{1}{ d^2_{\min}(\msf{X}^{+}_k+\msf{X}^{+}_{\bar{k}})} + \frac{1}{12}\right)-\frac{1}{2}\log_2\left( \frac{7}{3}\right) \label{eq:type1_rate_d} \\
\geq& r_k -|\mathcal{A}_k|- \frac{1}{2}\log_2 2\pi e \left(\frac{13}{12} \right)-\frac{1}{2}\log_2\left( \frac{7}{3}\right),
\end{align}
\end{subequations}
where \eqref{eq:type1_rate_d} follows from Proposition \ref{remark:type1} that $\sum_{i_k \in \mathcal{A}_k} \text{rank}(\boldsymbol{F}_{k,i_k})=\text{rank}(\boldsymbol{A}_k\boldsymbol{G}_k)=r_k$ in the D-IC, and $|\mathcal{A}_k|$ is due to the ``$-1$'' on the cardinality of $\msf{X}^{+}_k$ and $|\mathcal{A}_k|  \leq L_k$ with $L_k$ upper bounded by a constant as discussed in Remark \ref{remark_bound_lk}.

\begin{remark}
For some cases, it is possible to reduce the gap in \eqref{eq:type1_rate} by using less guard bits. Consider a specific example where $\beta_{11}<\beta_{12}$ and let $\Lambda_l$ and $\Lambda_{l+1}$ be the supports of $\text{PAM}(2^{m_l},1)$ and $\text{PAM}(2^{m_{l+1}},1)$, respectively, which belong to user 1 and user 2, respectively, for some $l \in [1:L]$. According to $(21b)$ from \cite[Proposition 2]{7451210}, we have $d_{\min}(2^{\sum_{i=1}^l\alpha_i+m_{i-1}+\beta_{11}}\Lambda_l+2^{\sum_{i=1}^{l+1}\alpha_i+m_{i-1}+\beta_{12}}\Lambda_{l+1}) = 2^{\sum_{i=1}^l\alpha_i+m_{i-1}+\beta_{11}}d_{\min}(\Lambda_l)$, which is a crucial condition in \eqref{prop:dmin3_con1} of Lemma \ref{prop:dmin3} in Appendix \ref{appendix:lemma} for lower bounding $d_{\min}(\Lambda_{\Sigma})$. Hence, there is no need to reduce $m_l$ (i.e., no guard bits). Moreover, when $\beta_{11}>\beta_{12}$ and $\alpha_{l+1} \geq 1$, the above condition still holds. The introduction of the `$-1$' to the cardinalities of all PAM signals is to universally lower bound $d_{\min}(\Lambda_{\Sigma})$ by a constant regardless of the values of $\beta_l$ and $\alpha_l$.
\end{remark}

\subsubsection{Bounding the Minimum Distance Under Type II Scheme}\label{sec:type2_dmin}
Following Sec. \ref{sec:tyep1_dmin}, when a type II scheme is used, $(\msf{X}^{+}_k+\msf{X}^{+}_{\bar{k}})$ can be written in the following form
\begin{subequations}\label{eq:type2_XA}
\begin{align}
\msf{X}^{+}_k+\msf{X}^{+}_{\bar{k}}=& \sum_{l=1}^LP_l\rho_l\msf{V}_l+\sum_{l' \in \Phi}P_{l''}\rho_{l'}\msf{V}_{l'} \nonumber \\
&= \sum_{l=1}^LP_l\rho_l\msf{V}_l+\sum_{l' \in \Phi}2^{\beta_{l'}-\beta_{\bar{l'}}}P_{\bar{l'}}\rho_{l'}\msf{V}_{l'} \label{eq:type2_XA_temp1} \\
\in& \sum_{l=1}^L 2^{\sum_{i=1}^l(\alpha_i+m_{i-1})+\beta_l}\rho_l \Lambda_l
+\sum_{l' \in \Phi}2^{\sum_{i=1}^{\bar{l'}}(\alpha_i+m_{i-1})+\beta_{l'}}\rho_{l'}\Lambda_{l'}\triangleq \Lambda_{\Sigma},
\end{align}
\end{subequations}
where $\msf{V}_l\sim\text{PAM}(2^{m_l-2},1)$ with support $\Lambda_l$ and $\msf{V}_1,\ldots,\msf{V}_L$ are $L$ \emph{independent} discrete random variables with their respective power coefficient $P_1,\ldots,P_L$, and \eqref{eq:type2_XA_temp1} follows that the power coefficient of $\msf{V}_{l'}$ is $P_{l'}+P_{l''}$ such that $\frac{P_{l''}}{2^{\beta_{l'}}}=\frac{P_{\bar{l'}}}{2^{\beta_{\bar{l'}}}}$, where $\exists l' \in \Phi \subset [1:L],\exists! \bar{l'} \in [1:L] \setminus \Phi$. Specifically, $\msf{V}_{l'}$ and $\msf{V}_{\bar{l'}}$ are a pair of random variables whose corresponding signals in the D-IC occupy the same power level\footnote{\label{footnote1}
As shown in \eqref{eq:rho} and the discussion below, in some of our design, $\msf{V}_{\bar{l'}} = \msf{F}_{k,i_k+1}+ 2^{\text{rank}(\boldsymbol{F}_{k,i_k+1})}\msf{F}_{k,i_k} \in \Lambda_a$ with $\msf{F}_{k,i_k} \sim \text{PAM}(2^{\text{rank}(\boldsymbol{F}_{k,i_k})-2},1)$ and $\msf{F}_{k,i_k+1} \sim \text{PAM}(2^{\text{rank}(\boldsymbol{F}_{k,i_k+1})-2},1)$. To obtain the lower bound on the achievable rate for this case, we always consider $\msf{V}_{\bar{l'}} \sim \text{PAM}(2^{\text{rank}([\boldsymbol{F}_{k,i_k}^T,\boldsymbol{F}_{k,i_k+1}^T]^T)-2},1)$ with support $ \Lambda_b$, such that $\Lambda_a \subset \Lambda_b$, $d_{\min}(\Lambda_a) = d_{\min}(\Lambda_b)=1$, $\max\{\Lambda_a\} < \max\{\Lambda_b\}$, and $4|\Lambda_a| = |\Lambda_b|$. Then for any $\Lambda_c \neq \emptyset$, we have $d_{\min}(\Lambda_c+\Lambda_a)\geq d_{\min}(\Lambda_c+\Lambda_b)$ and $|\Lambda_c+\Lambda_a| \geq \frac{1}{4}|\Lambda_c+\Lambda_b|$.} and their modulation orders satisfy $m_{l'}=m_{\bar{l'}}$, and $|\Phi|\in \{1,2\}$ is by design as in Appendix \ref{app:proof}. Following \eqref{eq:rho}, $(\rho_{l'},\rho_{\bar{l'}}) = (2^{\max\{\beta_{kk},\beta_{k\bar{k}}\}-\beta_{l'}},2^{\max\{\beta_{kk},\beta_{k\bar{k}}\}-\beta_{\bar{l'}}})$ when $P_{l'}<P_{l''}$ and $\rho_l = 1$ otherwise. Note that the power adjustments are to ensure perfect alignment, i.e., $P_{\bar{l'}}\rho_{\bar{l'}} = P_{l''}\rho_{l'}$ in the case of $P_{l'}<P_{l''}$ only when the channel gains are not powers of 2 and do not matter in the case of $P_{l'}>P_{l''}$.

In what follows, we analyze the minimum distance and the cardinality of $(\msf{X}^{+}_k+\msf{X}^{+}_{\bar{k}})$ under three scenarios in \emph{4a)} with $|\Phi|=1$ and $P_{l'}>P_{l''}$, \emph{4b)} with $|\Phi|=1$ and $P_{l'}<P_{l''}$, and in \emph{4c)} with $|\Phi|=2$, respectively. Note that scenarios \emph{4a)} and \emph{4b)} cover the type II schemes translated from \emph{all} the corner point-achieving type II schemes (as well as some integer rate pair-achieving type II schemes) in the D-IC. Thus, the achievable rate analysis of these scenarios together with time-sharing already suffices for deriving the constant-gap result. For the sake of completeness, we still present scenario \emph{4c)} which further covers one type II scheme translated from a particular integer rate pair-achieving type II scheme in the D-IC. These three scenarios cover all the type II schemes in this work.


\emph{4a)} Consider $P_{l'}>P_{l''}$ which implies $P_{l'}>P_{\bar{l'}}$ and $l'>\bar{l'}$. An example of such a scenario can be found in Example \ref{example:1}, where $k=2$, $(\msf{V}_{l'},\msf{V}_{\bar{l'}}) = (\msf{F}_{1,3},\msf{F}_{2,4})$, $P_{l'} = 2^{n_{22}-n_{12}+\beta_{21}}$, $P_{l''} = 2^{\beta_{21}}$, $P_{\bar{l'}} = 2^{\beta_{22}}$ and $(\rho_{l'},\rho_{\bar{l'}})=(\rho_{1,3},\rho_{2,4})=(1,1)$, which is the scheme translated from that in the D-IC for user 2 in \eqref{eq:case5ba1_user2} in Appendix \ref{sec:weak_12_example}a. The superimposed signal in \eqref{eq:type2_XA} becomes
\begin{align}
\msf{X}^{+}_k+\msf{X}^{+}_{\bar{k}}&= \sum_{l=1}^{\bar{l'}-1}\rho_{l}P_l\msf{V}_l+P_{\bar{l'}}\rho_{\bar{l'}}\msf{V}_{\bar{l'}}+\sum_{l=\bar{l'}+1}^{l'-1}\rho_{l}P_l\msf{V}_l + (P_{l''}+P_{l'})\rho_{l'}\msf{V}_{l'}+\sum_{l=l'+1}^L\rho_{l}P_l\rho_l\msf{V}_l  \\
&\in \Lambda_{\Sigma,1}
+ (2^{\sum_{i=1}^{\bar{l'}}\alpha_i+m_{i-1}}+2^{\sum_{i=1}^{l'}\alpha_i+m_{i-1}})2^{\beta_{l'}}\rho_{l'}\Lambda_{l'} 
+ \Lambda_{\Sigma,2}\triangleq \Lambda_{\Sigma}, \label{eq:align_dmin_1a}
\end{align}
where in \eqref{eq:align_dmin_1a} we have used the relationship $P_{l''}=P_{\bar{l'}}2^{\beta_{l'}-\beta_{\bar{l'}}}$ and
\begin{align}
\sum_{l=1}^{\bar{l'}-1}P_l\rho_l\msf{V}_l+P_{\bar{l'}}\rho_{\bar{l'}}\msf{V}_{\bar{l'}}&+\sum_{l=\bar{l'}+1}^{l'-1}P_l\rho_{l}\msf{V}_l \in \Lambda_{\Sigma,1} \triangleq \sum_{l=1}^{l'-1} 2^{\sum_{i=1}^l(\alpha_i+m_{i-1})+\beta_l}\rho_l\Lambda_l,\label{lambda_1} \\
\sum_{l=l'+1}^LP_l\rho_l\msf{V}_l\in\Lambda_{\Sigma,2} &\triangleq \sum_{l=l'+1}^{L} 2^{\sum_{i=1}^l(\alpha_i+m_{i-1})+\beta_l}\rho_l\Lambda_l  \nonumber \\
&= 2^{\sum_{i=1}^{l'}\alpha_i+m_{i-1}}\sum_{l=1}^{L-l'} 2^{\sum_{i=1}^l(\alpha_{l'+i}+m_{l'+i-1})+\beta_{l+l'}}\rho_{l+l'}\Lambda_{l+l'}. \label{lambda_3}
\end{align}
Then, we have the following proposition.
\begin{proposition}\label{prop:type2dmin1}
$(\msf{X}^{+}_k+\msf{X}^{+}_{\bar{k}})$ is uniformly distributed over $\Lambda_{\Sigma}$ defined in \eqref{eq:align_dmin_1a}, and $\Lambda_{\Sigma}$ satisfying
\begin{align}
&\text{i)}\;|\Lambda_{\Sigma}| =  2^{\sum_{l=1}^Lm_l-2L}, \nonumber \\
&\text{ii)}\;d_{\min}(\Lambda_{\Sigma}) =  2^{\alpha_1+\beta_1} \geq 1. \nonumber
\end{align}
\end{proposition}
\begin{IEEEproof}
See Appendix \ref{appD}.
\end{IEEEproof}

\begin{figure}[t!]
	\centering
\includegraphics[width=3.43in,clip,keepaspectratio]{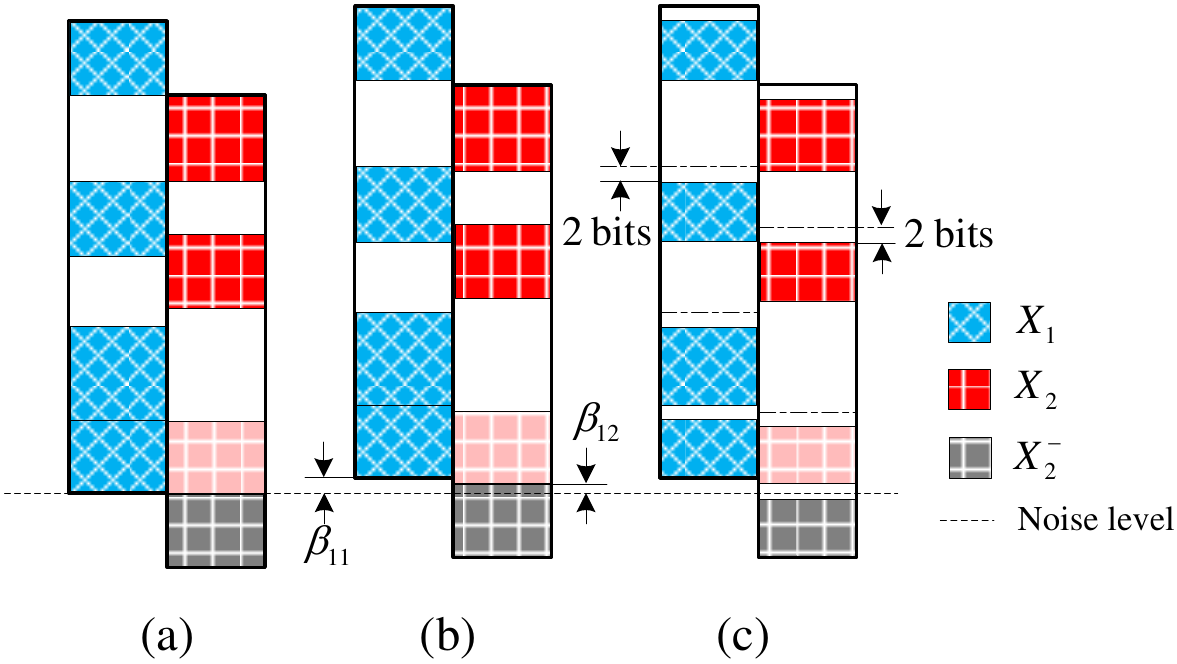}
\caption{The desired and interference signals observed by receiver 1 when (a) each power level is a power of 2 and no reduction on the cardinality, (b) each power level is not a power of 2 and no reduction on the cardinality, (c) each power level is not a power of 2 and with `$-2$' reduction on the cardinality of each signal.}
\label{fig:type_IIa_rx}
\end{figure}

In Fig. \ref{fig:type_IIa_rx}, we give a visual illustration on the effect of `$-2$' bits. As it can be seen from Fig. \ref{fig:type_IIa_rx}(a)-(b), due to the replicated signal occupying the pink power levels, a part of the desired signal and a part of the interference signal occupy the same power level, regardless of whether the channel parameters are powers of 2. 
Hence, the `$-2$' bits serves a larger guard interval to avoid the carry over from the signal with two power levels to the signal above as shown in Fig. \ref{fig:type_IIa_rx}(c).

\begin{remark}
We introduce the `$-2$ bits' reduction to the cardinalities of all the PAM signals for a type II scheme to simplify the presentation of our scheme under G-IC without losing the constant-gap optimality. In practice, the reduction on the cardinality can be less. 
\end{remark}

\emph{4b)} Consider $P_{l'}<P_{l''}$, which implies $P_{l'}<P_{\bar{l'}}$ and $l'<\bar{l'}$. An example of such a scenario can be found in Example \ref{example:1}, where $k=1$, $(\msf{V}_{l'},\msf{V}_{\bar{l'}}) = (\msf{F}_{1,3},\msf{F}_{2,1})$, $P_{l'} = 2^{n_{11}-n_{21}+\beta_{11}}$, $P_{l''} = 2^{n_{11}+n_{22}-n_{12}-n_{21}+\beta_{11}}$, $P_{\bar{l'}} = 2^{n_{11}+n_{22}-n_{12}-n_{21}+\beta_{12}}$ and $(\rho_{l'},\rho_{\bar{l'}})=(\rho_{1,3},\rho_{2,1})=(1,2^{\beta_{11}-\beta_{12}})$, which is the scheme translated from that in the D-IC for user 1 in \eqref{eq:case5ba1} in Appendix \ref{sec:weak_12_example}a. The superimposed signal in \eqref{eq:type2_XA} becomes
\begin{align}
\msf{X}^{+}_k+\msf{X}^{+}_{\bar{k}}=& \sum_{l=1}^{l'-1}P_l\rho_l\msf{V}_l+P_{l'}\rho_{l'}\msf{V}_{l'}+\sum_{l=l'+1}^{\bar{l'}-1}P_l\rho_l\msf{V}_l + P_{\bar{l'}}\rho_{\bar{l'}}\msf{V}_{\bar{l'}}+ P_{l''}\rho_{l'}\msf{V}_{l'}+\sum_{l=\bar{l'}+1}^LP_l\rho_l\msf{V}_l \label{eq:align_dmin_4b} \\
\in& \Lambda_{\Sigma,3}+2^{\sum_{i=1}^{l'}\alpha_i+m_{i-1}+\max\{\beta_{kk},\beta_{k\bar{k}}\}}\Lambda_{l'}
+\Lambda_{\Sigma,4} \nonumber \\
&+ 2^{\sum_{i=1}^{\bar{l'}}\alpha_i+m_{i-1}+\max\{\beta_{kk},\beta_{k\bar{k}}\}}(\Lambda_{\bar{l'}}+\Lambda_{l'}) 
+ \Lambda_{\Sigma,5}\triangleq \Lambda_{\Sigma}, \label{eq:align_dmin_1b}
\end{align}
where in \eqref{eq:align_dmin_1b} we have used the relationships $(\rho_{l'},\rho_{\bar{l'}}) = (2^{\max\{\beta_{kk},\beta_{k\bar{k}}\}-\beta_{l'}},2^{\max\{\beta_{kk},\beta_{k\bar{k}}\}-\beta_{\bar{l'}}})$ and $P_{l''}\rho_{l'}=P_{\bar{l'}}\rho_{\bar{l'}}$, as well as
\begin{align}
\sum_{l=1}^{l'-1}P_l\rho_1\msf{V}_l \in \Lambda_{\Sigma,3} \triangleq& \sum_{l=1}^{l'-1} 2^{\sum_{i=1}^l(\alpha_i+m_{i-1})+\beta_l}\rho_1\Lambda_l, \label{lambda_3aaa}\\
\sum_{l=l'+1}^{l'-1}P_l\rho_1\msf{V}_l\in \Lambda_{\Sigma,4} \triangleq& \sum_{l=l'+1}^{l'-1} 2^{\sum_{i=1}^l(\alpha_i+m_{i-1})+\beta_l}\rho_1\Lambda_l \nonumber\\
= &2^{\sum_{i=1}^{l'}\alpha_i+m_{i-1}}\sum_{l=1}^{\bar{l'}-l'-1} 2^{\sum_{i=1}^l(\alpha_{l'+i}+m_{l'+i-1})+\beta_{l+l'}}\rho_{l+l'}\Lambda_{l+l'}, \label{lambda_2}\\
\sum_{l=\bar{l'}+1}^LP_l\rho_1\msf{V}_l\in\Lambda_{\Sigma,5} \triangleq& \sum_{l=\bar{l'}+1}^{L} 2^{\sum_{i=1}^l(\alpha_i+m_{i-1})+\beta_l}\rho_1\Lambda_l \nonumber\\
= &2^{\sum_{i=1}^{\bar{l'}}\alpha_i+m_{i-1}}\sum_{l=1}^{L-\bar{l'}} 2^{\sum_{i=1}^l(\alpha_{\bar{l'}+i}+m_{\bar{l'}+i-1})+\beta_{l+\bar{l'}}}\rho_{l+\bar{l'}}\Lambda_{l+\bar{l'}}.
\end{align}
We then have the following proposition.
\begin{proposition}\label{prop:type2dmin2}
$(\msf{X}^{+}_k+\msf{X}^{+}_{\bar{k}})$ is uniformly distributed over $\Lambda_{\Sigma}$ defined in \eqref{eq:align_dmin_1b}, and $\Lambda_{\Sigma}$ satisfies the two properties in Proposition \ref{prop:type2dmin1}.
\end{proposition}
\begin{IEEEproof}
See Appendix \ref{appE}.
\end{IEEEproof}

\begin{remark}
The power adjustments are designed to enforce perfect alignment, i.e., $P_{\bar{l'}}\rho_{\bar{l'}} = P_{l''}\rho_{l'}$ in the case of $P_{l'}<P_{l''}$ only. There are many other ways to satisfy this requirement. For example, in Example \ref{example:1}, one can simply let $(\rho_{1,3},\rho_{2,1})=(2^{-\beta_{11}},2^{-\beta_{12}})\in (0,1]^2$. However, at receiver 2 (i.e., scenario $4a)$) we end up having $(2^{\beta_{21}}\rho_{1,3},2^{\beta_{22}}\rho_{2,1})=(2^{\beta_{21}-\beta_{11}},2^{\beta_{22}-\beta_{12}})$ and the $d_{\min}(\Lambda_{\Sigma})$ for user 2 can be reduced if $\beta_{21}<\beta_{11}$ or $\beta_{22}<\beta_{12}$. On the other hand, if the power adjustment is too large, the ``$-2$'' guard bits may not be sufficient to guarantee a constant lower bound on the minimum distance. Therefore, our design in \eqref{eq:rho} ensures that the power adjustments are neither too small nor too large such that $d_{\min}(\Lambda_{\Sigma})$ is lower bounded by a constant under the ``$-2$'' guard bits for all users.
\end{remark}

\emph{4c)} Consider the case of $|\Phi|=2$. The only instance of such a case is the type II scheme in Appendix. \ref{sec:weak_12_example}b. Notice that for both users, $\Lambda_{\Sigma}$ is a mixed of two superimposed constellation from \eqref{eq:align_dmin_1a} and \eqref{eq:align_dmin_1b}, where the signals with one replica are $\msf{F}_{1,5}$ and $\msf{F}_{2,5}$. We prove in Appendix \ref{app:F} that scenario $4c)$ can be equivalently treated as $4b)$ and thus Proposition \ref{prop:type2dmin2} is also applied in this scenario.

\subsubsection{Bounding User $k$'s Achievable Rate Under Type II Scheme}
Equipping with the lower bound on $d_{\min}(\Lambda_{\Sigma})$, we obtain the achievable rate of user $k$ under a type II scheme by following \eqref{eq:type1_rate}
\begin{align}\label{eq:u1_rate_final_1}
I(\msf{X}_k;\msf{Y}_k) \geq  r_k -2|\mathcal{A}_k|- \frac{1}{2}\log_2 2\pi e \left(\frac{13}{12} \right)-\frac{1}{2}\log_2\left( \frac{7}{3}\right),
\end{align}
where 2$|\mathcal{A}_k|$ is due to the $-2$ on the cardinality of user $k$'s signals above the noise level.

\subsubsection{Constant Gap}
As it can be seen from \eqref{eq:type1_rate} and \eqref{eq:u1_rate_final_1} that $r_1-I(\msf{X}_1;\msf{Y}_1)$ and $r_2-I(\msf{X}_2;\msf{Y}_2)$ are upper bounded by two constants, respectively. Hence, our scheme is able to achieve every corner point of $\mathcal{C}_{\text{G-IC}}$ to within a constant gap. This, together with time-sharing, shows that our proposed scheme achieves a rate region $\mathcal{R}_{\text{G-IC}}^{\text{TIN}}$ satisfying $\mathcal{C}_{\text{G-IC}} \subseteq \mathcal{R}_{\text{G-IC}}^{\text{TIN}}+c''$ for some constant $c''>0$. This completes the proof of Theorem \ref{the:main1}.

\section{The Complex Gaussian Interference Channel}\label{sec:G-IC_trans}\label{sec:CGIC}
In this section, we translate the proposed scheme from the D-IC to the discrete input distribution for the complex G-IC and analyze the achievable rate pair.

\subsection{Proposed Scheme}
For the complex G-IC, we similarly denote the difference between the actual channel value and the corresponding quantized value (in the D-IC) as $\beta_{kk} \triangleq \log_2\SNR_k-n_{kk}$ and $\beta_{k\bar{k}} \triangleq \log_2\INR_{k}-n_{k\bar{k}}$. Thus, $\beta_{kk},\beta_{k\bar{k}} \in [0,1)$. We then translate our input distributions for the D-IC in Section \ref{sec:proposed} to obtain the following proposed input signaling, which is a multi-layer superposition of QAM constellation for each user
\begin{align}\label{eq:X1X2_complex}
\msf{X}_k = 2^{-\frac{q}{2}}\sum_{i_k=1}^{L_k}2^{\frac{\sum_{i=j_k+1}^{M_k}\text{row}(\boldsymbol{E}_{k,i})}{2}}\msf{F}_{k,i_k}.
\end{align}
The notations and their definitions here follow from those in \eqref{eq:X1}, except that $2^{-\frac{q}{2}}$ is the normalization factor to ensure $\mathbb{E}[\|\msf{X}_k\|^2] \leq 1$, and $\msf{F}_{k,i_k} \sim \text{QAM}(2^{\text{rank}(\boldsymbol{F}_{k,i_k})},1)$. When using discrete signaling in the complex G-IC, the most difficult problem to deal with is the channel phase distortions in two direct links and cross links. Thus, we ignore the fine tunes, including $-1$ or $-2$ to the cardinality and the power adjustment $\rho_{k,i_k}$, as they may not be effective.

To analyze the achievable rate of user $k$ under our scheme, we start by writing the received signal following \eqref{eq:gic_1}
\begin{align}\label{eq:XA_formal_complex}
&h_{kk}\msf{X}_k+h_{k\bar{k}}\msf{X}_{\bar{k}} =|h_{kk}|e^{j\theta_{kk}}\msf{X}_k+|h_{k\bar{k}}|e^{j\theta_{k\bar{k}}}\msf{X}_{\bar{k}} \nonumber \\
&= \underbrace{2^{\frac{n_{kk}+\beta_{kk}-q}{2}}\sum_{i_k=1}^{L_k}2^{\frac{\sum_{i=j_k+1}^{M_k}\text{row}(\boldsymbol{E}_{k,i})}{2}}e^{j\theta_{kk}}\msf{F}_{k,i_k}}_{\msf{X}_k^{+}+\msf{X}_k^{-}} 
+\underbrace{2^{\frac{n_{k\bar{k}}+\beta_{k\bar{k}}-q}{2}}\sum_{i_{\bar{k}}=1}^{L_{\bar{k}}}2^{\frac{\sum_{i=j_{\bar{k}}+1}^{M_{\bar{k}}}\text{row}(\boldsymbol{E}_{{\bar{k}},i})}{2}}e^{j\theta_{\bar{k}\bar{k}}}\msf{F}_{\bar{k},i_{\bar{k}}}}_{\msf{X}_{\bar{k}}^{+}+\msf{X}_{\bar{k}}^{-}}  ,
\end{align}
where we adopt the same the definitions in \eqref{eq:decompose_X} - \eqref{eq:XB_set_k1} to characterize the signals above and below the noise level.

Similar to the rate analysis for real setting in \eqref{eq:I1_part1} and \eqref{eq:I1_part2}, user $k$'s achievable rate in the complex G-IC is lower bounded by
\begin{subequations}\label{eq:user1_rate_QAM}
\begin{align}
I(\msf{X}_k;\msf{Y}_k)
&\geq I(\msf{X}_k^{+}+\msf{X}_{\bar{k}}^{+};\msf{X}_k^{+}+\msf{X}_{\bar{k}}^{+}+\msf{Z}_k)-H(\msf{X}^{+}_{\bar{k}})-\log_2\left(\frac{7}{3}\right) \nonumber \\
&\geq H(\msf{X}_k^{+}+\msf{X}_{\bar{k}}^{+}) - \log_2 2\pi e \left(\frac{4}{\pi d^2_{\min}(\msf{X}_k^{+}+\msf{X}_{\bar{k}}^{+})} + \frac{1}{4}\right)-H(\msf{X}^{+}_{\bar{k}})-\log_2\left(\frac{7}{3}\right)  \label{eq:user1_rate_QAM_a}\\
&= r_k - \log_2 2\pi e \left(\frac{4}{\pi d^2_{\min}(\msf{X}_k^{+}+\msf{X}_{\bar{k}}^{+})} + \frac{1}{4}\right)+\log_2\left(\frac{7}{3}\right),\label{eq:user1_rate_QAM_b}
\end{align}
\end{subequations}
where \eqref{eq:user1_rate_QAM_a} follows by applying an Ozarow-type bound \cite{ozarow90} for a uniform input distribution over a two-dimensional constellation in Lemma \ref{lma:gap_lattice} in Appendix \ref{appendix:lemma}, and \eqref{eq:user1_rate_QAM_b} follows the same argument used in \eqref{eq:type1_rate_d} under the condition that $|\msf{X}_k^{+}+\msf{X}_{\bar{k}}^{+}| = 2^{r_k}$. In Proposition \ref{prop:zero}, we will show that the condition is true almost everywhere, i.e., the set of phases resulting in overlapping in $(\msf{X}_k^{+}+\msf{X}_{\bar{k}}^{+})$ has Lebesgue measure zero. Then the gap between the capacity and the achievable rate solely depends on the minimum distance of the signal above the noise level from user $k$'s point of view.

\subsection{Minimum Euclidean Distance Analysis}
The signals above the noise level for the complex G-IC in \eqref{eq:gic_1} can be written in the following form based on \eqref{eq:XA_formal_complex}, \eqref{eq:XA_set_k} and \eqref{eq:XA_set_k1}
\begin{align}
\msf{X}_k^{+}+\msf{X}_{\bar{k}}^{+}&= 2^{\frac{\beta_{kk}}{2}}e^{j \theta_{kk}}\sum_{i_k \in\mathcal{A}_k}P_{k,i_k}\msf{F}_{k,i_k} + 2^{\frac{\beta_{k\bar{k}}}{2}}e^{j \theta_{k\bar{k}}}\sum_{i_{\bar{k}} \in\mathcal{A}_{\bar{k}}}P_{{\bar{k}},i_{\bar{k}}}\msf{F}_{{\bar{k}},i_{\bar{k}}},
\end{align}
where $P_{k,i_k}\triangleq 2^{\frac{n_{kk}-q+\sum_{i=j_k+1}^{M_k}\text{row}(\boldsymbol{E}_{k,i})}{2}}$ and $\msf{F}_{k,i_k}\sim \text{QAM}(2^{\text{rank}(\boldsymbol{F}_{k,i_k})},1)$ with support $\Lambda_{k,i_k}$. Moreover, since any phase rotation at the receiver does not lose information, we equivalently consider $e^{-j \theta_{kk}}(\msf{X}_k^{+}+\msf{X}_{\bar{k}}^{+})$, where
\begin{equation}\label{eq:const_XA_equivalent}
    e^{-j \theta_{kk}}(\msf{X}_k^{+}+\msf{X}_{\bar{k}}^{+}) \in 2^{\frac{\beta_{kk}}{2}}\sum_{i_k \in\mathcal{A}_k}P_{k,i_k}\Lambda_{k,i_k} + 2^{\frac{\beta_{k\bar{k}}}{2}}e^{j \theta}\sum_{i_{\bar{k}} \in\mathcal{A}_{\bar{k}}}P_{{\bar{k}},i_{\bar{k}}}\Lambda_{{\bar{k}},i_{\bar{k}}} \triangleq \Lambda_{\Sigma},
\end{equation}
where $\theta\triangleq \theta_{\bar{k}}-\theta_{kk} \in[0,2\pi]$ is the phase difference between $h_{kk}$ and $h_{k\bar{k}}$.

For any $\lambda_{k,i_k} \in \Lambda_{k,i_k}$, we let $\Re(\lambda_{k,i_k}),\Im(\lambda_{k,i_k}) \in \{\pm\frac{1}{2},\ldots,\pm 2^{\frac{\log_2|\Lambda_{k,i_k}|}{2}-1}-\frac{1}{2}\}$ represent the real and imaginary part of a constellation point $\lambda_{k,i_k}$, respectively. $\forall \lambda ,\lambda' \in \Lambda_{\Sigma}$ and $\lambda \neq \lambda'$, the square Euclidean distance between $\lambda$ and $\lambda'$ is
\begin{align}\label{eq:dmin2L}
d^2(\lambda,\lambda')=&\left(\Re\left(2^{\frac{\beta_{kk}}{2}}\sum_{i_k \in\mathcal{A}_k}P_{k,i_k}(\lambda_{k,i_k}-\lambda'_{k,i_k}) + 2^{\frac{\beta_{k\bar{k}}}{2}}e^{j \theta}\sum_{i_{\bar{k}} \in\mathcal{A}_{\bar{k}}}P_{{\bar{k}},i_{\bar{k}}}(\lambda_{{\bar{k}},i_{\bar{k}}}-\lambda'_{{\bar{k}},i_{\bar{k}}})\right)\right)^2\nonumber \\
&+\left(\Im\left(2^{\frac{\beta_{kk}}{2}}\sum_{i_k \in\mathcal{A}_k}P_{k,i_k}(\lambda_{k,i_k}-\lambda'_{k,i_k}) + 2^{\frac{\beta_{k\bar{k}}}{2}}e^{j \theta}\sum_{i_{\bar{k}} \in\mathcal{A}_{\bar{k}}}P_{{\bar{k}},i_{\bar{k}}}(\lambda_{{\bar{k}},i_{\bar{k}}}-\lambda'_{{\bar{k}},i_{\bar{k}}})\right)\right)^2 \nonumber \\
=& 2^{\beta_{kk}}(\Delta_{R_k}^2+ \Delta_{I_k}^2) + 2^{\beta_{k\bar{k}}}(\Delta_{R_{\bar{k}}}^2+ \Delta_{I_{\bar{k}}}^2) 
+2^{\frac{\beta_{kk}+\beta_{k\bar{k}}}{2}+1}\cos\theta(\Delta_{R_k}\Delta_{R_{\bar{k}}}+\Delta_{I_k}\Delta_{I_{\bar{k}}})\nonumber \\
&+2^{\frac{\beta_{kk}+\beta_{k\bar{k}}}{2}+1}\sin\theta(\Delta_{R_{\bar{k}}}\Delta_{I_k}-\Delta_{R_k}\Delta_{I_{\bar{k}}}),\nonumber \\
=&(2^{\frac{\beta_{k{\bar{k}}}}{2}}\Delta_{R_{\bar{k}}}+2^{\frac{\beta_{kk}}{2}}\Delta_{R_k}\cos\theta+2^{\frac{\beta_{kk}}{2}}\Delta_{I_k}\sin\theta)^2\nonumber \\
&+(2^{\frac{\beta_{k\bar{k}}}{2}}\Delta_{I_{\bar{k}}} -2^{\frac{\beta_{kk}}{2}}\Delta_{R_k}\sin\theta+2^{\frac{\beta_{kk}}{2}}\Delta_{I_k}\cos\theta)^2,
\end{align}
where we have used the following definitions
\begin{align}
\Delta_{R_k} &\triangleq \sum_{i_k \in\mathcal{A}_k}P_{k,i_k}\Re(\lambda_{k,i_k}-\lambda'_{k,i_k}) \in \mathbb{Z} , \label{eq:delta_R}\\
\Delta_{I_k} &\triangleq \sum_{i_k \in\mathcal{A}_k}P_{k,i_k}\Im(\lambda_{k,i_k}-\lambda'_{k,i_k}) \in \mathbb{Z},\label{eq:delta_I}
\end{align}
and $\Re(\lambda_{k,i_k}-\lambda'_{k,i_k}),\Im(\lambda_{k,i_k}-\lambda'_{k,i_k}) \in \{0,\pm 1,\ldots, \pm 2^{\frac{\log_2|\Lambda_{k,i_k}|}{2}}-1\}$. The condition $\lambda \neq \lambda'$ guarantees that $\Delta_{R_k}^2+\Delta_{I_k}^2+\Delta_{R_{\bar{k}}}^2+\Delta_{I_{\bar{k}}}^2 \neq 0$. The minimum Euclidean distance of $(\msf{X}_k^{+}+\msf{X}_{\bar{k}}^{+})$ is $d_{\min}(\Lambda_{\Sigma}) = \min\{d(\lambda,\lambda')\}$.

We define the outage probability $\eta\triangleq \text{Pr}\{d_{\min}(\Lambda_{\Sigma}) < d_{\delta} \}$ for a target minimum distance $d_{\delta}>0$.
According to \eqref{eq:dmin2L}, it is obvious that $d_{\min}(\Lambda_{\Sigma}) =0$ if and only if
\begin{align}
2^{\frac{\beta_{k{\bar{k}}}}{2}}\Delta_{R_{\bar{k}}}+2^{\frac{\beta_{kk}}{2}}\Delta_{R_k}\cos\theta+2^{\frac{\beta_{kk}}{2}}\Delta_{I_k}\sin\theta = 0, \label{sol_1a} \\
2^{\frac{\beta_{k\bar{k}}}{2}}\Delta_{I_{\bar{k}}} -2^{\frac{\beta_{kk}}{2}}\Delta_{R_k}\sin\theta+2^{\frac{\beta_{kk}}{2}}\Delta_{I_k}\cos\theta = 0. \label{sol_1b}
\end{align}

In what follows, we show that this event has measure zero.

\begin{proposition}\label{prop:zero}
For the complex G-IC with $(h_{kk},h_{k\bar{k}}) \in \mathbb{C}^2$ with $\theta_{kk},\theta_{k\bar{k}}\in [0,2\pi]$ and by using the scheme in \eqref{eq:X1X2_complex}, the channels such that $d_{\min}(\Lambda_{\Sigma})=0$ for $\Lambda_{\Sigma}$ in \eqref{eq:const_XA_equivalent} have Lebesgue measure zero.
\end{proposition}

\begin{IEEEproof}
Following \eqref{sol_1a} and \eqref{sol_1b}, the conditions that $d_{\min}(\Lambda_{\Sigma}) =0$ are
\begin{align}
\sin\theta &= 2^{\frac{\beta_{k\bar{k}}-\beta_{kk}}{2}}\frac{\Delta_{R_k}\Delta_{I_{\bar{k}}}-\Delta_{I_k}\Delta_{R_{\bar{k}}}}{\Delta^2_{R_k}+\Delta^2_{I_k}} \in [-1,1], \label{sol_2a} \\
\cos\theta &= -2^{\frac{\beta_{k\bar{k}}-\beta_{kk}}{2}}\frac{\Delta_{R_k}\Delta_{R_{\bar{k}}}+\Delta_{I_k}\Delta_{I_{\bar{k}}}}{\Delta^2_{R_k}+\Delta^2_{I_k}} \in [-1,1]  \label{sol_2b}.
\end{align}
Note that when $\Delta^2_{R_k}+\Delta^2_{I_k}=0$, we have that $d_{\min}(\Lambda_{\Sigma})=\min\{2^{\beta_{k\bar{k}}}(\Delta^2_{R_{\bar{k}}}+\Delta^2_{I_{\bar{k}}})\}\geq1$ based on \eqref{eq:dmin2L}-\eqref{eq:delta_I}. Since \eqref{sol_2a} and \eqref{sol_2b} need to satisfy $\sin^2\theta+\cos^2\theta = 1$, we thus obtain
\begin{align}\label{eq:beta1112_solve}
2^{\beta_{k\bar{k}}-\beta_{kk}}=\frac{\Delta^2_{R_k}+\Delta^2_{I_k}}{\Delta^2_{R_{\bar{k}}}+\Delta^2_{I_{\bar{k}}}},
\end{align}
where we note that $\Delta^2_{R_{\bar{k}}}+\Delta^2_{I_{\bar{k}}} = 0$ leads to $d_{\min}(\Lambda_{\Sigma})=\min\{2^{\beta_{kk}}(\Delta^2_{R_{k}}+\Delta^2_{I_{k}})\}\geq1$ based on \eqref{eq:dmin2L}-\eqref{eq:delta_I}.
Substituting \eqref{eq:beta1112_solve} into \eqref{sol_2a} and \eqref{sol_2b} gives
\begin{align}
\sin\theta &= \frac{\Delta_{R_k}\Delta_{I_{\bar{k}}}-\Delta_{I_k}\Delta_{R_{\bar{k}}}}{\sqrt{(\Delta^2_{R_k}+\Delta^2_{I_k})(\Delta^2_{R_{\bar{k}}}+\Delta^2_{I_{\bar{k}}})}}\in [-1,1],  \\
\cos\theta &= -\frac{\Delta_{R_k}\Delta_{R_{\bar{k}}}+\Delta_{I_k}\Delta_{I_{\bar{k}}}}{\sqrt{(\Delta^2_{R_k}+\Delta^2_{I_k})(\Delta^2_{R_{\bar{k}}}+\Delta^2_{I_{\bar{k}}})}}\in [-1,1].
\end{align}
Again since $\Delta_{R_k}$ and $\Delta_{I_k}$ only take value from a subset of the integer set as shown in \eqref{eq:delta_R} and \eqref{eq:delta_I}, the solution set of $\theta$ to \eqref{eq:dmin2L} is a discrete set and thus countable. Hence, $d_{\min}(\Lambda_{\Sigma})=0$ has measure zero.
\end{IEEEproof}


To obtain a closed-form expression for $d_{\min}(\Lambda_{\Sigma})$ is difficult. In what follows, we use some examples to show the values of $d_{\min}(\Lambda_{\Sigma})$ for a number of channel settings.

\begin{example}
Consider a superimposed constellation $\Lambda_{\Sigma} = \sum_{l=1}^3e^{j\theta_l}P_l\Lambda_l$, where $\theta_1 = \theta_3 = \theta_{11},\theta_2 = \theta_{12}$, for $l\in\{1,3\}$, $\Lambda_l$ is the support of $\text{QAM}(2^{m_l},1)$, and $(P_1,P_2,P_3) = (1,2^{\frac{m_1}{2}},2^{\frac{m_1+m_2}{2}})$. The outage probability $\eta$ versus target minimum distance $d_{\delta}$ for various channel settings are shown in Fig. \ref{fig:dmin_1}, where the legend shows the values of $(m_1,m_2,m_3)$.

\begin{figure}[t!]
	\centering
\includegraphics[width=3.43in,clip,keepaspectratio]{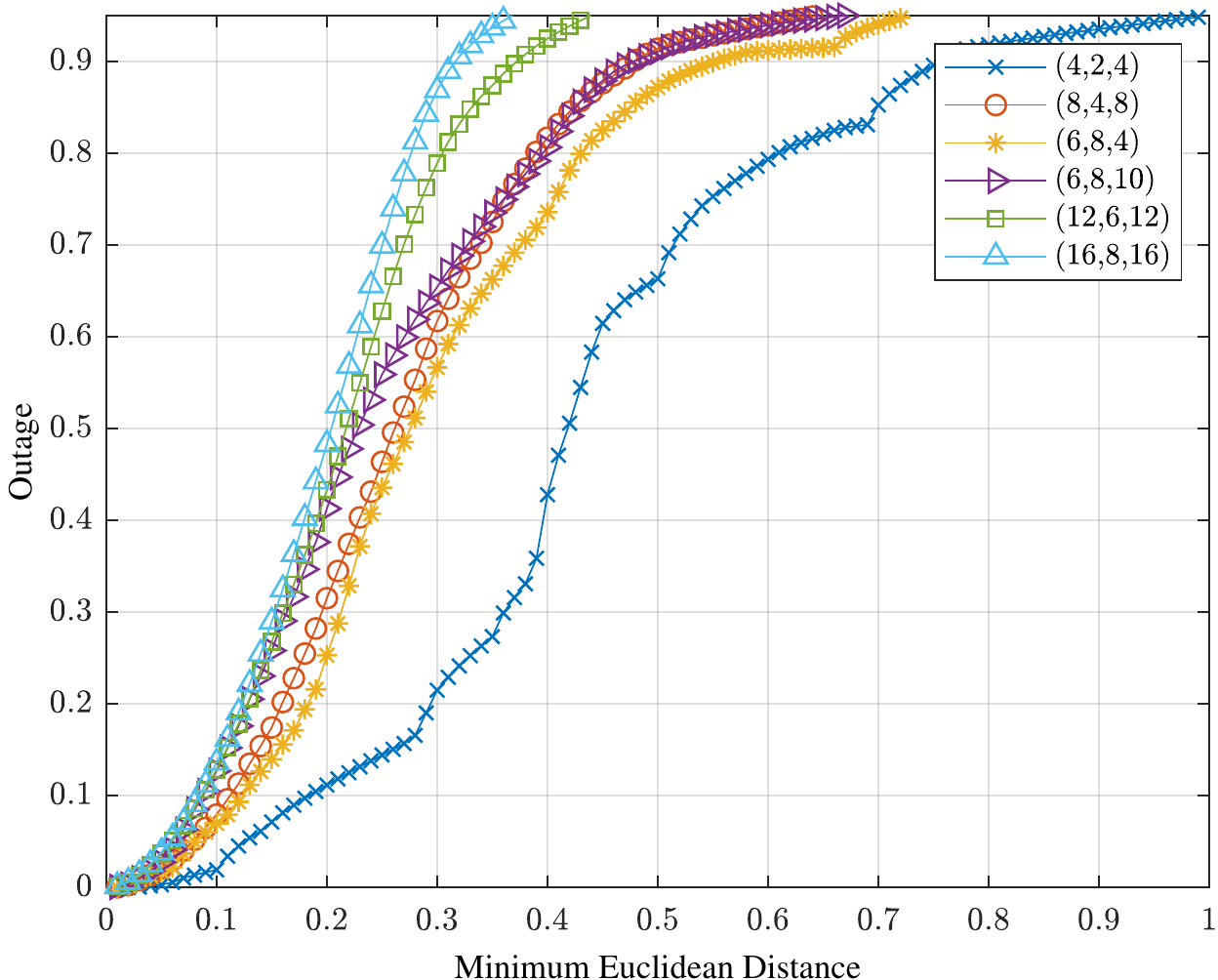}
\caption{Minimum distance and the outage.}
\label{fig:dmin_1}
\end{figure}

From the figure, it can be seen that for a given target outage probability $\eta$, $d_{\min}(\Lambda_{\Sigma})$ is reduced by \emph{at most} about a factor of 2 when the superimposed constellation size $|\Lambda_{\Sigma}| = 2^{m_1+m_2+m_3}$ is increased from $2^{10}$ to $2^{20}$, which is equivalent to about doubling $\max\{\SNR_k,\INR_k\}$ in dB. This is because $|\Lambda_{\Sigma}|$ is at most $\max\{n_{kk},n_{k\bar{k}}\} = \max\{\log_2\SNR_k - \beta_{kk},\log_2\INR_k - \beta_{k\bar{k}}\}$. Moreover, the minimum distance does not reduced much when the overall constellation size is increased from $2^{30}$ to $2^{40}$.
 %
%
%
\end{example}

\subsection{Achievable Rate Pairs Simulation}\label{SIM}

We consider two cases: $(\SNR_1,\INR_1,\SNR_2,\INR_2) = (49,37,43,31)$ and $(25,30,13,17)$ dB, corresponding to $(n_{11},n_{12},n_{22},n_{21})=(16,12,14,10)$ and $(8,10,4,2)$, respectively. The first case belongs to the case of Weak 1-2 in Appendix \ref{sec:weak_12_example} and the second case is Mixed 5-1 in Table \ref{table:mix5} in Appendix \ref{app:corner}. The achievable rate pairs $(I(\msf{X}_1;\msf{Y}_1),I(\msf{X}_2;\msf{Y}_2))$ are averaged over 50000 samples of random channel phases. To put the results of the proposed scheme in context, we also include the capacity outer bound of the complex G-IC from \cite{4675741}, the capacity region of complex D-IC from \cite{doi:10.1002/ett.1287}, the Han–Kobayashi achievable region with Gaussian signaling from \cite{4675741} and the achievable rate of Gaussian signaling with TIN.
\begin{figure}[t!]
	\centering
\includegraphics[width=3.43in,clip,keepaspectratio]{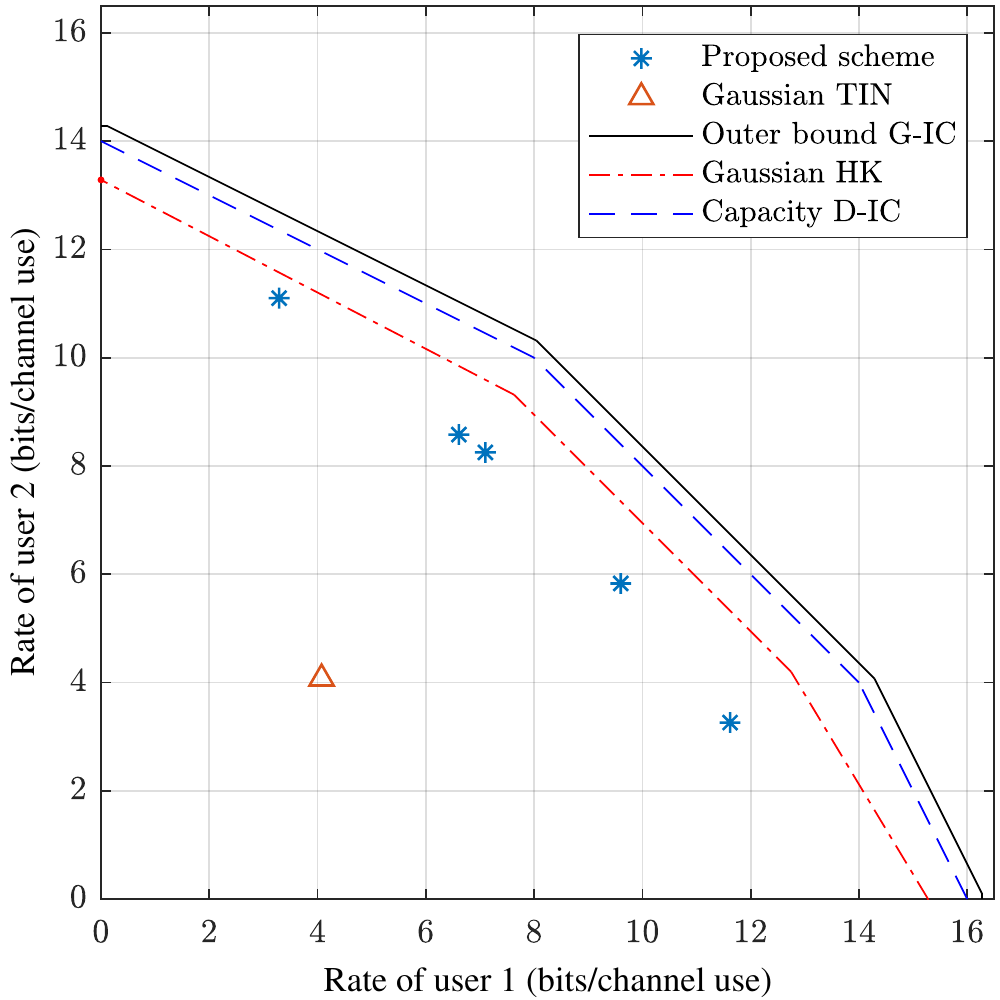}
\caption{Achievable rate pairs for the weak interference regime.}
\label{fig:sim_1}
\end{figure}

\begin{figure}[t!]
	\centering
\includegraphics[width=3.43in,clip,keepaspectratio]{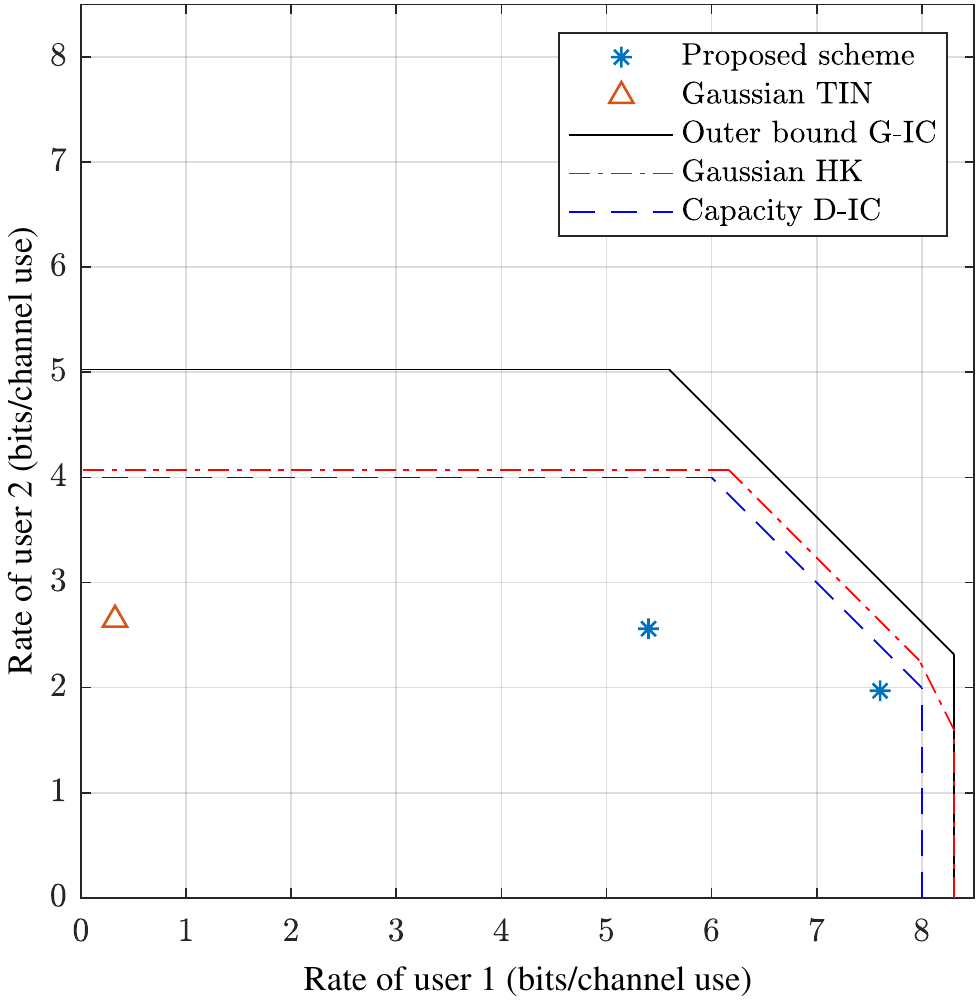}
\caption{Achievable rate pairs for the mixed interference regime.}
\label{fig:sim_2}
\end{figure}

As shown in Fig. \ref{fig:sim_1} and Fig. \ref{fig:sim_2}, our scheme with purely discrete inputs and single-user TIN decoding can operate close to the outer bound of the capacity region of the complex G-IC and that of the complex D-IC for both cases. Notably, our scheme significantly outperforms the conventional approach using Gaussian signaling with single-user TIN decoding in the weak interference regime (Fig. \ref{fig:sim_1}) and for user 1 in the mixed interference regime (Fig. \ref{fig:sim_2}). The reason that user 2's achievable rate is similar to that of the Gaussian TIN is because the interference experience by user 2 is already very weak. In summary, our results indicate that although suffering from random phase distortion introduced by complex channel coefficients, by properly designing the input distributions, the proposed scheme with low-complexity TIN decoding can still be very promising.


\section{Concluding Remarks}\label{sec:conclude}
In this work, we have studied the problem of using discrete signaling with TIN for the real and complex G-IC. Most importantly, we have constructed coding schemes with TIN to achieve the \emph{entire} capacity region of the D-IC for all cases under weak, strong and mixed interference regimes. We have then translated the schemes from the D-IC into the G-IC and provided a systematic way to design discrete input signaling. 
For the real G-IC, we have proved that our scheme is able to achieve any rate pair lying inside the capacity region to within a constant gap, regardless of all channel parameters. For the complex G-IC, the impact of phase distortions on the minimum distance of the proposed scheme has been investigated. Simulation results have been provided to demonstrate substantial gains of the proposed scheme over the existing scheme with TIN. We remark that when translating schemes in the D-IC to that for the G-IC, there are many parameters one can tune to get improved results. However, since our motivation is to showcase the usefulness of discrete input signalling under TIN, we focus on a simple and systematic translation that is analytically tractable and we leave meticulous optimization for future work.

\appendices

\section{Proof of Theorem \ref{the:main} cont.}\label{app:proof}
To clearly express the idea of our approach and in the interest of space, we first give the full achievability proof by showing that the proposed schemes are capable of achieving all the integer rate pairs inside the capacity region of the D-IC for two cases under the weak interference regime in Appendix \ref{sec:weak1}. For the rest of the proof, we focus on achieving all the non-trivial corner points on the capacity region (excluding the trivial points $(n_{11},0),(0,n_{22})$) and provide our choices of $\mathbf{G}_1$ and $\mathbf{G}_2$ in Appendix \ref{app:corner}.


\subsection{Achievability Proof for the Weak Interference Regime}\label{sec:weak1}
We consider a subregime of the weak interference regime, which is defined by $n_{11}> n_{22}>n_{12}> n_{21}$. 


\subsubsection{}\label{weak01}
We further consider the case $n_{11}> n_{12}+n_{21}> n_{22}$. The corner points of this regime are $(n_{11},0), (n_{11}+n_{12}+n_{21}-2n_{22},2n_{22} - n_{12}- n_{21} ),(n_{11}-n_{12}-n_{21},n_{22}), (0,n_{22})$.

\emph{1a) }To achieve the rate pairs between point $(n_{11},0)$ and $(n_{11}+n_{12}+n_{21}-2n_{22},2n_{22} - n_{12}- n_{21} )$, we propose
\begin{align}\label{eq:1aE}
\boldsymbol{G}_1 = \begin{bmatrix}
\boldsymbol{F}_{1,1}\\
\boldsymbol{0}^{t_1,r_1} \\
\boldsymbol{F}_{1,2}\\
\boldsymbol{0}^{t_2,r_1}\\
\boldsymbol{F}_{1,3}
\end{bmatrix},
\boldsymbol{G}_2 = \begin{bmatrix}
\boldsymbol{F}_{2,1}\\
\boldsymbol{0}^{n_{22}-n_{21}-t_2,r_2}\\
\boldsymbol{0}^{n_{12}+n_{21}-n_{22},r_2}\\
\boldsymbol{0}^{n_{22}-n_{12}-t_1,r_2}\\
\boldsymbol{F}_{2,2} \\
\boldsymbol{0}^{n_{11}-n_{22},r_2}
\end{bmatrix}
\end{align}
where $\boldsymbol{F}_{1,1} \in \mathbb{F}_2^{n_{21}-t_1,r_1},\boldsymbol{F}_{1,2} \in \mathbb{F}_2^{n_{11}-n_{21}-n_{12},r_1},\boldsymbol{F}_{1,3}\in \mathbb{F}_2^{n_{12}-t_2,r_1},\boldsymbol{F}_{2,1}\in \mathbb{F}_2^{t_2,r_2},\boldsymbol{F}_{2,2}\in \mathbb{F}_2^{t_1,r_2}$
and $t_1 \in [0:n_{22}-n_{12}],t_2 \in [0:n_{22}-n_{21}]$ are two \emph{independent} variables that are tunable parameters allowing our scheme to achieve all the integer rate pairs between corner points $(n_{11},0)$ and $(n_{11}+n_{12}+n_{21}-2n_{22},2n_{22} - n_{12}- n_{21} )$.

Substituting $\boldsymbol{G}_1$ and $\boldsymbol{G}_2$ into the first term of \eqref{eq:I1} leads to
\begin{align}\label{eq:case5ba1111}
\text{rank}([\boldsymbol{A}_1\boldsymbol{G}_1 \; \boldsymbol{B}_1\boldsymbol{G}_2]) &=\text{rank}\left(\begin{bmatrix}
\left. {\begin{array}{*{20}{c}}
\boldsymbol{F}_{1,1} \\
\boldsymbol{0}^{t_1,r_1} \\
\boldsymbol{F}_{1,2}  \\
\end{array}} \right\} & \boldsymbol{0}^{n_{11}-n_{12},r_2}\\
\boldsymbol{0}^{t_2,r_1} &  \boldsymbol{F}_{2,1} \\
\boldsymbol{F}_{1,3} &  \left\{ {\begin{array}{*{20}{c}}
\boldsymbol{0}^{n_{22}-n_{21}-t_2,r_2}\\
\boldsymbol{0}^{n_{12}+n_{21}-n_{22},r_2}\\
\end{array}} \right.
\end{bmatrix}
\right) 
= \min\{n_{11}-t_1,r_1+r_2\}.
\end{align}
Note that for our proposed $\boldsymbol{G}_1$, it holds that $\text{rank}(\boldsymbol{A}_1\boldsymbol{G}_1)=r_1$ by design. We then have that
\begin{align}\label{eq:criteria:u1}
r_1 = n_{11}-t_1-t_2.
\end{align}
Substituting \eqref{eq:criteria:u1} into \eqref{eq:case5ba1111} gives
\begin{align}\label{eq:case5ba1112}
\text{rank}([\boldsymbol{A}_1\boldsymbol{G}_1 \; \boldsymbol{B}_1\boldsymbol{G}_2]) = \min\{n_{11}-t_1 ,n_{11}-t_1-t_2+r_2\} = n_{11}-t_1.
\end{align}
The last term of \eqref{eq:I1} becomes
\begin{align}\label{eq:case5ba1a111}
\text{rank}(\boldsymbol{B}_1\boldsymbol{G}_2) = \text{rank}(\boldsymbol{F}_{2,1}) = \min\{t_2,r_2\}=t_2.
\end{align}

Substituting \eqref{eq:case5ba1111}, \eqref{eq:case5ba1112} and \eqref{eq:case5ba1a111} into \eqref{eq:I1}, we obtain user 1's rate as
\begin{align}
I(\msf{X}_1;\msf{Y}_1) = n_{11}-t_1-t_2.
\end{align}

For user 2, substituting $\boldsymbol{G}_1$ and $\boldsymbol{G}_2$ into \eqref{eq:I1} gives
\begin{align}
\text{rank}([\boldsymbol{A}_2\boldsymbol{G}_2 \; \boldsymbol{B}_2\boldsymbol{G}_1]) &=\text{rank}\left(\begin{bmatrix}
\left. {\begin{array}{*{20}{c}}
\boldsymbol{0}^{n_{11}-n_{22},r_2} \\
\boldsymbol{F}_{2,1}  \\
\boldsymbol{0}^{n_{22}-n_{21}-t_2,r_2} \\
\end{array}} \right\} & \boldsymbol{0}^{n_{11}-n_{21},r_1}\\
\left. {\begin{array}{*{20}{c}}
\boldsymbol{0}^{n_{12}+n_{21}-n_{22},r_2}\\
\boldsymbol{0}^{n_{22}-n_{12}-t_1,r_2}\\
\end{array}} \right\} & \boldsymbol{F}_{1,1}\\
\boldsymbol{F}_{2,2} & \boldsymbol{0}^{t_1,r_1}\\
\end{bmatrix}
\right) 
= \min\{n_{21}+t_2,r_2+r_1\}.
\end{align}
Similar to the case for user 1, for our designed $\boldsymbol{G}_2$, it holds that
\begin{align}\label{eq:criteria:u2}
\text{rank}(\boldsymbol{A}_2\boldsymbol{G}_2) =  t_1+t_2= r_2.
\end{align}
Note that our design of $\boldsymbol{G}_1$ and $\boldsymbol{G}_2$ ensures that conditions \eqref{eq:criteria:u1} and \eqref{eq:criteria:u2} are satisfied \emph{simultaneously}. Due to \eqref{eq:criteria:u2}, the rank of $[\boldsymbol{A}_2\boldsymbol{G}_2 \; \boldsymbol{B}_2\boldsymbol{G}_1]$ can be written as
\begin{align}
\text{rank}([\boldsymbol{A}_2\boldsymbol{G}_2 \; \boldsymbol{B}_2\boldsymbol{G}_1]) = \min\{n_{21}+t_2,t_1+t_2+r_1\} = n_{21}+t_2.
\end{align}
The last term of \eqref{eq:I1} becomes
\begin{align}
\text{rank}(\boldsymbol{B}_2\boldsymbol{G}_1) = \text{rank}(\boldsymbol{F}_{1,1}) = \min\{n_{21}-t_1,r_1\} = n_{21}-t_1.
\end{align}

Hence, user 2's rate is obtained as
\begin{align}
I(\msf{X}_2;\msf{Y}_2) = t_1+t_2.
\end{align}

It can be easily verified that the above achievable rate pair achieves the corner points $(n_{11},0)$ and $(n_{11}+n_{12}+n_{21}-2n_{22},2n_{22} - n_{12}- n_{21} )$ as well as all the integer rate pairs between them.

To further simplify the proof, we use the following proposition.
\begin{proposition}\label{remark:type1}
User $k$'s mutual information in \eqref{eq:I1} achieves the target rate $r_k$ if $\boldsymbol{G}_k$ and $\boldsymbol{G}_{\bar{k}}$ satisfy the following two properties \\
\textbf{P1.} Each binary submatrix $\boldsymbol{F}_{k,i_k}$ is in a subset of the rows of matrix $[\boldsymbol{A}_k\boldsymbol{G}_k \; \boldsymbol{B}_k\boldsymbol{G}_{\bar{k}}]$ that is not occupied by $\boldsymbol{F}_{\bar{k},i_{\bar{k}}}$; \\
\textbf{P2.} $\text{rank}(\boldsymbol{A}_k\boldsymbol{G}_k) =\text{rank}(\boldsymbol{G}_k)= r_k$.
\end{proposition}
\begin{IEEEproof}
We start from \eqref{eq:I1}
\begin{align}\label{eq:I1_sim_type1}
I(\msf{X}_k;\msf{Y}_k)=&\text{rank}([\boldsymbol{A}_k\boldsymbol{G}_k \; \boldsymbol{B}_k\boldsymbol{G}_{\bar{k}}]) - \text{rank}(\boldsymbol{B}_k\boldsymbol{G}_{\bar{k}}) \nonumber \\
\overset{\textbf{P1.}}{=}&\min\{\text{rank}(\boldsymbol{A}_k\boldsymbol{G}_k)+\text{rank}(\boldsymbol{B}_k\boldsymbol{G}_{\bar{k}}),r_k+r_{\bar{k}}\} 
-\min\{\text{rank}(\boldsymbol{B}_k\boldsymbol{G}_{\bar{k}}),r_{\bar{k}}\} \nonumber \\
\overset{\textbf{P2.}}{=}&\text{rank}(\boldsymbol{A}_1\boldsymbol{G}_k)+\text{rank}(\boldsymbol{B}_k\boldsymbol{G}_{\bar{k}})
-\text{rank}(\boldsymbol{B}_k\boldsymbol{G}_{\bar{k}})  \nonumber  \\
=&\text{rank}(\boldsymbol{A}_k\boldsymbol{G}_{\bar{k}})= r_k.
\end{align}
\end{IEEEproof}


\emph{1b)} To achieve the capacity region between points $(n_{11}+n_{12}+n_{21}-2n_{22},2n_{22} - n_{12}- n_{21})$ and $(n_{11}-n_{12}-n_{21},n_{22})$, we propose
\begin{align}
\boldsymbol{G}_1 = \begin{bmatrix}
\boldsymbol{0}^{t_3,r_1} \\
\boldsymbol{F}_{1,1}\\
\boldsymbol{0}^{n_{22} - n_{12},r_1} \\
\boldsymbol{F}_{1,2}\\
\boldsymbol{0}^{n_{22}-n_{21}+t_3,r_1}\\
\boldsymbol{F}_{1,3}
\end{bmatrix},\boldsymbol{G}_2 = \begin{bmatrix}
\boldsymbol{F}_{2,1}\\
\boldsymbol{0}^{n_{12} +n_{21}- n_{22}-t_3,r_2} \\
\boldsymbol{F}_{2,2}\\
\boldsymbol{0}^{n_{11}-n_{22},r_2}
\end{bmatrix},
\end{align}
where $\boldsymbol{F}_{1,1}\in\mathbb{F}_2^{n_{12}+n_{21}-n_{22}-t_3,r_1},\boldsymbol{F}_{1,2}\in\mathbb{F}_2^{n_{11} -n_{12}-n_{21} ,r_1},\boldsymbol{F}_{1,3}\in\mathbb{F}_2^{n_{12} +n_{21}-n_{22}-t_3,r_1}$, $\boldsymbol{F}_{2,1}\in\mathbb{F}_2^{n_{22} - n_{21}+t_3,r_2}$, \\$\boldsymbol{F}_{2,2}\in\mathbb{F}_2^{n_{22} - n_{12},r_2},t_3 \in[0:n_{12} + n_{21} - n_{22} ]$.

We design $\boldsymbol{G}_1$ and $\boldsymbol{G}_2$ such that {\bf P1.} and {\bf P2.} hold, which can be seen by noting that user 1 and user 2's binary submatrices are disjoint in $[\boldsymbol{A}_1\boldsymbol{G}_1 \; \boldsymbol{B}_1\boldsymbol{G}_2]$, i.e.,
\begin{align}\label{eq:weak1b_u1}
[\boldsymbol{A}_1\boldsymbol{G}_1 \; \boldsymbol{B}_1\boldsymbol{G}_2]=\begin{bmatrix}
\left. {\begin{array}{*{20}{c}}
\boldsymbol{0}^{t_3,r_1} \\
\boldsymbol{F}_{1,1}\\
\boldsymbol{0}^{n_{22} - n_{12},r_1} \\
\boldsymbol{F}_{1,2}  \\
\end{array}} \right\} & \boldsymbol{0}^{n_{11} - n_{12},r_2}\\
\boldsymbol{0}^{n_{22}-n_{21}+t_3,r_1} & \boldsymbol{F}_{2,1}\\
\boldsymbol{F}_{1,3} & \boldsymbol{0}^{n_{12} +n_{21}- n_{22}-t_3,r_2} \\
\end{bmatrix},
\end{align}
and
\begin{align}
\text{rank}(\boldsymbol{A}_1\boldsymbol{G}_1) = n_{11}+n_{12}+n_{21}-2n_{22} - 2t_3 = r_1.
\end{align}

As a result, user 1's rate can be directly obtained by using Proposition \ref{remark:type1} as
\begin{align}
I(\msf{X}_1;\msf{Y}_1) = \text{rank}(\boldsymbol{A}_1\boldsymbol{G}_1) = n_{11}+n_{12}+n_{21}-2n_{22} - 2t_3.
\end{align}

For user 2, notice that
\begin{align}
[\boldsymbol{A}_2\boldsymbol{G}_2 \; \boldsymbol{B}_2\boldsymbol{G}_1] &=\begin{bmatrix}
\left. {\begin{array}{*{20}{c}}
 \boldsymbol{0}^{n_{11} - n_{22},r_1}\\
\boldsymbol{F}_{2,1}\\
 \end{array}} \right\}& \boldsymbol{0}^{n_{11} - n_{21}+t_3,r_2}\\
\boldsymbol{0}^{n_{12} +n_{21}- n_{22}-t_3,r_1} & \boldsymbol{F}_{1,1}\\
\boldsymbol{F}_{2,2} & \boldsymbol{0}^{n_{22} - n_{12},r_2}\\
\end{bmatrix},
\end{align}
and
\begin{align}
\text{rank}(\boldsymbol{A}_2\boldsymbol{G}_2) &= 2n_{22}-n_{12}-n_{21}+t_3 = r_2.
\end{align}
Hence, {\bf P1.} and {\bf P2.} still hold. User 2's rate can then be directly obtained by using Proposition \ref{remark:type1} as
\begin{align}
I(\msf{X}_2;\msf{Y}_2) =\text{rank}(\boldsymbol{A}_2\boldsymbol{G}_2) = 2n_{22}-n_{12}-n_{21}+t_3.
\end{align}

%

\emph{1c) }To achieve the capacity region between $(n_{11}-n_{12}-n_{21},n_{22})$ and $(0,n_{22})$, we propose
\begin{align}
\boldsymbol{G}_1 = \begin{bmatrix}
\boldsymbol{0}^{n_{21},r_1} \\
\boldsymbol{0}^{t_4,r_1} \\
\boldsymbol{F}_{1,1}\\
\boldsymbol{0}^{n_{12} ,r_1}\\
\end{bmatrix},
\boldsymbol{G}_2 = \begin{bmatrix}
\boldsymbol{F}_{2,1}\\
\boldsymbol{F}_{2,2}\\
\boldsymbol{0}^{n_{11}-n_{22},r_2}
\end{bmatrix},
\end{align}
where $\boldsymbol{F}_{1,1}\in\mathbb{F}_2^{n_{11}-n_{12}-n_{21}-t_4,r_1},\boldsymbol{F}_{2,1}\in\mathbb{F}_2^{n_{12},r_2},\boldsymbol{F}_{2,2}\in\mathbb{F}_2^{n_{22} - n_{12},r_2},t_4 \in[0:n_{11} - n_{12} - n_{21} ]$.

For user 1, we note that {\bf P1.} and {\bf P2.} hold since
\begin{align}
[\boldsymbol{A}_1\boldsymbol{G}_1 \; \boldsymbol{B}_1\boldsymbol{G}_2] =\begin{bmatrix}
\left. {\begin{array}{*{20}{c}}
\boldsymbol{0}^{n_{21},r_1}  \\
\boldsymbol{0}^{t_4,r_1} \\
\boldsymbol{F}_{1,1}   \\
 \end{array}} \right\}& \boldsymbol{0}^{n_{11} - n_{12},r_2} \\
\boldsymbol{0}^{n_{12} ,r_1} & \boldsymbol{F}_{2,1}\\
\end{bmatrix}.
\end{align}

Thus, user 1's achievable rate can be obtained by using Proposition \ref{remark:type1} as
\begin{align}
I(\msf{X}_1;\msf{Y}_1) = n_{11}-n_{12}-n_{21} - t_4.
\end{align}

For user 2, {\bf P1.} and {\bf P2.} still hold since
\begin{align}
[\boldsymbol{A}_2\boldsymbol{G}_2 \; \boldsymbol{B}_2\boldsymbol{G}_1] =\begin{bmatrix}
\left. {\begin{array}{*{20}{c}}
\boldsymbol{0}^{n_{11}-n_{22},r_1} \\
\boldsymbol{F}_{2,1}\\
\boldsymbol{F}_{2,2}\\
\end{array}} \right\} & \boldsymbol{0}^{n_{11},r_2}
\end{bmatrix}.
\end{align}

Thus, user 2's achievable rate can be obtained by using Proposition \ref{remark:type1} as
\begin{align}
I(\msf{X}_2;\msf{Y}_2) = n_{22}.
\end{align}

\subsubsection{}\label{sec:weak_12_example}
We now consider $n_{11}<n_{12}+n_{21}$ and $n_{11}+n_{22} - n_{12} - 2n_{21}<0$. The corner points on the corresponding capacity region are $(n_{11},0),(n_{11}+n_{12} - n_{22},2(n_{22}-n_{12})),(2(n_{11}-n_{12}),n_{22}+n_{12}-n_{11}),(0,n_{22})$. We further consider $2(n_{12}+n_{21}-n_{22})-n_{11}<0$ which implies that $2n_{11}+n_{22}-2n_{12}-2n_{21}>0$ because $n_{11}>n_{22}$. To achieve the capacity for this subregime, we use a type II scheme.

\emph{2a)} To achieve the region between $(n_{11},0),(n_{11}+n_{12} - n_{22},2(n_{22}-n_{12}))$, we propose
\begin{align}\label{eq:weak1_2a_E}
\boldsymbol{G}_1=
\begin{bmatrix}
\boldsymbol{F}_{1,1}\\
\boldsymbol{F}_{1,2} \\
\boldsymbol{F}_{1,3}\\
\boldsymbol{F}_{1,4}\\
\boldsymbol{0}^{2n_{22}+n_{11}-2n_{12}-2n_{21}-t_2 ,r_1}\\
\boldsymbol{F}_{1,5} \\
\boldsymbol{F}_{1,3} \\
\boldsymbol{F}_{1,6}
\end{bmatrix},
\boldsymbol{G}_2= \begin{bmatrix}
\boldsymbol{0}^{n_{12}-n_{21},r_2} \\
\boldsymbol{0}^{t_1,r_2} \\
\boldsymbol{F}_{2,1} \\
\boldsymbol{0}^{t_2,r_2} \\
\boldsymbol{F}_{2,2} \\
\boldsymbol{0}^{n_{12}+n_{21}-n_{22},r_2}\\
\boldsymbol{0}^{t_2,r_2} \\
\boldsymbol{F}_{2,3} \\
\boldsymbol{0}^{t_1,r_2} \\
\boldsymbol{F}_{2,4} \\
\boldsymbol{0}^{n_{11}-n_{22},r_2}
\end{bmatrix},
\end{align}
where $\boldsymbol{F}_{1,1},\boldsymbol{F}_{1,6}\in \mathbb{F}_2^{n_{11}-n_{21},r_1},\boldsymbol{F}_{1,2},\boldsymbol{F}_{1,5}\in \mathbb{F}_2^{t_1 ,r_1},\boldsymbol{F}_{1,3}\in \mathbb{F}_2^{2n_{21}+n_{12}-n_{11}-n_{22}-t_1,r_1},\boldsymbol{F}_{1,4}\in \mathbb{F}_2^{t_2 ,r_1}$, \\$\boldsymbol{F}_{2,1},\boldsymbol{F}_{2,4}\in \mathbb{F}_2^{2n_{21}+n_{12}-n_{11}-n_{22}-t_1,r_2},\boldsymbol{F}_{2,2},\boldsymbol{F}_{2,3}\in \mathbb{F}_2^{2n_{22}+n_{11}-2n_{12}-2n_{21}-t_2 ,r_2}$, and $t_1\in [0:2n_{21}+n_{12}-n_{11}-n_{22}],t_2\in[0:2n_{22}+n_{11}-2n_{12}-2n_{21}]$.

Substituting $\boldsymbol{G}_1$ and $\boldsymbol{G}_2$ into the first term of \eqref{eq:I1} leads to
\begin{align}\label{eq:case5ba1}
\text{rank}([\boldsymbol{A}_1\boldsymbol{G}_1 \; \boldsymbol{B}_1\boldsymbol{G}_2])&= 
\text{rank}\left(\begin{bmatrix}
\boldsymbol{F}_{1,1} & \left\{ {\begin{array}{*{20}{c}}
\boldsymbol{0}^{n_{11}-n_{12},r_2}\\
\boldsymbol{0}^{n_{12}-n_{21},r_2}\\
\end{array}} \right.\\
\boldsymbol{F}_{1,2} & \boldsymbol{0}^{t_1,r_2}\\
\boldsymbol{F}_{1,3}& \boldsymbol{F}_{2,1}\\
\boldsymbol{F}_{1,4} & \boldsymbol{0}^{t_2,r_2}\\
\boldsymbol{0}^{2n_{22}+n_{11}-2n_{12}-2n_{21}-t_2 ,r_1}& \boldsymbol{F}_{2,2}\\
\left. {\begin{array}{*{20}{c}}
\boldsymbol{F}_{1,5} \\
\boldsymbol{F}_{1,3} \\
\boldsymbol{F}_{1,6}
\end{array}} \right\} &\boldsymbol{0}^{n_{12}+n_{21}-n_{22},r_2}\\
\end{bmatrix}
\right)
= \min\{n_{11},r_1+r_2\}.
\end{align}
In this type II scheme, we introduce a correlation in $\boldsymbol{G}_1$ such that $\boldsymbol{F}_{1,3}$ appears in two different power levels. Moreover, we still design the generator matrices such that \textbf{P2.} holds and thus
\begin{align}
\text{rank}(\boldsymbol{A}_1\boldsymbol{G}_1)&= \text{rank}\left(\begin{bmatrix}
\boldsymbol{F}_{1,1}\\
\boldsymbol{F}_{1,2} \\
\boldsymbol{F}_{1,3}\\
\boldsymbol{F}_{1,4}\\
\boldsymbol{0}^{2n_{22}+n_{11}-2n_{12}-2n_{21}-t_2 ,r_1}\\
\boldsymbol{F}_{1,5} \\
\boldsymbol{F}_{1,3} \\
\boldsymbol{F}_{1,6}
\end{bmatrix}\right)  \\
& = n_{11}+n_{12}-n_{22}+t_1+t_2 = r_1,\label{eq:a}
\end{align}
where $\eqref{eq:a}$ follows that these two matrices $\boldsymbol{F}_{1,3}$ are exactly the same matrix (linearly dependent). Then, the rank of $[\boldsymbol{A}_1\boldsymbol{G}_1 \; \boldsymbol{B}_1\boldsymbol{G}_2]$ in \eqref{eq:case5ba1} becomes
\begin{align}
\text{rank}([\boldsymbol{A}_1\boldsymbol{G}_1 \; \boldsymbol{B}_1\boldsymbol{G}_2]) = \min\{n_{11},n_{11}+n_{12}-n_{22}+t_1+t_2+r_2\}=n_{11}.
\end{align}
The last term of \eqref{eq:I1} becomes
\begin{align}\label{eq:case5ba1a1}
\text{rank}(\boldsymbol{B}_1\boldsymbol{G}_2) &= \text{rank}\left(\begin{bmatrix}
\boldsymbol{F}_{2,1}\\
\boldsymbol{F}_{2,2}\\
\end{bmatrix}
\right) 
=n_{22}-n_{12}-t_1-t_2.
\end{align}

Substituting \eqref{eq:case5ba1}-\eqref{eq:case5ba1a1} into \eqref{eq:I1} gives
\begin{align}
I(\msf{X}_1;\msf{Y}_1) = n_{11}+n_{12}-n_{22}+t_1+t_2.
\end{align}

For user 2, substituting $\boldsymbol{G}_1$ and $\boldsymbol{G}_2$ into \eqref{eq:I1} gives
\begin{align}\label{eq:case5ba1_user2}
\text{rank}([\boldsymbol{A}_2\boldsymbol{G}_2 \; \boldsymbol{B}_2\boldsymbol{G}_1]) &=
\text{rank}\left(\begin{bmatrix}
\left. {\begin{array}{*{20}{c}}
\boldsymbol{0}^{n_{11}-n_{22},r_2} \\
\boldsymbol{0}^{n_{12}-n_{21},r_2}\\
\boldsymbol{0}^{t_1,r_2}\\
\boldsymbol{F}_{2,1} \\
\boldsymbol{0}^{t_2,r_2}\\
\boldsymbol{F}_{2,2}
\end{array}} \right\} & \boldsymbol{0}^{n_{11}-n_{21},r_1} \\
\boldsymbol{0}^{n_{12}+n_{21}-n_{22},r_2}& \left\{ {\begin{array}{*{20}{c}}
\boldsymbol{F}_{1,1}\\
\boldsymbol{F}_{1,2}\\
\boldsymbol{F}_{1,3}
\end{array}} \right.\\
\boldsymbol{0}^{t_2,r_2} &\boldsymbol{F}_{1,4}\\
\boldsymbol{F}_{2,3} &\boldsymbol{0}^{2n_{22}+n_{11}-2n_{12}-2n_{21}-t_2 ,r_1}\\
\boldsymbol{0}^{t_1,r_2} &\boldsymbol{F}_{1,5}\\
\boldsymbol{F}_{2,4} &\boldsymbol{F}_{1,3}
\end{bmatrix}
\right)\nonumber \\
&= \min\{n_{22}+n_{21}-n_{12}-t_1-t_2,r_2+r_1\}.
\end{align}
Similar to user 1, we design the generator matrices such that {\bf P2.} holds. 
As a result, the rank of $[\boldsymbol{A}_2\boldsymbol{G}_2 \; \boldsymbol{B}_2\boldsymbol{G}_1]$ in \eqref{eq:case5ba1_user2} becomes
\begin{align}
\text{rank}([\boldsymbol{A}_2\boldsymbol{G}_2 \; \boldsymbol{B}_2\boldsymbol{G}_1])= n_{22}+n_{21}-n_{12}-t_1-t_2.
\end{align}
The last term of \eqref{eq:I1} becomes
\begin{align}
\text{rank}(\boldsymbol{B}_2\boldsymbol{G}_1) &= \text{rank}\left(\begin{bmatrix}
\boldsymbol{F}_{1,1} \\
\boldsymbol{F}_{1,2} \\
\boldsymbol{F}_{1,3} \\
\boldsymbol{F}_{1,4} \\
\boldsymbol{F}_{1,5} \\
\boldsymbol{F}_{1,3}
\end{bmatrix}
\right)=  n_{21}+n_{12}-n_{22}+t_1+t_2. \label{eq:case5ba1_user2a}
\end{align}


Hence, user 2's rate is obtained as
\begin{align}
I(\msf{X}_2;\msf{Y}_2) =2(n_{22}-n_{12}-t_1-t_2).
\end{align}

Notice that the rank of the replicated $\boldsymbol{F}_{k,i_k}$ does not contribute to the rank of $\boldsymbol{G}_k$. Moreover, \emph{either} $\boldsymbol{F}_{k,i_k}$ \emph{or} its replica and a binary submatrix of user $\bar{k}$ occupy the same subset of rows of matrix $[\boldsymbol{A}_k\boldsymbol{G}_k \; \boldsymbol{B}_k\boldsymbol{G}_{\bar{k}}]$. As a result, the calculation of $\text{rank}([\boldsymbol{A}_1\boldsymbol{G}_k \; \boldsymbol{B}_k\boldsymbol{G}_{\bar{k}}])$ under a type II scheme can be made equivalently to that of a type I scheme as if the aligned $\boldsymbol{F}_{k,i_k}$ is replaced by $\boldsymbol{0}$. This, with property {\bf P2.} induced by our design, guarantees that Proposition \ref{remark:type1} still holds for all type II schemes.

\emph{2b)} To achieve the region between $(n_{11}+n_{12} - n_{22},2(n_{22}-n_{12}))$ and $(2(n_{11}-n_{12}),n_{22}+n_{12}-n_{11})$, we propose
\begin{align}\label{eq:ex_two_align}
\boldsymbol{G}_1 = \begin{bmatrix}
\boldsymbol{F}_{1,1}\\
\boldsymbol{F}_{1,2} \\
\boldsymbol{F}_{1,3} \\
\boldsymbol{0}^{t_3,r_1}\\
\boldsymbol{F}_{1,4} \\
\boldsymbol{0}^{t_4,r_1}\\
\boldsymbol{F}_{1,5} \\
\boldsymbol{0}^{2n_{22}+n_{11}-2n_{12}-2n_{21},r_1}\\
\boldsymbol{0}^{t_4,r_1}\\
\boldsymbol{F}_{1,5} \\
\boldsymbol{F}_{1,6} \\
\boldsymbol{0}^{t_5,r_1}\\
\boldsymbol{F}_{1,7} \\
\boldsymbol{F}_{1,8}\\
\boldsymbol{F}_{1,9}
\end{bmatrix},
\boldsymbol{G}_2 = \begin{bmatrix}
\boldsymbol{F}_{2,1} \\
\boldsymbol{0}^{n_{12}-n_{21}-t_3,r_2} \\
\boldsymbol{F}_{2,2} \\
\boldsymbol{F}_{2,3} \\
\boldsymbol{F}_{2,4} \\
\boldsymbol{F}_{2,5} \\
\boldsymbol{0}^{2n_{21}+n_{12}-n_{11}-n_{22}-t_4,r_2} \\
\boldsymbol{0}^{2n_{11}+n_{22}-2n_{12}-2n_{21},r_2} \\
\boldsymbol{F}_{2,6} \\
\boldsymbol{0}^{n_{12}-n_{21}-t_5,r_2}\\
\boldsymbol{F}_{2,5} \\
\boldsymbol{0}^{2n_{21}+n_{12}-n_{11}-n_{22}-t_4,r_2} \\
\boldsymbol{F}_{2,7} \\
\boldsymbol{F}_{2,8} \\
\boldsymbol{F}_{2,9} \\
\boldsymbol{0}^{n_{11}-n_{22},r_2}
\end{bmatrix},
\end{align}
where $\boldsymbol{F}_{1,1}\in\mathbb{F}_2^{t_4,r_1},\boldsymbol{F}_{1,2},\boldsymbol{F}_{1,5},\boldsymbol{F}_{1,9} \in \mathbb{F}_2^{2n_{21}+n_{12}-n_{11}-n_{22}-t_4,r_1},\boldsymbol{F}_{1,3},\boldsymbol{F}_{1,6} \in \mathbb{F}_2^{2n_{11}+n_{22}-2n_{12}-2n_{21},r_1} ,\boldsymbol{F}_{1,4}\in \mathbb{F}_2^{n_{12}-n_{21}-t_3,r_1},\boldsymbol{F}_{1,7} \in \mathbb{F}_2^{n_{12}-n_{21}-t_5,r_1},\boldsymbol{F}_{1,8}\in \mathbb{F}_2^{t_4,r_1},\boldsymbol{F}_{2,1} \in \mathbb{F}_2^{t_3,r_2},\boldsymbol{F}_{2,2},\boldsymbol{F}_{2,5},\boldsymbol{F}_{2,8}\in \mathbb{F}_2^{t_4,r_2},\boldsymbol{F}_{2,3},\boldsymbol{F}_{2,9} \in \mathbb{F}_2^{2n_{21}+n_{12}-n_{11}-n_{22}-t_4,r_2},\boldsymbol{F}_{2,4},\boldsymbol{F}_{2,7}\in \mathbb{F}_2^{2n_{22}+n_{11}-2n_{12}-2n_{21},r_2},\boldsymbol{F}_{2,6} \in \mathbb{F}_2^{t_5,r_2}$, and $t_3,t_5\in [0:n_{12}-n_{21}],t_4 \in [0:2n_{21}+n_{12}-n_{11}-n_{22}]$. Here, the value of $t_5$ depends on the value of $t_3$, i.e., $t_5 = 0$ when $t_3<n_{12}-n_{21}$ and $t_5$ can take any value from $[0:n_{12}-n_{21}]$ when $t_3=n_{12}-n_{21}$. The dependence of $t_5$ on $t_3$ ensures that $\boldsymbol{F}_{2,6}$ and $\boldsymbol{F}_{1,4}$ are disjoint at receiver 2, which will be shown in \eqref{eq:type2_example2}.

For user 1, note that
\begin{align}\label{demonstrate1}
&\text{rank}([\boldsymbol{A}_1\boldsymbol{G}_1 \; \boldsymbol{B}_1\boldsymbol{G}_2]) =\text{rank}\left(
\begin{bmatrix}
\left. {\begin{array}{*{20}{c}}
\boldsymbol{F}_{1,1} \\
\boldsymbol{F}_{1,2} \\
\boldsymbol{F}_{1,3} \\
\end{array}} \right\} & \boldsymbol{0}^{n_{11}-n_{12},r_2}\\
\boldsymbol{0}^{t_3,r_1}& \boldsymbol{F}_{2,1}\\
\boldsymbol{F}_{1,4} & \boldsymbol{0}^{n_{12}-n_{21}-t_3,r_2}\\
\boldsymbol{0}^{t_4,r_1} & \boldsymbol{F}_{2,2}\\
\boldsymbol{F}_{1,5} & \boldsymbol{F}_{2,3} \\
\boldsymbol{0}^{2n_{22}+n_{11}-2n_{12}-2n_{21},r_1} & \boldsymbol{F}_{2,4}\\
\boldsymbol{0}^{t_4,r_1}&\boldsymbol{F}_{2,5}\\
\boldsymbol{F}_{1,5} &\boldsymbol{0}^{2n_{21}+n_{122}-n_{11}-n_{22}-t_4,r_2} \\
\boldsymbol{F}_{1,6} & \boldsymbol{0}^{2n_{11}+n_{22}-2n_{12}-2n_{21},r_2}  \\
\boldsymbol{0}^{t_5,r_1} & \boldsymbol{F}_{2,6}\\
\boldsymbol{F}_{1,7} & \boldsymbol{0}^{n_{12}-n_{21}-t_5,r_2} \\
\boldsymbol{F}_{1,8} & \boldsymbol{F}_{2,5} \\
\boldsymbol{F}_{1,9} & \boldsymbol{0}^{2n_{21}+n_{12}-n_{11}-n_{22}-t_4,r_2}
\end{bmatrix}\right).
\end{align}
It is easy to see that the above rank is equal to the rank of the above matrix with the upper $\boldsymbol{F}_{1,5}$ and the lower $\boldsymbol{F}_{2,5}$ replaced by $\boldsymbol{0}$\footnote{This can be easily seen by the fact that Gaussian elimination does not alter the rank. However, we opt not to use the term ``Gaussian elimination" deliberately to avoid causing the confusion that we are doing SIC, which we do not.}
Moreover, we have that
\begin{align}
\text{rank}(\boldsymbol{B}_1\boldsymbol{G}_2) = \text{rank}\left(\begin{bmatrix}
\boldsymbol{F}_{2,1}\\
\boldsymbol{F}_{2,2}\\
\boldsymbol{F}_{2,3} \\
\boldsymbol{F}_{2,4}\\
\boldsymbol{F}_{2,5} \\
\boldsymbol{F}_{2,6} \\
\boldsymbol{F}_{2,5} \\
\end{bmatrix}
\right)
= \text{rank}\left(\begin{bmatrix}
\boldsymbol{F}_{2,1}\\
\boldsymbol{F}_{2,2}\\
\boldsymbol{F}_{2,3} \\
\boldsymbol{F}_{2,4}\\
\boldsymbol{F}_{2,5} \\
\boldsymbol{F}_{2,6} \\
\boldsymbol{0} \\
\end{bmatrix}
\right).
\end{align}
These indicate that evaluating the rate of this type II scheme is equivalent to evaluate a corresponding type I scheme with the upper $\boldsymbol{F}_{1,5}$ and the lower $\boldsymbol{F}_{2,5}$ replaced by $\boldsymbol{0}$. Hence, we can again use the property \textbf{P2.} to obtain user 1's achievable rate by using Proposition \ref{remark:type1} as
\begin{align}
I(\msf{X}_1;\msf{Y}_1) = \text{rank}(\boldsymbol{A}_1\boldsymbol{G}_1)= n_{11}+n_{12}-n_{22}-t_3-t_4-t_5.
\end{align}
For user 2, the following rank equals to that with the lower $\boldsymbol{F}_{1,5}$ and upper $\boldsymbol{F}_{2,5}$ replaced by $\boldsymbol{0}$.
\vspace{-0.4mm}
\begin{align}\label{eq:type2_example2}
\text{rank}([\boldsymbol{A}_2\boldsymbol{G}_2 \; \boldsymbol{B}_2\boldsymbol{G}_1]) &=
\text{rank}\left(
\begin{bmatrix}
\left. {\begin{array}{*{20}{c}}
\boldsymbol{0}^{n_{11}-n_{22},r_2} \\
\boldsymbol{F}_{2,1} \\
\boldsymbol{0}^{n_{12}-n_{21}-t_3,r_2} \\
\boldsymbol{F}_{2,2} \\
\boldsymbol{F}_{2,3} \\
\boldsymbol{F}_{2,4} \\
\end{array}} \right\} & \boldsymbol{0}^{n_{11}-n_{21},r_1} \\
\boldsymbol{F}_{2,5} & \boldsymbol{F}_{1,1}\\
\boldsymbol{0}^{2n_{21}+n_{12}-n_{11}-n_{22}-t_4,r_2} & \boldsymbol{F}_{1,2}\\
\boldsymbol{0}^{2n_{11}+n_{22}-2n_{12}-2n_{21},r_2} & \boldsymbol{F}_{1,3}\\
\left. {\begin{array}{*{20}{c}}
\boldsymbol{F}_{2,6} \\
\boldsymbol{0}^{n_{12}-n_{21}-t_5,r_2}
\end{array}} \right\} &\left\{ {\begin{array}{*{20}{c}}
\boldsymbol{0}^{t_3,r_1}\\
\boldsymbol{F}_{1,4}\\
\end{array}} \right.\\
\boldsymbol{F}_{2,5} & \boldsymbol{0}^{t_4,r_1}\\
\boldsymbol{0}^{2n_{21}+n_{12}-n_{11}-n_{22}-t_4,r_2} & \boldsymbol{F}_{1,5} \\
\boldsymbol{F}_{2,7} &\boldsymbol{0}^{2n_{22}+n_{11}-2n_{12}-2n_{21},r_1}\\
\boldsymbol{F}_{2,8} &\boldsymbol{0}^{t_4,r_1}\\
\boldsymbol{F}_{2,9} &\boldsymbol{F}_{1,5} \\
\end{bmatrix}
\right).
\end{align}
This with property \textbf{P2.} allows us to use Proposition \ref{remark:type1} to obtain that
\begin{align}
I(\msf{X}_2;\msf{Y}_2)= \left\{ {\begin{array}{*{20}{c}}
&2(n_{22}-n_{12})+t_3+t_4,&t_5 = 0,&t_3<n_{12}-n_{21}\\
&2n_{22}-n_{21}-n_{12}+t_5+t_4,&t_5\geq0, &t_3=n_{12}-n_{21}\\
\end{array}} \right..
\end{align}

\emph{2c)} To achieve the region between $(2(n_{11}-n_{12}),n_{22}+n_{12}-n_{11})$ and $(0,n_{22})$, we propose
\begin{align}\label{eq:E1_case3}
\boldsymbol{G}_1 = \begin{bmatrix}
\boldsymbol{0}^{t_6,r_1}\\
\boldsymbol{F}_{1,1}\\
\boldsymbol{0}^{t_7,r_1} \\
\boldsymbol{F}_{1,2} \\
\boldsymbol{0}^{n_{22}-n_{21},r_1}\\
\boldsymbol{0}^{2n_{21}+n_{12}-n_{11}-n_{22},r_1}\\
\boldsymbol{0}^{t_7,r_1}\\
\boldsymbol{F}_{1,3} \\
\boldsymbol{0}^{t_6,r_1}\\
\boldsymbol{F}_{1,4}\\
\boldsymbol{0}^{n_{12}-n_{21},r_1}
\end{bmatrix},
\boldsymbol{G}_2 = \begin{bmatrix}
\boldsymbol{F}_{2,1} \\
\boldsymbol{F}_{2,2} \\
\boldsymbol{F}_{2,3} \\
\boldsymbol{F}_{2,4} \\
\boldsymbol{0}^{2n_{11}+n_{22}-2n_{12}-2n_{21}-t_7,r_2} \\
\boldsymbol{F}_{2,5} \\
\boldsymbol{F}_{2,3} \\
\boldsymbol{F}_{2,6} \\
\boldsymbol{F}_{2,7} \\
\boldsymbol{0}^{n_{11}-n_{22},r_2}
\end{bmatrix},
\end{align}
where $\boldsymbol{F}_{1,1}, \boldsymbol{F}_{1,4}\in \mathbb{F}_2^{2n_{21}+n_{12}-n_{11}-n_{22}-t_6,r_1},\boldsymbol{F}_{1,2},\boldsymbol{F}_{1,3} \in \mathbb{F}_2^{2n_{11}+n_{22}-2n_{12}-2n_{21}-t_7,r_1},\boldsymbol{F}_{2,1}\in \mathbb{F}_2^{n_{22}-n_{21},r_5}$, \\$\boldsymbol{F}_{2,2},\boldsymbol{F}_{2,5}\in \mathbb{F}_2^{t_6,r_2},\boldsymbol{F}_{2,3}\in \mathbb{F}_2^{2n_{21}+n_{12}-n_{11}-n_{22}-t_6,r_2},\boldsymbol{F}_{2,4}\in \mathbb{F}_2^{t_7,r_2},\boldsymbol{F}_{2,6} \in \mathbb{F}_2^{n_{12}-n_{21},r_2},\boldsymbol{F}_{2,7} \in \mathbb{F}_2^{n_{22}-n_{12},r_2}$, and $t_6 \in [0:2n_{21}+n_{12}-n_{11}-n_{22}],t_7 \in [0:2n_{11}+n_{22}-2n_{12}-2n_{21}]$.

For user 1, we design the generator matrices such that {\bf P2.} holds. Moreover, by noting that
\begin{align}\label{eq:weak1_2c_u1}
\text{rank}([\boldsymbol{A}_1\boldsymbol{G}_1 \; \boldsymbol{B}_1\boldsymbol{G}_2]) =\text{rank}\left(
\begin{bmatrix}
\left. {\begin{array}{*{20}{c}}
\boldsymbol{0}^{t_6,r_1}\\
\boldsymbol{F}_{1,1}\\
\boldsymbol{0}^{t_7,r_1}\\
\boldsymbol{F}_{1,2} \\
\end{array}} \right\} & \boldsymbol{0}^{n_{11}-n_{12},r_2} \\
\boldsymbol{0}^{n_{22}-n_{21},r_1} & \boldsymbol{F}_{2,1} \\
\boldsymbol{0}^{2n_{21}+n_{12}-n_{11}-n_{22},r_1}  & \left\{ {\begin{array}{*{20}{c}}
\boldsymbol{F}_{2,2} \\
\boldsymbol{F}_{2,3} \\
\end{array}} \right.\\
\boldsymbol{0}^{t_7,r_1} & \boldsymbol{F}_{2,4} \\
\boldsymbol{F}_{1,3} & \boldsymbol{0}^{2n_{11}+n_{22}-2n_{12}-2n_{21}-t_7,r_2}\\
\boldsymbol{0}^{t_6,r_1} & \boldsymbol{F}_{2,5}\\
\boldsymbol{F}_{1,4}& \boldsymbol{F}_{2,3}\\
\boldsymbol{0}^{n_{12}-n_{21},r_1} &\boldsymbol{F}_{2,6}
\end{bmatrix}\right).
\end{align}
Since the above rank is equal to the rank of the above matrix with the lower $\boldsymbol{F}_{2,3}$ replaced by $\boldsymbol{0}$, user 1's achievable rate is then obtained as
\begin{align}
I(\msf{X}_1;\msf{Y}_1) = \text{rank}(\boldsymbol{A}_1\boldsymbol{G}_1) = 2(n_{11}-n_{12}-t_6-t_7).
\end{align}

For user 2, we design the generator matrices to ensure that that {\bf P2.} also holds. Moreover, we note that
\begin{align}
\text{rank}([\boldsymbol{A}_2\boldsymbol{G}_2 \; \boldsymbol{B}_2\boldsymbol{G}_1]) =\text{rank}\left(
\begin{bmatrix}
\left. {\begin{array}{*{20}{c}}
\boldsymbol{0}^{n_{11}-n_{22},r_2} \\
\boldsymbol{F}_{2,1} \\
\end{array}} \right\} & \boldsymbol{0}^{n_{11}-n_{21},r_1} \\
\boldsymbol{F}_{2,2} & \boldsymbol{0}^{t_6,r_1} \\
\boldsymbol{F}_{2,3} & \boldsymbol{F}_{1,1}\\
\boldsymbol{F}_{2,4} & \boldsymbol{0}^{t_7,r_1} \\
\boldsymbol{0}^{2n_{11}+n_{22}-2n_{12}-2n_{21}-t_7,r_2} & \boldsymbol{F}_{1,2} \\
\left. {\begin{array}{*{20}{c}}
\boldsymbol{F}_{2,5} \\
\boldsymbol{F}_{2,3} \\
\boldsymbol{F}_{2,6} \\
\boldsymbol{F}_{2,7}
\end{array}} \right\} & \left\{ {\begin{array}{*{20}{c}}
\boldsymbol{0}^{n_{22}-n_{21},r_1}\\
\boldsymbol{0}^{2n_{21}+n_{12}-n_{11}-n_{22},r_1}\\
\end{array}} \right.\\
\end{bmatrix}
\right),
\end{align}
By using Proposition \ref{remark:type1}, user 2's achievable rate is obtained as
\begin{align}
I(\msf{X}_2;\msf{Y}_2)= \text{rank}(\boldsymbol{A}_2\boldsymbol{G}_2) =  n_{22}+n_{12}-n_{11}+t_6+t_7.
\end{align}


\subsection{Achievability Proof for Other Subregimes}\label{app:corner}
We report the generator matrices for each of the remaining subregimes in a table.

\footnotesize
\setlength\tabcolsep{2.0pt}
\begin{longtable}{|l|Y|l|}  
\caption{Achievable schemes for Weak 1$:n_{11}>n_{22}>n_{12}>n_{21}$ cont.} \label{table:weak1}\\
\hline
  Interference subregime &Generator matrices  & Target corner point \\
\hline
\endfirsthead
 \hline
\endhead
 \hline
\endfoot
\multirow{8}{*}{ $\begin{aligned}(1)&: n_{22}<n_{11}<n_{12}+n_{21},\\n_{11}&+n_{22} - 2n_{12} - n_{21}>0\\
\Rightarrow n_{11}&+n_{22} - n_{12}- 2n_{21}>0.\end{aligned}$} & \makecell{$\boldsymbol{G}_1 = \begin{bmatrix}
\boldsymbol{F}_{1,1} \\
\boldsymbol{0}^{n_{12}+n_{21}-n_{11},r_1} \\
\boldsymbol{F}_{1,2} \\
\end{bmatrix},
\boldsymbol{G}_2 = \begin{bmatrix}
\boldsymbol{F}_{2,1}\\
\boldsymbol{0}^{n_{11}+n_{22}-n_{12}-2n_{21},r_2}\\
\boldsymbol{0}^{n_{12}+n_{21}-n_{22},r_2}\\
\boldsymbol{0}^{n_{11}+n_{22}-2n_{12}-n_{21},r_2}\\
\boldsymbol{F}_{2,2} \\
\boldsymbol{0}^{n_{11}-n_{22},r_2}\\
\end{bmatrix}$,\\
$\boldsymbol{F}_{1,1} \in\mathbb{F}_2^{n_{11}-n_{12} ,r_1},\boldsymbol{F}_{1,2} \in \mathbb{F}_2^{n_{11}-n_{21},r_1}$, \\
$\boldsymbol{F}_{2,1},\boldsymbol{F}_{2,2} \in \mathbb{F}_2^{n_{12}+n_{21}-n_{11},r_2}$.}
 & $\begin{aligned}r_1&=2n_{11}- n_{12}-n_{21},\\
 r_2&=2(n_{21}+n_{12}-n_{11}).\end{aligned}$ \\
\hhline{~--}
 & \makecell{$\boldsymbol{G}_1 = \begin{bmatrix}
\boldsymbol{F}_{1,1} \\
\boldsymbol{0}^{n_{11}+n_{22}-2n_{12}-n_{21},r_1} \\
\boldsymbol{0}^{n_{12}+n_{21}-n_{11},r_1} \\
\boldsymbol{0}^{n_{11}+n_{22}-n_{12}-2n_{21} ,r_1} \\
\boldsymbol{F}_{1,2} \\
\end{bmatrix},
\boldsymbol{G}_2 = \begin{bmatrix}
\boldsymbol{F}_{2,1}\\
\boldsymbol{F}_{2,2}\\
\boldsymbol{0}^{n_{12}+n_{21}-n_{22},r_2}\\
\boldsymbol{F}_{2,3} \\
\boldsymbol{0}^{n_{11}-n_{22},r_2}\\
\end{bmatrix}$,\\
$\boldsymbol{F}_{1,1},\boldsymbol{F}_{1,2} \in \mathbb{F}_2^{n_{12}+n_{21}-n_{22},r_1}, \boldsymbol{F}_{2,1} \in \mathbb{F}_2^{n_{12}+n_{21}-n_{11},r_2}$, \\
$\boldsymbol{F}_{2,2} \in \mathbb{F}_2^{n_{11}+n_{22}-n_{12}-2n_{21},r_2},\boldsymbol{F}_{2,3} \in \mathbb{F}_2^{n_{22}-n_{12},r_2}$.}
 & $\begin{aligned}r_1&=2(n_{12}+n_{21}-n_{22}),\\
 r_2&=2n_{22}-n_{12}-n_{21}.\end{aligned}$ \\
\hline
\end{longtable}

\footnotesize
\setlength\tabcolsep{2.0pt}
\begin{longtable}{|l|Y|l|}  
\caption{Achievable schemes for Weak 2$:n_{11}>n_{22}>n_{21}>n_{12}$} \label{table:weak2}\\
\hline
  Interference subregime &Generator matrices  & Target corner point \\
\hline
\endfirsthead
 \hline
\endhead
 \hline
\endfoot
 $(1):n_{11}>n_{12}+n_{21}>n_{22}$ & Same as in Appendix \ref{weak01}
 & Same as in Appendix \ref{weak01} \\
\hhline{---}
 $\begin{aligned}(2)&:n_{22}<n_{11}<n_{12}+n_{21},\\n_{11}&+n_{22} - 2n_{21} - n_{21}>0\\
\Rightarrow n_{11}&+n_{22} - 2n_{12}- n_{21}<0.\end{aligned}$ & Same as in Table \ref{table:weak1}(1)
 & Same as in Table \ref{table:weak1}(1) \\
\hhline{---}
 $\begin{aligned}(3a1)&:n_{22}<n_{11}<n_{12}+n_{21},\\n_{11}&+n_{22} - 2n_{21} - n_{12}<0,\\
 3n_{21}&+n_{12} - 2n_{22}- n_{11}>0.\end{aligned}$ & \makecell{
 $\boldsymbol{G}_1 \hspace{-1mm}=\hspace{-1mm}\begin{bmatrix}
\boldsymbol{F}_{1,1}\\
\boldsymbol{F}_{1,2}\\
\boldsymbol{0}^{n_{11}+2n_{22}-n_{12}-3n_{21},r_1}\\
\boldsymbol{F}_{1,2}\\
\boldsymbol{F}_{1,3}\\
\end{bmatrix}\hspace{-1mm},
\boldsymbol{G}_2 \hspace{-1mm}=\hspace{-1mm}\begin{bmatrix}
\boldsymbol{F}_{2,1} \\
\boldsymbol{F}_{2,2} \\
\boldsymbol{0}^{n_{12}+n_{21}-n_{22},r_2} \\
\boldsymbol{0}^{n_{21}-n_{12},r_2}\\
\boldsymbol{F}_{2,3} \\
\boldsymbol{F}_{2,4} \\
\boldsymbol{0}^{n_{11}-n_{22},r_2}
\end{bmatrix}$,\\
$\boldsymbol{F}_{1,1}\in \mathbb{F}_2^{n_{11}-n_{12},r_1},\boldsymbol{F}_{1,2}\in \mathbb{F}_2^{2n_{21}+n_{12}-n_{11}-n_{22},r_1}$, \\$\boldsymbol{F}_{1,3} \in \mathbb{F}_2^{n_{11}-n_{21},r_1},\boldsymbol{F}_{2,1},\boldsymbol{F}_{2,4}\in \mathbb{F}_2^{2n_{21}+n_{12}-n_{11}-n_{22},r_2}$,\\$\boldsymbol{F}_{2,2},\boldsymbol{F}_{2,3} \in \mathbb{F}_2^{n_{11}+2n_{22}-n_{12}-3n_{21},r_2}.$}
 & \multirow{10}{*}{$\begin{aligned}r_1&=n_{11}+n_{12} - n_{22},\\
 r_2&=2(n_{22}-n_{12}).\end{aligned}$} \\
\hline
\end{longtable} 

\footnotesize
\setlength\tabcolsep{2.0pt}
\begin{longtable}{|l|Y|l|}  
\caption{Achievable schemes for Weak 3$:n_{11}>n_{21}>n_{22}>n_{12}$} \label{table:weak3}\\
\hline
  Interference subregime &Generator matrices  & Target corner point \\
\hline
\endfirsthead
\hline
\endhead
\hline
\endfoot
\multirow{8}{*}{ $(1): n_{11}>n_{12}+n_{21}>n_{22}.$} & \makecell{$\boldsymbol{G}_1=\begin{bmatrix}
\boldsymbol{F}_{1,1}\\
\boldsymbol{F}_{1,2}\\
\boldsymbol{0}^{n_{22}-n_{12},r_1}\\
\boldsymbol{F}_{1,3}
\end{bmatrix},
\boldsymbol{G}_2=\begin{bmatrix}
\boldsymbol{0}^{n_{12},r_2} \\
\boldsymbol{F}_{2,1} \\
\boldsymbol{0}^{n_{11}-n_{22},r_2}
\end{bmatrix}$,\\
$\boldsymbol{F}_{1,1}\in \mathbb{F}_2^{n_{21}-n_{22},r_1},\boldsymbol{F}_{1,2}\in \mathbb{F}_2^{n_{12},r_1}$,\\$\boldsymbol{F}_{1,3} \in \mathbb{F}_2^{n_{11}-n_{21},r_1},\boldsymbol{F}_{2,1} \in \mathbb{F}_2^{n_{22}-n_{12},r_2}$.}
 & $\begin{aligned}r_1&=n_{11}+n_{12}-n_{22},\\
 r_2&=n_{22}-n_{12}.\end{aligned}$ \\
\hhline{~--}
 & \makecell{$\boldsymbol{G}_1=\begin{bmatrix}
\boldsymbol{F}_{1,1}\\
\boldsymbol{0}^{n_{22},r_1}\\
\boldsymbol{F}_{1,2}\\
\boldsymbol{0}^{n_{12},r_1}
\end{bmatrix},
\boldsymbol{G}_2=\begin{bmatrix}
\boldsymbol{F}_{2,1} \\
\boldsymbol{F}_{2,2} \\
\boldsymbol{0}^{n_{11}-n_{22},r_2}
\end{bmatrix}$,\\
$\boldsymbol{F}_{1,1}\in \mathbb{F}_2^{n_{21}-n_{22},r_1},\boldsymbol{F}_{1,2}\in \mathbb{F}_2^{n_{11}-n_{12}-n_{21},r_1}$,\\$\boldsymbol{F}_{2,1} \in \mathbb{F}_2^{n_{12},r_2},\boldsymbol{F}_{2,2} \in \mathbb{F}_2^{n_{22}-n_{12},r_2}$.}
 & $\begin{aligned}r_1&=n_{11}-n_{22}-n_{12},\\
 r_2&=n_{22}.\end{aligned}$ \\
\hline
\end{longtable} 

\footnotesize
\setlength\tabcolsep{2.0pt}
\begin{longtable}{|l|Y|l|}  
\caption{Achievable schemes for Strong 1$:n_{12}>n_{21}>n_{11}>n_{22}$} \label{table:strong1}\\
\hline
  Interference subregime &Generator matrices  & Target corner point \\
\hline
\endfirsthead
\hline
\endhead
\hline
\endfoot
\multirow{8}{*}{ $(1):n_{12}>n_{11}+n_{22}>n_{21}.$} & \makecell{$\boldsymbol{G}_1 = \begin{bmatrix}
\boldsymbol{F}_{1,1}\\
\boldsymbol{F}_{1,2}\\
\boldsymbol{0}^{n_{21}-n_{11},r_1} \\
\boldsymbol{0}^{n_{12}-n_{21},r_1}
\end{bmatrix},
\boldsymbol{G}_2 = \begin{bmatrix}
\boldsymbol{0}^{n_{11}+n_{22}-n_{21},r_2} \\
\boldsymbol{F}_{2,1} \\
\boldsymbol{0}^{n_{12}-n_{22}-n_{11},r_2} \\
\boldsymbol{0}^{n_{11},r_2}
\end{bmatrix}$,\\
$\boldsymbol{F}_{1,1}\in \mathbb{F}_2^{n_{21}-n_{22},r_1},\boldsymbol{F}_{1,2} \in \mathbb{F}_2^{n_{11}+n_{22}-n_{21},r_1}$,\\$\boldsymbol{F}_{2,1}, \in \mathbb{F}_2^{n_{21}-n_{11},r_2}$.}
 & $\begin{aligned}r_1&=n_{11},\\
 r_2&=n_{21}-n_{11}.\end{aligned}$ \\
 \hhline{~--}
 &  \makecell{$\boldsymbol{G}_1 = \begin{bmatrix}
\boldsymbol{F}_{1,1}\\
\boldsymbol{0}^{n_{12}+n_{22}-n_{21},r_1} \\
\end{bmatrix},
\boldsymbol{G}_2 = \begin{bmatrix}
\boldsymbol{F}_{2,1} \\
\boldsymbol{F}_{2,2} \\
\boldsymbol{0}^{n_{21}-n_{22},n_{12}} \\
\boldsymbol{0}^{n_{11}+n_{22}-n_{21},n_{12}}
\end{bmatrix}$,\\
$\boldsymbol{F}_{1,1},\in \mathbb{F}_2^{n_{21}-n_{22},n_{12}},\boldsymbol{F}_{2,1}, \in \mathbb{F}_2^{n_{22},n_{12}}$,\\$\boldsymbol{F}_{2,2}, \in \mathbb{F}_2^{n_{12}-n_{11}-n_{22},n_{12}}$.}
 & $\begin{aligned}r_1&=n_{21}-n_{22},\\
 r_2&=n_{22}.\end{aligned}$ \\
\hline 
\end{longtable} 

\footnotesize
\setlength\tabcolsep{2.0pt}
\begin{longtable}{|l|Y|l|}  
\caption{Achievable schemes for Strong 2$:n_{21}>n_{12}>n_{11}>n_{22}$} \label{table:strong2}\\
\hline
  Interference subregime &Generator matrices  & Target corner point \\
\hline
\endfirsthead
\hline
\endhead
\hline
\endfoot
\multirow{6}{*}{ $(1):n_{21}>n_{11}+n_{22}>n_{12}.$} & \makecell{$\boldsymbol{G}_1=\begin{bmatrix}
\boldsymbol{F}_{1,1}\\
\boldsymbol{0}^{n_{21}-n_{11}-n_{22},r_1}\\
\boldsymbol{0}^{n_{22},n_{21}}\\
\end{bmatrix},
\boldsymbol{G}_2=\begin{bmatrix}
\boldsymbol{F}_{2,1} \\
\boldsymbol{0}^{n_{11}+n_{22}-n_{12},r_2} \\
\boldsymbol{0}^{n_{12}-n_{22},n_{21}}\\
\boldsymbol{0}^{n_{21}-n_{12},n_{21}}\\
\end{bmatrix}$,\\
$\boldsymbol{F}_{1,1}\in \mathbb{F}_2^{n_{11},r_1},\boldsymbol{F}_{2,1}\in \mathbb{F}_2^{n_{12}-n_{11},r_2}$.}
 & $\begin{aligned}r_1&=n_{11},\\
 r_2&=n_{12}-n_{11}.\end{aligned}$ \\
 \hhline{~--}
 &  \makecell{$\boldsymbol{G}_1=\begin{bmatrix}
\boldsymbol{0}^{n_{11}+n_{22}-n_{12},r_1}\\
\boldsymbol{F}_{1,1}\\
\boldsymbol{0}^{n_{21}-n_{11}-n_{22},r_1}\\
\boldsymbol{0}^{n_{22},n_{21}}\\
\end{bmatrix},
\boldsymbol{G}_2=\begin{bmatrix}
\boldsymbol{F}_{2,1} \\
\boldsymbol{F}_{2,2} \\
\boldsymbol{0}^{n_{12}-n_{22},r_2}\\
\boldsymbol{0}^{n_{21}-n_{12},r_2}\\
\end{bmatrix}$,\\
$\boldsymbol{F}_{1,1}\in \mathbb{F}_2^{n_{12}-n_{22},r_1},\boldsymbol{F}_{2,1}\in \mathbb{F}_2^{n_{12}-n_{11},r_2},\boldsymbol{F}_{2,2}\in \mathbb{F}_2^{n_{11}+n_{22}-n_{12},r_2}$.}
 & $\begin{aligned}r_1&=n_{12}-n_{22},\\
 r_2&=n_{22}.\end{aligned}$ \\
\hline
\end{longtable} 

\footnotesize
\setlength\tabcolsep{2.0pt}
\begin{longtable}{|l|Y|l|}  
\caption{Achievable schemes for Strong 3$:n_{21}>n_{11}>n_{12}>n_{22}$} \label{table:strong3}\\
\hline
  Interference subregime &Generator matrices  & Target corner point \\
\hline
\endfirsthead
\hline
\endhead
\hline
\endfoot
 $(1):n_{21}>n_{11}+n_{22}>n_{12}.$ & \makecell{$\boldsymbol{G}_1=\begin{bmatrix}
\boldsymbol{F}_{1,1}\\
\boldsymbol{0}^{n_{22},r_1}\\
\boldsymbol{F}_{1,2}\\
\boldsymbol{0}^{n_{21}-n_{11}-n_{22},r_1}\\
\boldsymbol{0}^{n_{22},r_1}\\
\end{bmatrix},
\boldsymbol{G}_2=\begin{bmatrix}
\boldsymbol{F}_{2,1} \\
\boldsymbol{0}^{n_{12}-n_{22},r_2}\\
\boldsymbol{0}^{n_{21}-n_{12},r_2}\\
\end{bmatrix}$,\\
$\boldsymbol{F}_{1,1}\in \mathbb{F}_2^{n_{11}-n_{12},r_1},\boldsymbol{F}_{1,2}\in \mathbb{F}_2^{n_{12}-n_{22},r_2},\boldsymbol{F}_{2,1}\in \mathbb{F}_2^{n_{22},r_2}$.}
 &  \\
 \hhline{--~}
 $\begin{aligned}(2)&:n_{11}+n_{22}>n_{21}>n_{12},\\
 n_{12}&+n_{21}-n_{11}-2n_{22}>0.\end{aligned}$&  \makecell{$\boldsymbol{G}_1=\begin{bmatrix}
\boldsymbol{F}_{1,1}\\
\boldsymbol{F}_{1,2}\\
\boldsymbol{0}^{n_{21}-n_{11},r_1}\\
\boldsymbol{F}_{1,3}\\
\boldsymbol{F}_{1,2}\\
\boldsymbol{0}^{n_{21}-n_{11},r_1}\\
\end{bmatrix},
\boldsymbol{G}_2=\begin{bmatrix}
\boldsymbol{F}_{2,1} \\
\boldsymbol{F}_{2,2} \\
\boldsymbol{0}^{n_{12}-n_{22},r_2}\\
\boldsymbol{0}^{n_{21}-n_{12},r_2}\\
\end{bmatrix}$,\\
$\boldsymbol{F}_{1,1}\in \mathbb{F}_2^{n_{11}-n_{12},r_1},\boldsymbol{F}_{1,2}\in \mathbb{F}_2^{n_{11}+n_{22}-n_{21},r_1},\boldsymbol{F}_{2,2}\in \mathbb{F}_2^{n_{21}-n_{11},r_2}$,\\$\boldsymbol{F}_{1,3}\in \mathbb{F}_2^{n_{12}+n_{21}-n_{11}-2n_{22},r_1},\boldsymbol{F}_{2,1}\in \mathbb{F}_2^{n_{11}+n_{22}-n_{21},r_2}$.}
 &  \multirow{16}{*}{$\begin{aligned}r_1&=n_{11}-n_{22},\\
 r_2&=n_{22}.\end{aligned}$}\\
\hline
\end{longtable} 

\footnotesize
\setlength\tabcolsep{2.0pt}
\begin{longtable}{|l|Y|l|}  
\caption{Achievable schemes for Mixed 1$:n_{11}>n_{12}>n_{22}>n_{21}$} \label{table:mix1}\\
\hline
  Interference subregime &Generator matrices  & Target corner point \\
\hline
\endfirsthead
 \hline
\endhead
 \hline
\endfoot
 \multirow{6}{*}{$(1):n_{11} > n_{12}+n_{21}>n_{22}$} &  \makecell{$\boldsymbol{G}_1=\begin{bmatrix}
\boldsymbol{F}_{1,1} \\
\boldsymbol{F}_{1,2}\\
\boldsymbol{0}^{n_{22}-n_{21},r_1}\\
\boldsymbol{F}_{1,3} \\
\end{bmatrix},
\boldsymbol{G}_2 = \begin{bmatrix}
\boldsymbol{F}_{2,1}\\
\boldsymbol{0}^{n_{21},r_2}\\
\boldsymbol{0}^{n_{12}-n_{22},r_2}\\
\boldsymbol{0}^{n_{11}-n_{12},r_2}\\
\end{bmatrix}$,\\
$\boldsymbol{F}_{1,1} \in \mathbb{F}_2^{n_{21} ,r_1}, \boldsymbol{F}_{1,2} \in \mathbb{F}_2^{n_{11} - n_{21} - n_{12},r_1}$, \\$\boldsymbol{F}_{1,3} \in \mathbb{F}_2^{n_{12}+n_{21}-n_{22},r_1},\boldsymbol{F}_{2,1} \in \mathbb{F}_2^{n_{22}-n_{21},r_2}$.}
 & $\begin{aligned}r_1&=n_{11}+n_{21}-n_{22},\\
 r_2&=n_{22}-n_{21}.\end{aligned}$ \\
\hhline{~--}
  & \makecell{$\boldsymbol{G}_1=\begin{bmatrix}
\boldsymbol{0}^{n_{21},r_1} \\
\boldsymbol{F}_{1,1} \\
\boldsymbol{0}^{n_{22},r_1}\\
\boldsymbol{F}_{1,2} \\
\end{bmatrix},
\boldsymbol{G}_2 = \begin{bmatrix}
\boldsymbol{F}_{2,1}\\
\boldsymbol{0}^{n_{12}-n_{22},r_2}\\
\boldsymbol{F}_{2,2}
\end{bmatrix}$,\\
$\boldsymbol{F}_{1,1} \in  \mathbb{F}_2^{n_{11} - n_{21} - n_{12},r_1},\boldsymbol{F}_{1,2} \in \mathbb{F}_2^{n_{12}-n_{22},r_1}$,\\$\boldsymbol{F}_{2,1} \in \mathbb{F}_2^{n_{22},r_2},\boldsymbol{F}_{2,2} \in \mathbb{F}_2^{n_{11}- n_{12},r_2}$..}
 & $\begin{aligned}r_1&=n_{11}-n_{21}-n_{22},\\
 r_2&=n_{22}.\end{aligned}$ \\
\hline
\end{longtable} 

\footnotesize
\setlength\tabcolsep{2.0pt}
\begin{longtable}{|l|Y|l|}  
\caption{Achievable schemes for Mixed 2$:n_{11}>n_{21}>n_{12}>n_{22}$} \label{table:mix2}\\
\hline
  Interference subregime &Generator matrices  & Target corner point \\
\hline
\endfirsthead
 \hline
\endhead
 \hline
\endfoot
 $\begin{aligned}(1)&:n_{22}<n_{11}<n_{12}+n_{21},\\
 n_{12}&+n_{21}<n_{11}+n_{22},\\
 2(n_{12}&+n_{21}-n_{11})-n_{22}>0.\end{aligned}$ &  \makecell{$\boldsymbol{G}_1=\hspace{30mm}\boldsymbol{G}_2=$\\
$\begin{bmatrix}
\boldsymbol{F}_{1,1}\\
\boldsymbol{F}_{1,2}\\
\boldsymbol{0}^{2(n_{12}+n_{21}-n_{11})-n_{22},r_1}\\
\boldsymbol{F}_{1,2}\\
\boldsymbol{F}_{1,3}\\
\boldsymbol{F}_{1,4}\\
\end{bmatrix},
\begin{bmatrix}
\boldsymbol{F}_{2,1} \\
\boldsymbol{F}_{2,2} \\
\boldsymbol{F}_{2,3} \\
\boldsymbol{0}^{n_{11}+n_{22}-n_{12}-n_{21},r_2} \\
\boldsymbol{0}^{n_{12}-n_{22},r_2}\\
\boldsymbol{0}^{n_{11}-n_{12},r_2}\\
\end{bmatrix}$,\\
$\boldsymbol{F}_{1,1}\in \mathbb{F}_2^{n_{21}-n_{22},r_1},\boldsymbol{F}_{1,2},\boldsymbol{F}_{1,3}\in \mathbb{F}_2^{n_{11}+n_{22}-n_{12}-n_{21},r_1}$,\\$\boldsymbol{F}_{1,4}\in \mathbb{F}_2^{n_{12}-n_{22},r_1},\boldsymbol{F}_{2,2} \in \mathbb{F}_2^{|3n_{12}+3n_{21}-3n_{11}-2n_{22}|,r_2}$,\\$\boldsymbol{F}_{2,1},\boldsymbol{F}_{2,3}\in \mathbb{F}_2^{\min\{2(n_{12}+n_{21}-n_{11})-n_{22},n_{11}+n_{22}-n_{12}-n_{21}\},r_2}$.}
 & $\begin{aligned}r_1&=2n_{11}-n_{12}-n_{21},\\
 r_2&=n_{12}+n_{21}-n_{11}.\end{aligned}$ \\
\hline
\end{longtable} 

\footnotesize
\setlength\tabcolsep{2.0pt}
\begin{longtable}{|l|Y|l|}  
\caption{Achievable schemes for Mixed 4$:n_{12}>n_{11}>n_{21}>n_{22}$} \label{table:mix4}\\
\hline
  Interference subregime &Generator matrices  & Target corner point \\
\hline
\endfirsthead
\hline
\endhead
\hline
\endfoot
 $(1):n_{12}>n_{11}+n_{22}>n_{21}.$ & \makecell{$\boldsymbol{G}_1=\begin{bmatrix}
\boldsymbol{F}_{1,1}\\
\boldsymbol{0}^{n_{22},r_1} \\
\boldsymbol{F}_{1,2}\\
\boldsymbol{0}^{n_{12}-n_{11},r_1}\\
\end{bmatrix},
\boldsymbol{G}_2 = \begin{bmatrix}
\boldsymbol{F}_{2,1}\\
\boldsymbol{0}^{n_{12}-n_{11}-n_{22},r_2} \\
\boldsymbol{0}^{n_{11},r_2} \\
\end{bmatrix}$,\\
$\boldsymbol{F}_1 \in \mathbb{F}_2^{n_{21}-n_{22},r_1},\boldsymbol{F}_{1,2} \in \mathbb{F}_2^{n_{11}-n_{21},r_1},\boldsymbol{F}_{2,1} \in \mathbb{F}_2^{n_{22},r_2}$.}
 &  \\
 \hhline{--~}
 $\begin{aligned}(2)&:n_{11}+n_{22}>n_{12}>n_{21},\\
 2n_{22}&+n_{11}-n_{12}-n_{21}<0.\end{aligned}$&  \makecell{$\boldsymbol{G}_1=\begin{bmatrix}
\boldsymbol{F}_{1,1}\\
\boldsymbol{F}_{1,2}\\
\boldsymbol{0}^{n_{22},r_1}\\
\boldsymbol{F}_{1,3}\\
\boldsymbol{0}^{n_{12}-n_{11},r_1}
\end{bmatrix},
\boldsymbol{G}_2 = \begin{bmatrix}
\boldsymbol{F}_{2,1} \\
\boldsymbol{F}_{2,2} \\
\boldsymbol{0}^{n_{12}+n_{21}-2n_{22}-n_{11},r_2} \\
\boldsymbol{F}_{2,2} \\
\boldsymbol{0}^{n_{12}-n_{11},r_2} \\
\boldsymbol{0}^{n_{11}-n_{21},r_2}
\end{bmatrix}$,\\
$\boldsymbol{F}_{1,1}\in \mathbb{F}_2^{n_{22}+n_{11}-n_{12},r_1},\boldsymbol{F}_{1,3},\in \mathbb{F}_2^{n_{11}-n_{21},r_1},\boldsymbol{F}_{2,1} \in \mathbb{F}_2^{n_{12}-n_{11},r_2}$,\\$\boldsymbol{F}_{1,2} \in \mathbb{F}_2^{n_{12}+n_{21}-2n_{22}-n_{11},r_1},\boldsymbol{F}_{2,2}\in \mathbb{F}_2^{n_{22}+n_{11}-n_{12},r_2}$.}
 &  \multirow{8}{*}{$\begin{aligned}r_1&=n_{11}-n_{22},\\
 r_2&=n_{22}.\end{aligned}$}\\
\hline
\end{longtable} 

\footnotesize
\setlength\tabcolsep{2.0pt}
\begin{longtable}{|l|Y|l|}  
\caption{Achievable schemes for Mixed 5$:n_{12}>n_{11}>n_{22}>n_{21}$} \label{table:mix5}\\
\hline
  Interference subregime &Generator matrices  & Target corner point \\
\hline
\endfirsthead
 \hline
\endhead
 \hline
\endfoot
\multirow{6}{*}{ $(1): n_{12}+n_{21}<n_{11}+n_{22}.$} & \makecell{$\boldsymbol{G}_1 = \begin{bmatrix}
\boldsymbol{F}_{1,1}\\
\boldsymbol{F}_{1,2}\\
\boldsymbol{0}^{n_{12}-n_{11},r_1}
\end{bmatrix},
\boldsymbol{G}_2 = \begin{bmatrix}
\boldsymbol{F}_{2,1} \\
\boldsymbol{0}^{n_{11}+n_{22}-n_{12}-n_{21},r_2} \\
\boldsymbol{0}^{n_{21},n_{12}} \\
\boldsymbol{0}^{n_{12}-n_{22},r_2}
\end{bmatrix}$,\\
$\boldsymbol{F}_{1,1}\in \mathbb{F}_2^{n_{21},r_1},\boldsymbol{F}_{1,2} \in \mathbb{F}_2^{n_{11}-n_{21},r_1},\boldsymbol{F}_{2,1}, \in \mathbb{F}_2^{n_{12}-n_{11},r_2}$.}
 & $\begin{aligned}r_1&=n_{11},\\
 r_2&=n_{12}-n_{11}.\end{aligned}$ \\
\hhline{~--}
 & \makecell{$\boldsymbol{G}_1 = \begin{bmatrix}
\boldsymbol{0}^{n_{21},r_1} \\
\boldsymbol{0}^{n_{11}+n_{22}-n_{12}-n_{21},r_1} \\
\boldsymbol{F}_{1,1}\\
\boldsymbol{0}^{n_{12}-n_{11},r_1}
\end{bmatrix},
\boldsymbol{G}_2 = \begin{bmatrix}
\boldsymbol{F}_{2,1} \\
\boldsymbol{F}_{2,2} \\
\boldsymbol{0}^{n_{12}-n_{22},r_2}
\end{bmatrix}$,\\
$\boldsymbol{F}_{1,1},\in \mathbb{F}_2^{n_{12}-n_{22},r_1}, \boldsymbol{F}_{2,1} \in \mathbb{F}_2^{n_{22}-n_{21},r_2},\boldsymbol{F}_{2,2}\in \mathbb{F}_2^{n_{21},r_2}$.}
 & $\begin{aligned}r_1&=n_{12}-n_{22},\\
 r_2&=n_{22}.\end{aligned}$ \\
\hhline{---}
\multirow{4}{*}{ $(2):n_{12}+n_{21}>n_{11}+n_{22}.$} & \makecell{$\boldsymbol{G}_1 = \begin{bmatrix}
\boldsymbol{F}_{1,1}\\
\boldsymbol{F}_{1,2}\\
\boldsymbol{0}^{n_{12}-n_{11},r_1}\\
\end{bmatrix},
\boldsymbol{G}_2=\begin{bmatrix}
\boldsymbol{F}_{2,1} \\
\boldsymbol{0}^{n_{12}+n_{21}-n_{22},r_2}
\end{bmatrix}$,\\
$\boldsymbol{F}_{1,1}\in \mathbb{F}_2^{n_{21},r_1},\boldsymbol{F}_{1,2}\in \mathbb{F}_2^{n_{11}-n_{21},r_1},\boldsymbol{F}_{2,1} \in \mathbb{F}_2^{n_{22}-n_{21},r_2}$.}
 & $\begin{aligned}r_1&=n_{11},\\
 r_2&=n_{22}-n_{21}.\end{aligned}$ \\
\hline
\end{longtable}

\footnotesize
\setlength\tabcolsep{2.0pt}
\begin{longtable}{|l|Y|l|}  
\caption{Achievable schemes for Mixed 6$:n_{21}>n_{11}>n_{22}>n_{12}$} \label{table:mix6}\\
\hline
  Interference subregime &Generator matrices  & Target corner point \\
\hline
\endfirsthead
 \hline
\endhead
 \hline
\endfoot
\multirow{6}{*}{ $(1): n_{12}+n_{21}<n_{11}+n_{22}.$} & \makecell{$\boldsymbol{G}_1 = \begin{bmatrix}
\boldsymbol{F}_{1,1}\\
\boldsymbol{F}_{1,2}\\
\boldsymbol{0}^{n_{21}-n_{11},r_1}\\
\end{bmatrix},
\boldsymbol{G}_2=\begin{bmatrix}
\boldsymbol{0}^{n_{12},r_2}\\
\boldsymbol{0}^{n_{11}+n_{22}-n_{12}-n_{21},r_2}\\
\boldsymbol{F}_{2,1} \\
\boldsymbol{0}^{n_{21}-n_{22},r_2}\\
\end{bmatrix}$,\\
$\boldsymbol{F}_{1,1}\in \mathbb{F}_2^{n_{11}-n_{12},r_1},\boldsymbol{F}_{1,2}\in \mathbb{F}_2^{n_{12},r_1},\boldsymbol{F}_{2,1}\in \mathbb{F}_2^{n_{21}-n_{11},r_1}$.}
 & $\begin{aligned}r_1&=n_{11},\\
 r_2&=n_{21}-n_{11}.\end{aligned}$ \\
\hhline{~--}
 & \makecell{$\boldsymbol{G}_1 = \begin{bmatrix}
\boldsymbol{F}_{1,1}\\
\boldsymbol{0}^{n_{11}+n_{22}-n_{12}-n_{21},r_1}\\
\boldsymbol{0}^{n_{12},r_1}\\
\boldsymbol{0}^{n_{21}-n_{11},r_1}\\
\end{bmatrix},
\boldsymbol{G}_2=\begin{bmatrix}
\boldsymbol{F}_{2,1} \\
\boldsymbol{F}_{2,2} \\
\boldsymbol{0}^{n_{21}-n_{22},r_2}\\
\end{bmatrix}$,\\
$\boldsymbol{F}_{1,1}\in \mathbb{F}_2^{n_{21}-n_{22},r_1},\boldsymbol{F}_{2,1}\in \mathbb{F}_2^{n_{12},r_2},\boldsymbol{F}_{2,2}\in \mathbb{F}_2^{n_{22}-n_{12},r_2}$.}
 & $\begin{aligned}r_1&=n_{21}-n_{22},\\
 r_2&=n_{22}.\end{aligned}$ \\
\hhline{---}
\multirow{4}{*}{ $(2):n_{12}+n_{21}>n_{11}+n_{22}.$} & \makecell{$\boldsymbol{G}_1 = \begin{bmatrix}
\boldsymbol{F}_{1,1}\\
\boldsymbol{F}_{1,2}\\
\boldsymbol{0}^{n_{21}-n_{11},r_1}\\
\end{bmatrix},
\boldsymbol{G}_2=\begin{bmatrix}
\boldsymbol{0}^{n_{12},n_{21}}\\
\boldsymbol{F}_{2,1} \\
\boldsymbol{0}^{n_{21}-n_{22},r_2}\\
\end{bmatrix}$,\\
$\boldsymbol{F}_{1,1}\in \mathbb{F}_2^{n_{11}-n_{12},r_1},\boldsymbol{F}_{1,2}\in \mathbb{F}_2^{n_{12},r_1},\boldsymbol{F}_{2,1}\in \mathbb{F}_2^{n_{22}-n_{12},r_2}$.}
 & $\begin{aligned}r_1&=n_{11},\\
 r_2&=n_{22}-n_{12}.\end{aligned}$ \\
\hline
\end{longtable}

\section{Useful lemmas}\label{appendix:lemma}
\begin{lemma}\label{lem:normalization}
The normalization factor $2^q$ in \eqref{eq:X1} satisfies $\E[\|\msf{x}_k\|^2] \leq 1,\forall k\in\{1,2\}$ .
\end{lemma}
\begin{IEEEproof}
Denote the generator matrix of user $k$ by $\boldsymbol{G}_k$. The actual normalization factor satisfies
\begin{subequations}
\begin{align}
 \frac{1}{\sqrt{\E[\|\msf{x}_k\|^2]}}=& \frac{1}{\sqrt{\E[\|\sum_{i_k=1}^{L_k}2^{\sum_{i=j_k+1}^{M_k}\text{row}(\boldsymbol{E}_{k,i})}\rho_{k,i_k}\msf{F}_{k,i_k}\|^2]}} \\
\geq& \frac{1}{\max\limits_{i_k \in [1:L_k]}\{\rho_{k,i_k} \}\sqrt{\E[\|\text{PAM}(2^q,1)\|^2]}} \label{eq:app_eqa} \\
>& \frac{1}{2}\sqrt{\frac{12}{2^{2q}-1}} > 2^{-q},
\end{align}
\end{subequations}
where $\eqref{eq:app_eqa}$ follows that the largest possible constellation generated by matrix $\boldsymbol{G}_k$ is $\text{PAM}(2^q,1)$.
\end{IEEEproof}

In what follows, we provide some properties of superimposed constellations. First, we define the inter-constellation distance, which will facilitate the analysis of minimum distance.
\begin{define}\label{def:dic}
Consider two one-dimensional constellations $(\Lambda_1,\Lambda_2)$ (not necessarily regular). If $\min\{\Lambda_2\} - \max\{\Lambda_1\}>0$, the inter-constellation distance between $\Lambda_2$ and $\Lambda_1$ is defined as
\begin{align}
d_{\text{IC}}(\Lambda_2,\Lambda_1) \triangleq \min\{\Lambda_2\} - \max\{\Lambda_1\}.
\end{align}
\end{define}
Definition \ref{def:dic} is also used as an indication in relation to the minimum distance of the joint constellation $\Lambda_1\cup\Lambda_2$, as shown in the following Corollary.
\begin{corollary}\label{corollary1}
Consider the two constellations defined in Definition \ref{def:dic} with $d_{\min}\{\Lambda_1\}>0$ and $d_{\min}\{\Lambda_2\}>0$. When $d_{\text{IC}}(\Lambda_2,\Lambda_1)>0$, we have that
\begin{align}
d_{\min}\{\Lambda_2  \cup \Lambda_1\} = \min\{d_{\text{IC}}(\Lambda_2,\Lambda_1), d_{\min}\{\Lambda_1\},d_{\min}\{\Lambda_2\}\}.
\end{align}
\end{corollary}

\begin{lemma}\label{prop:dmin2}
Let $(P_1,P_2)\in \mathbb{R}^2$ be two positive constants. Let $(\Lambda_1,\Lambda_2)$ be two one-dimensional constellations (not necessarily regular) with $d_{\min}(\Lambda_1)>0$ and $d_{\min}(\Lambda_2)>0$, respectively. Consider
\begin{align}\label{prop:dmin2_con1}
P_1d_{\min}(\Lambda_1) < P_2d_{\min}(\Lambda_2),
\end{align}
without loss of generality. Then
\begin{align}
d_{\min}(P_1\Lambda_1+P_2\Lambda_2) = \min\{P_1(\min\{\Lambda_1\}-\max\{\Lambda_1\})+P_2d_{\min}(\Lambda_2),P_1d_{\min}(\Lambda_1)\},
\end{align}
under the following condition
\begin{align}\label{prop:dmin2_184}
P_1(\min\{\Lambda_1\}-\max\{\Lambda_1\})+P_2d_{\min}(\Lambda_2) >0.
\end{align}
\end{lemma}
\begin{IEEEproof}
Let $\Lambda_2 \triangleq \{\lambda_1,\ldots,\lambda_{N}\}$ and $\lambda_{i+1}>\lambda_i$, $\forall i \in [1:,N-1]$ and some $N>1$. Notice that the minimum of the inter-constellation distance (Definition \ref{def:dic}) between $P_1\Lambda_1+P_2\lambda_{i+1}$ and $P_1\Lambda_1+P_2\lambda_i$ is
\begin{align}\label{prop:dmin2_185}
\min_{i \in [1:,N-1] }\{d_{\text{IC}}(P_1\Lambda_1&+P_2\lambda_{i+1},P_1\Lambda_1+P_2\lambda_i)\}\nonumber \\
=& \min_{i \in [1:,N-1]}\{P_1\min\{\Lambda_1\}+P_2\lambda_{i+1} - P_1\max\{\Lambda_1\}-P_2\lambda_i\} \nonumber \\
=&
P_1(\min\{\Lambda_1\}-\max\{\Lambda_1\})+P_2d_{\min}(\Lambda_2)>0.
\end{align}
With \eqref{prop:dmin2_185} and Corollary \ref{corollary1}, we obtain that
\begin{align}
d_{\min}(P_1\Lambda_1+P_2\Lambda_2) &= d_{\min}\left(\bigcup_{i=1}^{N}(P_1\Lambda_1+P_2\lambda_i) \right),\nonumber \\
&= \min\left\{\min_{i \in [1:,N-1] }\{d_{\text{IC}}(P_1\Lambda_1+P_2\lambda_{i+1},P_1\Lambda_1+P_2\lambda_i)\},P_1d_{\min}(\Lambda_1)\right\},\nonumber \\
&  = \min\{P_1(\min\{\Lambda_1\}-\max\{\Lambda_1\})+P_2d_{\min}(\Lambda_2),P_1d_{\min}(\Lambda_1)\}.\nonumber
\end{align}
This completes the proof.
\end{IEEEproof}
Lemma \ref{prop:dmin2} can be seen as a generalization of \cite[Prop. 2]{7451210} with $\Lambda_1$ and $\Lambda_2$ being irregular.

\begin{lemma}\label{prop:dmin3}
Consider a superimposed constellation $\Lambda_{\Sigma} = \sum_{l=1}^LP_l\Lambda_l$, where $L>1$, $\Lambda_l$ is a one-dimensional constellation (not necessarily regular) with $d_{\min}(\Lambda_l)>0$. Then
\begin{align}\label{prop:dmin3_con1a}
d_{\min}(\Lambda_{\Sigma}) = P_1d_{\min}(\Lambda_1),
\end{align}
under the following conditions $\forall l \in [1:L-1]$
\begin{align}
P_{l+1}d_{\min}(\Lambda_{l+1}) &> P_ld_{\min}(\Lambda_{l}), \label{prop:dmin3_con1} \\
d_{\min}(P_l\Lambda_l+P_{l+1}\Lambda_{l+1})&= P_ld_{\min}(\Lambda_l).\label{prop:dmin3_con2b}
\end{align}
\end{lemma}
\begin{IEEEproof}
We proof this lemma by induction.

First, it can be easily verified that \eqref{prop:dmin3_con1a} is true for $L=2$ by substituting $l=1$ into \eqref{prop:dmin3_con1}
\begin{align}
d_{\min}(P_1\Lambda_1+P_2\Lambda_2)= P_1d_{\min}(\Lambda_1).
\end{align}
Next, we assume \eqref{prop:dmin3_con1a} is true up to $l=L-1$
\begin{align}\label{prop:dmin3_con4}
d_{\min}\left(\sum_{l=1}^{L-1}P_l\Lambda_l\right)  = P_1d_{\min}(\Lambda_1).
\end{align}
With \eqref{prop:dmin3_con1} and Lemma \ref{prop:dmin2}, it must be true that
\begin{align}\label{prop:dmin3_con2}
P_l(\min\{\Lambda_l\}-\max\{\Lambda_l\})+P_{l+1}d_{\min}(\Lambda_{l+1}) \geq P_ld_{\min}(\Lambda_l).
\end{align}
By summing the inequality in \eqref{prop:dmin3_con2} from $l=1$ to $L-1$, we get
\begin{align}
&\sum_{l=1}^{L-1}\left(P_l(\min\{\Lambda_l\}-\max\{\Lambda_l\})+P_{l+1}d_{\min}(\Lambda_{l+1})\right)  \geq \sum_{l=1}^{L-1}P_ld_{\min}(\Lambda_l) \\
\Rightarrow & \min\left\{ \sum_{l=1}^{L-1}P_l\Lambda_l\right\}-\max\left\{ \sum_{l=1}^{L-1}P_l\Lambda_l\right\}+\sum_{l=2}^LP_ld_{\min}(\Lambda_l) \geq \sum_{l=1}^{L-1}P_ld_{\min}(\Lambda_l) \\
\Rightarrow & \min\left\{ \sum_{l=1}^{L-1}P_l\Lambda_l\right\}-\max\left\{ \sum_{l=1}^{L-1}P_l\Lambda_l\right\}+P_Ld_{\min}(\Lambda_L) \geq P_1d_{\min}(\Lambda_1).\label{prop:dmin3_con5}
\end{align}
With \eqref{prop:dmin3_con5} together with
\begin{align}
P_Ld_{\min}(\Lambda_L) \overset{\eqref{prop:dmin3_con1}}{>} P_1d_{\min}(\Lambda_1) \overset{\eqref{prop:dmin3_con4}}{=} d_{\min}\left(\sum_{l=1}^{L-1}P_l\Lambda_l\right),
\end{align}
we can use Lemma \ref{prop:dmin2} to show that \eqref{prop:dmin3_con1a} is true for $l=L$
\begin{align}
&d_{\min}\left(\sum_{l=1}^{L-1}P_l\Lambda_l+P_L\Lambda_L\right) \nonumber \\
&=\min \left\{ \min\left\{ \sum_{l=1}^{L-1}P_l\Lambda_l\right\}-\max\left\{ \sum_{l=1}^{L-1}P_l\Lambda_l\right\}+P_Ld_{\min}(\Lambda_L),d_{\min}\left(\sum_{l=1}^{L-1}P_l\Lambda_l\right)\right\} \nonumber \\
& \overset{\eqref{prop:dmin3_con4}}= \min \left\{ \min\left\{ \sum_{l=1}^{L-1}P_l\Lambda_l\right\}-\max\left\{ \sum_{l=1}^{L-1}P_l\Lambda_l\right\}+P_Ld_{\min}(\Lambda_L),P_1d_{\min}(\Lambda_1)\right\} \nonumber \\
&\overset{\eqref{prop:dmin3_con5}}= P_1d_{\min}(\Lambda_1).
\end{align}
This completes the proof.
\end{IEEEproof}

\begin{figure}[t!]
	\centering
\includegraphics[width=4.5in,clip,keepaspectratio]{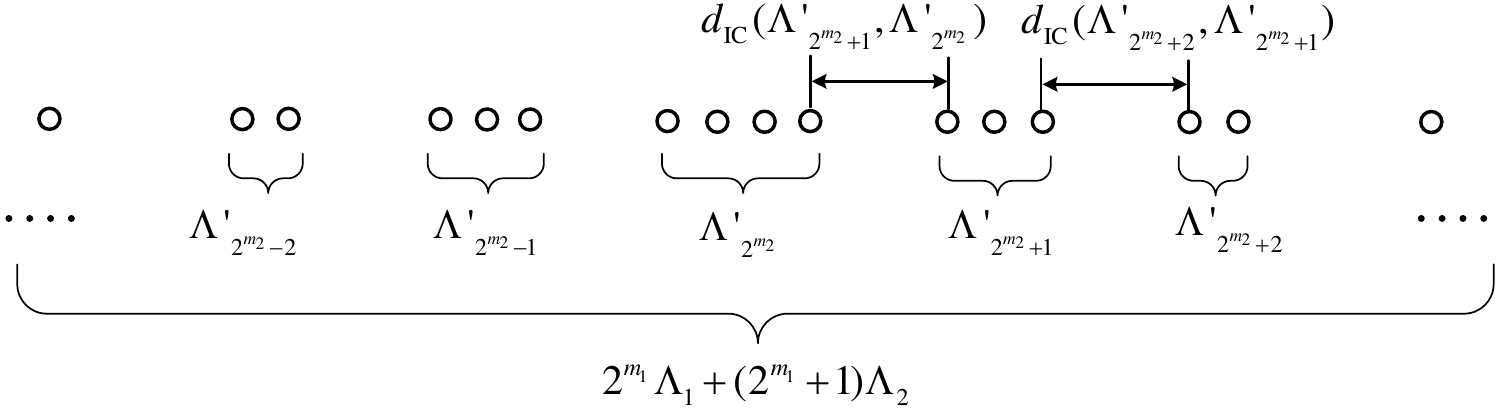}
\caption{Illustrations of set $2^{m_1}\Lambda_1+(2^{m_1}+1)\Lambda_2$ and its subset $\Lambda'_t$.}
\label{fig:illustration}
\end{figure}

\begin{lemma}\label{fact:1}
A superimposed constellation $2^{m_1}\Lambda_1+(2^{m_1}+1)\Lambda_2$, where $\Lambda_1=\Lambda_2$ are the same support of $\text{PAM}(2^{m_2},d)$ with $d>0$, $m_1\geq m_2$, and $m_1,m_2\in \mathbb{N}$, has the following properties:\\
i) the minimum distance of $2^{m_1}\Lambda_1+(2^{m_1}+1)\Lambda_2$ satisfies
\begin{align}
d_{\min}(2^{m_1}\Lambda_1+(2^{m_1}+1)\Lambda_2) = d,
\end{align}
ii) $2^{m_1}\Lambda_1+(2^{m_1}+1)\Lambda_2$ can be decomposed into the following $2^{m_2+1}-1$ subsets
\begin{align}\label{fact:1a}
2^{m_1}\Lambda_1+(2^{m_1}+1)\Lambda_2 = \bigcup_{t=1}^{2^{m_2+1}-1}\Lambda'_t,
\end{align}
where
\begin{align}
\Lambda'_t =&\{2^{m_1}(t-2^{m_2})d+\lambda_2:\lambda_2 \in \Psi_t\subseteq \Lambda_2\},\\
\Psi_t=&\left\{\begin{array}{ll}\left\{\left(-\frac{3\cdot2^{m_2}-1}{2}+t\right)d,\left(-\frac{3\cdot2^{m_2}-3}{2}+t\right)d,\ldots,\frac{2^{m_2}-1}{2}d\right\},&t\in [2^{m_2}:2^{m_2+1}-1],\\
\left\{-\frac{2^{m_2}-1}{2}d,-\frac{2^{m_2}-3}{2}d,\ldots,\left(-\frac{2^{m_2}+1}{2}+t\right)d\right\},&t \in[1 :2^{m_2}-1],
\end{array}\right. \label{eq:C_t_plus}
\end{align}
and the inter-constellation distance (Definition \ref{def:dic}) between $\Lambda_{t+1}$ and $\Lambda_{t}$ is
\begin{align}\label{dic_lambdat_result}
d_{\text{IC}}(\Lambda'_{t+1},\Lambda'_{t})    =\left\{\begin{array}{ll} (2+2^{m_1}-2^{m_2+1}+t)d,& t \in [2^{m_2}:2^{m_2+1}-2], \\
(1+2^{m_1}-t)d, &t \in [1:2^{m_2}-1].
\end{array}\right.
\end{align}
\end{lemma}
Before we proceed to the proof, as an example, we show the sketches of the superimposed constellation $2^{m_1}\Lambda_1+(2^{m_1}+1)\Lambda_2$ and its subset $\Lambda'_t$ in Fig. \ref{fig:illustration}.

\begin{IEEEproof}
Note that $\Lambda_1 = \Lambda_2=\{\pm\frac{1}{2}d,\ldots, \pm\frac{2^{m_2}-1}{2}d\}$. To prove property i), first consider any $\lambda_{1,1},\lambda_{1,2} \in \Lambda_1,\lambda_{2,1},\lambda_{2,2} \in \Lambda_2$ and $(\lambda_{1,1},\lambda_{2,1})\neq (\lambda_{1,2}, \lambda_{2,2})$ and define $\Delta_1 \triangleq \lambda_{1,1}-\lambda_{1,2},\Delta_2\triangleq \lambda_{2,1}-\lambda_{2,2}$. Thus, $\Delta_1,\Delta_2\in [(-2^{m_2}+1)d,:(2^{m_2}-1)d]$. The minimum distance is
\begin{subequations}
\begin{align}
d_{\min}(2^{m_1}\Lambda_1+(2^{m_1}+1)\Lambda_2) &= \min\{|2^{m_1}(\lambda_{1,1}-\lambda_{1,2})+(2^{m_1}+1)(\lambda_{2,1} -\lambda_{2,2})|\} \nonumber \\
& = \min\{|2^{m_1}\Delta_1+(2^{m_1}+1)\Delta_2|\} \label{eq:196} \\
&= d,\label{eq:197}
\end{align}
\end{subequations}
where \eqref{eq:197} follows that $|2^{m_1}\frac{\Delta_1}{d}+(2^{m_1}+1)\frac{\Delta_2}{d}| \in \mathbb{N}$ because $\frac{\Delta_1}{d},\frac{\Delta_2}{d} \in \mathbb{Z}$ and $\frac{\Delta_1}{d}<2^{m_1}$ and $\frac{\Delta_2}{d}<2^{m_1}$ and thus $\min\{|2^{m_1}\frac{\Delta_1}{d}+(2^{m_1}+1)\frac{\Delta_2}{d}|\}=1$. The minimum of \eqref{eq:196} can be obtained when $\Delta_1 = d,\Delta_2 = -d$ or $\Delta_1 = -d,\Delta_2 = d$.

To prove property ii), we let $\lambda_1 \in \Lambda_1,\lambda_2\in \Lambda_2$ and define $t\triangleq (\lambda_1+\lambda_2)\frac{1}{d} + 2^{m_2}\in[1:2^{m_2+1}-1]$, $\Lambda'_t \triangleq \{2^{m_1}(t-2^{m_2})d+\lambda_2 :\lambda_2 \in \Psi_t \}$ for some subset $\Psi_t \subseteq \Lambda_2$ such that
\begin{align}\label{eq:lem4_198}
\bigcup_{t=1}^{2^{m_2+1}-1}\Lambda'_t = 2^{m_1}\Lambda_1+(2^{m_1}+1)\Lambda_2.
\end{align}
Notice that
\begin{align}\label{eq:198}
&\min\{\Lambda'_{t+1}\} \geq \min \{2^{m_1}(t-2^{m_2}+1)d+\lambda_2 :\lambda_2 \in \Lambda_2 \} =2^{m_1}(t-2^{m_2}+1)d-\frac{2^{m_2}-1}{2}d \nonumber \\
&> 2^{m_1}(t-2^{m_2})d+\frac{2^{m_2}-1}{2}d = \max \{2^{m_1}(t-2^{m_2})d+\lambda_2 :\lambda_2 \in \Lambda_2 \} \geq \max\{\Lambda'_t\}.
\end{align}
Then $2^{m_1}\Lambda_1+(2^{m_1}+1)\Lambda_2$ can indeed be decomposed into $2^{m_2+1}-1$ subsets $\Lambda'_{1},\ldots,\Lambda'_{2^{m_2+1}-1}$, where the inter-constellation distance of each pair of neighboring subsets $\Lambda'_{t+1}$ and $\Lambda'_{t}$ satisfies $d_{\text{IC}}(\Lambda'_{t+1},\Lambda'_{t}) = \min\{\Lambda'_{t+1}\}- \max\{\Lambda'_t\} > (2^{m_1}-2^{m_2}+1)d$ according to \eqref{eq:198}.
It then remains to determine $\Psi_t$ in order to satisfy \eqref{eq:lem4_198}. By using the following, we obtain \eqref{eq:C_t_plus}.
\begin{align}
\Psi_t= & \{\lambda_2:\lambda_2 =(t-2^{m_2})d - \lambda_1,\lambda_1 \in \Lambda_1\}  \cap \Lambda_2,\label{phi1}\\
\min\{\Psi_t\} = & \max\{\min\{\lambda_2:\lambda_2=(t-2^{m_2})d-\lambda_1,\lambda \in\Lambda_1\},\min\{\Lambda_2\} \} \nonumber \\
=& \left\{\begin{array}{ll}-\frac{2^{m_2}-1}{2}d+(t-2^{m_2})d,t \in[2^{m_2} :2^{m_2+1}-1],\\
-\frac{2^{m_2}-1}{2}d,t \in[1 :2^{m_2}-1],
\end{array}\right. \label{phi2}\\
\max\{\Psi_t\} = & \min\{\max\{\lambda_2:\lambda_2 =(t-2^{m_2})d - \lambda_1,\lambda_1 \in \Lambda_1\},\max \{\Lambda_2\}\}\nonumber \\
=& \left\{\begin{array}{ll}\frac{2^{m_2}-1}{2}d,t \in[2^{m_2} :2^{m_2+1}-1],\\
\frac{2^{m_2}-1}{2}d+(t-2^{m_2})d,t \in[1 :2^{m_2}-1],
\end{array}\right. \label{phi3}
\end{align}
The exact inter-constellation distance between $\Lambda_t$ and $\Lambda_{t+1}$ is
\begin{align}
d_{\text{IC}}(\Lambda'_{t+1},\Lambda'_{t}) &= \min\{\Lambda'_{t+1}\}-\max\{\Lambda'_t\}  \nonumber \\
&=2^{m_1}(t+1-2^{m_2})d+\min\{\Psi_{t+1}\} -2^{m_1}(t-2^{m_2})d+\max\{\Psi_{t}\} \nonumber \\
& = 2^{m_1}d+\min\{\Psi_{t+1}\}-\max\{\Psi_{t}\}. \label{dic_lambda_t}
\end{align}
Plugging \eqref{phi2}-\eqref{phi3} into \eqref{dic_lambda_t} leads to \eqref{dic_lambdat_result}. This completes the proof.
\end{IEEEproof}

The following lemma is a modification of \cite[Prop. 2]{7451210} to encompass two-dimensional discrete constellations with {\it irregular shapes}.

\begin{lemma}\label{lma:gap_lattice}

%

Let $\msf{X}$ be a discrete random variable uniformly distributed over a two-dimensional constellation $\Lambda$ with minimum distance $d_{\min}(\Lambda)>0$. Let $\msf{Z}\sim \mathcal{CN}(0,1)$ and independent of $\msf{X}$. Then
    \begin{align}\label{eq:th1}
        I(\msf{X};\msf{X}+\msf{Z}) \geq H(\msf{X}) - \log_2 2\pi e \left(\frac{4}{\pi d^2_{\min}(\Lambda)} + \frac{1}{4}\right).
    \end{align}
\end{lemma}
\begin{IEEEproof}
Let $\msf{X'} = \msf{X} + \msf{U}$, where $\msf{U}$ is independent of $\msf{X}$ and $\msf{U} \sim \mathcal{B}(0,\frac{d_{\min}(\Lambda)}{2}) \triangleq \{t \in \mathbb{R}^2:|t | \leq \frac{d_{\min}(\Lambda)}{2}\}$. Clearly, $\msf{X'}, \msf{X}, \msf{Y}$ form a Markov chain in the following order
\begin{equation}
    \msf{X'} \rightarrow \msf{X} \rightarrow \msf{Y}.
\end{equation}
Therefore, from the data processing inequality \cite{Cover:2006:EIT:1146355}, we have
\begin{align}\label{eqn:mu_I}
    I(\msf{X};\msf{Y})&\geq I(\msf{X'};\msf{Y}) 
    = h(\msf{X'}) - h(\msf{X'}|\msf{Y}) 
    = H(\msf{X}) + h(\msf{U}) - h(\msf{X'}|\msf{Y}) \nonumber \\
     &= H(\msf{X}) + \log_2\left(\text{Vol}\left(\mathcal{B}\left(0,\frac{d_{\min}(\Lambda)}{2}\right)\right)\right) - h(\msf{X'}|\msf{Y}) \nonumber \\
     &= H(\msf{X}) + \log_2\left(\frac{d^2_{\min}(\Lambda)}{4}\pi\right) - h(\msf{X'}|\msf{Y}).
\end{align}
Note that
\begin{align}\label{eqn:h_x_y}
    h(\msf{X'}|\msf{Y}=y) &= -\int p(x'|y) \log_2 p(x'|y) \text{d}x' 
    \leq -\int p(x'|y) \log_2 q_{y}(x') \text{d}x',
\end{align}
for any valid distribution $q_{y}(x')$. We pick
\begin{equation}
    q_{y}(x') = \left(\frac{1}{\sqrt{2\pi}s}e^{-\frac{(x'-ly)^2}{2s^2}}\right).
\end{equation}
Plugging this choice into \eqref{eqn:h_x_y} gives
\begin{align}
    &h(\msf{X'}|\msf{Y}=y) 
    \leq \left(\ln 2\pi s^2 + \frac{1}{s^2}\mbb{E}\left[ \|\msf{X'}-ly\|^2|\msf{Y}=y\right] \right) \log_2 e. \nonumber \\
    \Rightarrow &h(\msf{X'}|\msf{Y}) \leq \left(\ln 2\pi s^2 + \frac{1}{s^2}\mbb{E} [\|\msf{X'}-l\msf{Y}\|^2] \right) \log_2 e.\label{eqn:h_x_y2}
\end{align}
Now, choosing $l=\frac{\E[\|\msf{X}\|^2]}{1+ \E[\|\msf{X}\|^2]}$, we have
\begin{subequations}
\begin{align}
    \mbb{E}[\| \msf{X'}-l\msf{Y}\|^2] &= \mbb{E}[\| \msf{X}+\msf{U}-l(\msf{X}+\msf{Z}) \|^2] \nonumber \\
    &= (1-l)^2 \E[\|\msf{X}\|^2] + \sigma^2\left(\mathcal{B}\left(0,\frac{d_{\min}(\Lambda)}{2}\right)\right) + l^2  \\
    & = \frac{\E[\|\msf{X}\|^2]}{1+\E[\|\msf{X}\|^2]} + \frac{d^2_{\min}(\Lambda)}{16}, \label{eq:EXKY}
\end{align}
\end{subequations}
where in \eqref{eq:EXKY} we have used \cite[Eq. (3)]{Ordentlich16}. Hence, \eqref{eqn:h_x_y2} becomes
\begin{align}
    h(\msf{X'}|\msf{Y}) \leq \left(\ln 2\pi s^2 + \frac{1}{s^2}  \left(\frac{\E[\|\msf{X}\|^2]}{1+\E[\|\msf{X}\|^2]} + \frac{d^2_{\min}(\Lambda)}{16}\right)\right) \log_2 e.
\end{align}
We choose $s^2 = \frac{\E[\|\msf{X}\|^2]}{1+\E[\|\msf{X}\|^2]} + \frac{d^2_{\min}(\Lambda)}{16}$ to obtain
\begin{align}\label{eqn:h_x_y3}
    h(\msf{X'}|\msf{Y}) &\leq \log_2 2\pi e \left(\frac{\E[\|\msf{X}\|^2]}{1+\E[\|\msf{X}\|^2]} + \frac{d^2_{\min}(\Lambda)}{16}\right) 
    \leq \log_2 2\pi e \left(1 + \frac{d^2_{\min}(\Lambda)}{16}\right).
\end{align}

Plugging \eqref{eqn:h_x_y3} into \eqref{eqn:mu_I} results in
\begin{align}
    I(\msf{X};\msf{Y})&\geq H(\msf{X}) + \log_2\left(\pi\frac{d^2_{\min}(\Lambda)}{4}\right) 
      - \log_2 2\pi e \left(1 + \frac{d^2_{\min}(\Lambda)}{16}\right) \nonumber \\
    &= H(\msf{X}) - \log_2 2\pi e \left(\frac{4}{\pi d^2_{\min}(\Lambda)} + \frac{1}{4}\right).
\end{align}
This completes the proof.
\end{IEEEproof}

\section{Proof of Proposition \ref{prop:dmin1}}\label{proof:prop:dmin1}
Given that $P_l \triangleq 2^{\sum_{i=1}^l(\alpha_i+m_{i-1})+\beta_l}$ and $\Lambda_l$ is the support of PAM$(2^{m_l-1},1)$ for $l \in [1:L]$, we have
\begin{align}
P_l|\Lambda_l|d_{\min}(\Lambda_l) &= 2^{\sum_{i=1}^l(\alpha_i+m_{i-1})+\beta_l}\cdot 2^{m_i-1} 
= 2^{-1+\sum_{i=1}^l(\alpha_i+m_i)+\beta_l} \nonumber \\
& \leq 2^{\sum_{i=1}^{l+1}(\alpha_i+m_{i-1})+\beta_{l+1}} = P_{l+1}d_{\min}(\Lambda_{l+1}). \label{eq:eq:proof_dmin1_216}
\end{align}
It should be noted that without the ``$-1$'' reduction in $m_l$, \eqref{eq:eq:proof_dmin1_216} may not hold (e.g., when $\alpha_{l+1} = \beta_{l+1} = 0$ and $\beta_{l}>0$). Now with \eqref{eq:eq:proof_dmin1_216}, we can use \cite[Prop. 2]{7451210} to obtain that
\begin{align}\label{eq:proof_dmin1_217}
d_{\min}(P_l\Lambda_l+P_{l+1}\Lambda_{l+1}) =\min\{P_ld_{\min}(\Lambda_l),P_{l+1}d_{\min}(\Lambda_{l+1}) \}= P_l d_{\min}(\Lambda_l).
\end{align}
Finally, with \eqref{eq:eq:proof_dmin1_216} and \eqref{eq:proof_dmin1_217}, we are able to use Lemma \ref{prop:dmin3} to obtain that
\begin{align}
d_{\min}\left(\sum_{i=1}^{L}P_i\Lambda_i\right) = d_{\min}(\Lambda_{\Sigma})= P_1d_{\min}(\Lambda_1) =2^{\alpha_1+\beta_1} \geq 1.
\end{align}
Since none of the constellation points are overlapped, hence
\begin{align}
|\Lambda_{\Sigma}| = \prod\limits_{l=1}^L|\Lambda_l| = 2^{\sum_{l=1}^Lm_l-L},
\end{align}
where $-L$ is due to the ``$-1$'' in the cardinality of $\Lambda_l$.

By considering the extreme case of $\beta_l=1$, we obtain an upper on $\max\{\Lambda_{\Sigma}\}$ and a lower bound on $\min\{\Lambda_{\Sigma}\}$ as
\begin{align}
\max\{\Lambda_{\Sigma}\} &\leq \sum_{l=1}^L2^{\sum_{i=1}^l(\alpha_i+m_{i-1})+1}\cdot \max\{\Lambda_l\}   =\sum_{l=1}^L2^{\sum_{i=1}^l(\alpha_i+m_{i-1})+1}\cdot\frac{2^{m_l-1}-1}{2} \nonumber \\
&<2^{\sum_{l=1}^L(\alpha_l+m_l)-1}-1,  \label{eq:typeI_max} \\
\min\{\Lambda_{\Sigma}\} & = -\max\{\Lambda_{\Sigma}\}
>1-2^{\sum_{l=1}^L(\alpha_l+m_l)-1}.
\end{align}

\section{Proof of Proposition \ref{prop:type2dmin1}}\label{appD}

Since $\Lambda_{\Sigma,1}$ has the same form as the $\Lambda_{\Sigma}$ in \eqref{eq:sc_const_temp}, one can directly use Proposition \ref{prop:dmin1} property i) to obtain that
\begin{align}
d_{\min}(\Lambda_{\Sigma,1})=2^{\alpha_1+\beta_1}\rho_1d_{\min}(\Lambda_1)  \geq 1.
\end{align}
Similarly, it can be easily checked that the following also holds
\begin{align}\label{eq:209}
d_{\min}\left(\Lambda_{\Sigma,1} +2^{\sum_{i=1}^{l'}\alpha_i+m_{i-1}+\beta_{l'}}\rho_{l'}\Lambda_{l'}\right) &=  d_{\min}\left(\sum_{l=1}^{l'} 2^{\sum_{i=1}^l(\alpha_i+m_{i-1})+\beta_l}\rho_l\Lambda_l \right) \nonumber \\
& =2^{\alpha_1+\beta_1}\rho_1d_{\min}(\Lambda_1)  \geq 1 .
\end{align}
With \eqref{eq:209} and the fact that
\begin{align}\label{eq:210a}
d_{\min}(\Lambda_{\Sigma,1})=2^{\alpha_1+\beta_1}\rho_1d_{\min}(\Lambda_1) < 2^{\sum_{i=1}^{l'}\alpha_i+m_{i-1}}2^{\beta_{l'}}\rho_{l'}d_{\min}(\Lambda_{l'}),
\end{align}
the following is true according to Lemma \ref{prop:dmin2}
\begin{align}\label{eq:210}
\min\{\Lambda_{\Sigma,1}\}-\max\{\Lambda_{\Sigma,1}\}+2^{\sum_{i=1}^{l'}\alpha_i+m_{i-1}}2^{\beta_{l'}}\rho_{l'}d_{\min}(\Lambda_{l'}) \geq d_{\min}(\Lambda_{\Sigma,1}).
\end{align}
As for $\Lambda_{\Sigma,2}$ in \eqref{lambda_3}, we obtain the following by directly using Proposition \ref{prop:dmin1} Property ii)
\begin{align}\label{eq:lambda_sigma3}
d_{\min}(\Lambda_{\Sigma,2}) = 2^{\sum_{i=1}^{l'}\alpha_i+m_{i-1}}\left( 2^{\alpha_{l'+1}+m_{l'}+\beta_{l'+1}}\rho_{l'+1}d_{\min}(\Lambda_{l'+1})\right)= 2^{\sum_{i=1}^{l'+1}\alpha_i+m_{i-1}+\beta_{l'+1}}\rho_{l'+1}.
\end{align}
Thanks to the ``$-2$'' reduction in $m_{l'}$, we have
\begin{subequations}\label{eq:215}
\begin{align}
\min&\{(2^{\sum_{i=1}^{\bar{l'}}\alpha_i+m_{i-1}}+2^{\sum_{i=1}^{l'}\alpha_i+m_{i-1}})2^{\beta_{l'}}\rho_{l'}\Lambda_{l'}\}-\max\{(2^{\sum_{i=1}^{\bar{l'}}\alpha_i+m_{i-1}}+2^{\sum_{i=1}^{l'}\alpha_i+m_{i-1}})2^{\beta_{l'}}\rho_{l'}\Lambda_{l'}\}\nonumber \\ & + d_{\min}(\Lambda_{\Sigma,2})\nonumber \\
 =& (2^{\sum_{i=1}^{\bar{l'}}\alpha_i+m_{i-1}}+2^{\sum_{i=1}^{l'}\alpha_i+m_{i-1}})2^{\beta_{l'}}\rho_{l'}(1-2^{m_{l'}-2})+d_{\min}(\Lambda_{\Sigma,2}) \nonumber \\
 >& -4(2^{\sum_{i=1}^{\bar{l'}}\alpha_i+m_{i-1}}+2^{\sum_{i=1}^{l'}\alpha_i+m_{i-1}})(2^{m_{l'}-2}-1)+2^{\sum_{i=1}^{l'+1}\alpha_i+m_{i-1}+\beta_{l'+1}} \label{eq:215a} \\
=& -2^{\sum_{i=1}^{\bar{l'}}\alpha_i+m_i}-2^{\sum_{i=1}^{l'}\alpha_i+m_i}+2^{\sum_{i=1}^{\bar{l'}}\alpha_i+m_{i-1}+2} \nonumber \\
&+2^{\sum_{i=1}^{l'}\alpha_i+m_{i-1}+2}+2^{\sum_{i=1}^{l'+1}\alpha_i+m_{i-1}+\beta_{l'+1}} \label{eq:215b} \\
\geq &2^{\sum_{i=1}^{\bar{l'}}\alpha_i+m_{i-1}+2}+3\cdot2^{\sum_{i=1}^{l'}\alpha_i+m_{i-1}}  \label{eq:215c}\\
>&d_{\min}\left((2^{\sum_{i=1}^{\bar{l'}}\alpha_i+m_{i-1}+\beta_{l'}}\rho_{l'}+2^{\sum_{i=1}^{l'}\alpha_i+m_{i-1}+\beta_{l'}})\Lambda_{l'}\right),
\end{align}
\end{subequations}
where in \eqref{eq:215a} we have used the fact that $2^{\beta_{l'}}\rho_{l'}< 4$, \eqref{eq:215b} follows that $2^{\sum_{i=1}^{\bar{l'}}\alpha_i+m_{i-1}}\cdot2^{m_{l'}} = 2^{\sum_{i=1}^{\bar{l'}}\alpha_i+m_{i}}$ because $m_{l'} = m_{\bar{l'}}$, and \eqref{eq:215c} is due to $2^{\sum_{i=1}^{l'}\alpha_i+m_{i-1}}\geq2^{\sum_{i=1}^{\bar{l'}}\alpha_i+m_{i}}$ since $l'>\bar{l'}$. Then, \eqref{eq:215} together with the fact that
\begin{align}\label{eq:217}
(2^{\sum_{i=1}^{\bar{l'}}\alpha_i+m_{i-1}}+2^{\sum_{i=1}^{l'}\alpha_i+m_{i-1}})2^{\beta_{l'}}\rho_{l'}d_{\min}(\Lambda_{l'})\overset{\eqref{eq:lambda_sigma3}}{<}d_{\min}(\Lambda_{\Sigma,2}),
\end{align}
allow us to use Lemma \ref{prop:dmin2} to obtain that
\begin{align}\label{eq:218}
d_{\min}&\left( (2^{\sum_{i=1}^{\bar{l'}}\alpha_i+m_{i-1}}+2^{\sum_{i=1}^{l'}\alpha_i+m_{i-1}})2^{\beta_{l'}}\rho_{l'}\Lambda_{l'}+ \Lambda_{\Sigma,2}\right) \nonumber \\
&= d_{\min}\left((2^{\sum_{i=1}^{\bar{l'}}\alpha_i+m_{i-1}+\beta_{l'}}\rho_{l'}+2^{\sum_{i=1}^{l'}\alpha_i+m_{i-1}+\beta_{l'}})\Lambda_{l'}\right).
\end{align}
Define $\Lambda_{\Sigma,2a}\triangleq (2^{\sum_{i=1}^{\bar{l'}}\alpha_i+m_{i-1}}+2^{\sum_{i=1}^{l'}\alpha_i+m_{i-1}})2^{\beta_{l'}}\rho_{l'}\Lambda_{l'}+ \Lambda_{\Sigma,2}$. Since \eqref{eq:210a} and \eqref{eq:210} imply that
\begin{align}
&d_{\min}(\Lambda_{\Sigma,1}) <2^{\sum_{i=1}^{l'}\alpha_i+m_{i-1}}2^{\beta_{l'}}\rho_{l'}\overset{\eqref{eq:218}}{<} d_{\min}(\Lambda_{\Sigma,2a}), \label{eq:213} \\
&\min\{\Lambda_{\Sigma,1}\}-\max\{\Lambda_{\Sigma,1}\}+d_{\min}(\Lambda_{\Sigma,2a})\overset{\eqref{eq:218}}{\geq} d_{\min}(\Lambda_{\Sigma,1}),
\end{align}
we thus directly use Lemma \ref{prop:dmin2} again to obtain that
\begin{align}\label{eq:214}
d_{\min}( \Lambda_{\Sigma}) =d_{\min}\left(\Lambda_{\Sigma,1} +\Lambda_{\Sigma,2a}\right)=d_{\min}(\Lambda_{\Sigma,1}) =2^{\alpha_1+\beta_1}\rho_1d_{\min}(\Lambda_1)  \geq 1.
\end{align}

%

The cardinality of $\Lambda_{\Sigma}$ is
\begin{align}\label{eq:lambda_sigma_card}
|\Lambda_{\Sigma}| = \prod\limits_{l=1}^L|\Lambda_l|=2^{\sum_{l=1}^L m_l - 2(L+1)},
\end{align}
where $-2L$ is due to the ``$-2$'' in the order of the cardinality of $\Lambda_l$.


\section{Proof of Proposition \ref{prop:type2dmin2}}\label{appE}
Let $\Lambda_{\Sigma} = \Lambda_{\Sigma,3}+\Lambda_{\Sigma,4}+\Lambda_\text{A}+\Lambda_{\Sigma,5}$, where
\begin{align}
&\Lambda_\text{A}\triangleq 2^{\sum_{i=1}^{l'}\alpha_i+m_{i-1}+\max\{\beta_{kk},\beta_{k\bar{k}}\}}\Lambda_{l'}
+ 2^{\sum_{i=1}^{\bar{l'}}\alpha_i+m_{i-1}+\max\{\beta_{kk},\beta_{k\bar{k}}\}}(\Lambda_{\bar{l'}}+\Lambda_{l'}) \nonumber \\
&=2^{\sum_{i=1}^{l'}\alpha_i+m_{i-1}+\max\{\beta_{kk},\beta_{k\bar{k}}\}}\left(2^{\sum_{i=1}^{\bar{l'}-l'}\alpha_{i+l'}+m_{i-1+l'}}\Lambda_{\bar{l'}}+  \left(2^{\sum_{i=1}^{\bar{l'}-l'}\alpha_{i+l'}+m_{i-1+l'}}+1 \right)\Lambda_{l'}\right). \label{eq:lambda_A}
\end{align}
First, by substituting $P_l\rho_l \rightarrow P_l$ and $\text{PAM}(2^{m_l-2},1) \rightarrow \text{PAM}(2^{m_l-1},1)$ into \eqref{eq:sc_const_temp}, we obtain the following by using Proposition \ref{prop:dmin1} property $ii)$
\begin{align}
d_{\min}(\Lambda_{\Sigma,4}) &= 2^{\sum_{i=1}^{l'+1}(\alpha_i+m_{i-1})+\beta_{l'+1}}\rho_{l'+1}, \label{eq:lambda_sigma2}\\
d_{\min}(\Lambda_{\Sigma,3}) &= 2^{\alpha_1+\beta_1}\rho_1d_{\min}(\Lambda_1) = 2^{\alpha_1+\beta_1}\rho_1, \label{eq:lambda_sigma3a}\\
d_{\min}(\Lambda_{\Sigma,5})& = 2^{\sum_{i=1}^{\bar{l'}+1}(\alpha_i+m_{i-1})+\beta_{\bar{l'}+1}}\rho_{\bar{l'}+1}.\label{eq:lambda_sigma5_dmin}
\end{align}
Next, using the fact that
\begin{align}\label{eq:upper_bound_lambda_l}
\max\{P_l\rho_l\Lambda_l \} = 2^{\sum_{i=1}^l\alpha_i+m_{i-1}+\beta_l}\rho_l\frac{2^{m_l-2}-1}{2}<2^{\sum_{i=1}^l\alpha_i+m_{i-1}}(2^{m_l-1}-1),
\end{align}
for $l \in [1:L] \setminus l'$, which is the same as the upper bound for $\max\{P_l\Lambda_l \}$ with $P_l\Lambda_l$ in \eqref{eq:sc_const_temp}, the following can be obtained by using Proposition \ref{prop:dmin1} property $iii)$
\begin{align}
\max\{\Lambda_{\Sigma,3}\} &<2^{\sum_{l=1}^{l'-1}(m_l+\alpha_l)-1}-1,\label{eq:lambda_sigma3b} \\
\max\{\Lambda_{\Sigma,4}\} &< 2^{\sum_{i=1}^{l'}\alpha_i+m_{i-1}}\left(2^{\sum_{l=1}^{\bar{l'}-l'-1}(m_{l+l'}+\alpha_{l+l'})-1}-1\right) .\label{eq:lambda_sigma2_max}
\end{align}

Since $\Lambda_{l'} = \Lambda_{\bar{l'}}$ are the same support of $\text{PAM}(2^{m_{l'}-2},1)$ as $m_{l'} = m_{\bar{l'}}$ and $m_{l'}-2<\sum_{i=1}^{\bar{l'}-l'}\alpha_{i+l'}+m_{i-1+l'}$, we can use Lemma \ref{fact:1} property i) to obtain that
\begin{align}\label{eq:dmin_CA}
d_{\min}(\Lambda_\text{A}) = 2^{\sum_{i=1}^{l'}\alpha_i+m_{i-1}+\max\{\beta_{kk},\beta_{k\bar{k}}\}}.
\end{align}
When deriving $d_{\min}(\Lambda_{\Sigma,4}+\Lambda_\text{A})$, we note that Lemma \ref{prop:dmin2} cannot be directly used because conditions \eqref{prop:dmin2_con1} and \eqref{prop:dmin2_184} cannot be satisfied simultaneously. Instead, we first use Lemma \ref{fact:1} property ii) to decompose $\Lambda_\text{A}$ into
\begin{align}\label{eq:break_CA_1}
\Lambda_\text{A}=& \bigcup_{t=1}^{2^{m_{l'}-1}-1}\Lambda'_t,\\
\Lambda'_t =&\{2^{\sum_{i=1}^{\bar{l'}-l'}\alpha_{i+l'}+m_{i-1+l'}}(t-2^{m_{l'}-2})d_{\min}(\Lambda_\text{A})+\lambda:\lambda \in \Psi_t \subseteq 2^{\sum_{i=1}^{l'}\alpha_i+m_{i-1}+\max\{\beta_{kk},\beta_{k\bar{k}}\}}\Lambda_{l'}\}, \label{eq:lambdat_pop4}\\
\Psi_t=&\left\{\begin{array}{ll}\left\{\left(-\frac{3\cdot2^{m_{l'}-2}-1}{2}+t\right)d_{\min}(\Lambda_\text{A}),\ldots,\frac{2^{m_{l'}-2}-1}{2}d_{\min}(\Lambda_\text{A})\right\},&t\in [2^{m_{l'}-2}:2^{m_{l'}-1}-1],\\
\left\{-\frac{2^{m_{l'}-2}-1}{2}d_{\min}(\Lambda_\text{A}),\ldots,\left(-\frac{2^{m_{l'}-2}+1}{2}+t\right)d_{\min}(\Lambda_\text{A})\right\},&t \in[1 :2^{m_{l'}-2}-1],
\end{array}\right.
\end{align}
and the inter-constellation distance between $\Lambda'_{t+1}$ and $\Lambda'_{t}$ is
\begin{align}\label{eq:edge_for_use}
d_{\text{IC}}&(\Lambda'_{t+1},\Lambda'_{t})   \nonumber\\
&=\left\{\begin{array}{ll} d_{\min}(\Lambda_\text{A})(2+2^{\sum_{i=1}^{\bar{l'}-l'}\alpha_{i+l'}+m_{i-1+l'}}-2^{m_{l'}-1}+t),& t \in [2^{m_{l'}-2},2^{m_{l'}-1}-2], \\
d_{\min}(\Lambda_\text{A})(1+2^{\sum_{i=1}^{\bar{l'}-l'}\alpha_{i+l'}+m_{i-1+l'}}-t), &t\in [1,2^{m_{l'}-2}-1].
\end{array}\right.
\end{align}

First, we use Lemma \ref{prop:dmin2} to obtain that
\begin{align}\label{eq:inter_distance_sub_1}
d_{\min}(\Lambda'_t+\Lambda_{\Sigma,4}) = d_{\min}(\Lambda_\text{A}),
\end{align}
because the following conditions hold by using \eqref{eq:lambda_sigma2} and \eqref{eq:lambdat_pop4}
\begin{align}
\min\{\Lambda'_t\}-\max\{\Lambda'_t\}+d_{\min}(\Lambda_{\Sigma,4})&\geq (1-2^{m_{l'}-2})d_{\min}(\Lambda_\text{A}) +2^{\sum_{i=1}^{\bar{l'}+1}\alpha_i+m_{i-1}+\beta_{\bar{l'}+1}}\rho_{l'+1} \nonumber \\
&\geq 2^{\sum_{i=1}^{l'}\alpha_i+m_{i-1}+1} >d_{\min}(\Lambda_\text{A}), \\
d_{\min}(\Lambda_{\Sigma,4})&>d_{\min}(\Lambda_\text{A}).
\end{align}
Next, we compute the minimum of the inter-constellation distance (Definition \ref{def:dic}) between $\Lambda'_{t+1}+\Lambda_{\Sigma,4}$ and $\Lambda'_t+\Lambda_{\Sigma,4}$ for $t\in [1:2^{m_{l'}-1}-2]$
\begin{align}\label{eq:edge_distance}
&\min_t\{d_{\text{IC}}(\Lambda'_{t+1}+\Lambda_{\Sigma,4},\Lambda'_{t}+\Lambda_{\Sigma,4})\} =\min_t\{\min\{\Lambda'_{t+1}+\Lambda_{\Sigma,4}\} - \max\{\Lambda'_t+\Lambda_{\Sigma,4}\}\} \nonumber \\
& =\min_t\{\min\{\Lambda'_{t+1}\}-\max\{\Lambda'_t\}\}+\min\{\Lambda_{\Sigma,4}\}-\max\{\Lambda_{\Sigma,4}\} \nonumber \\
 \overset{\eqref{eq:edge_for_use}}&{=}\min_t\{d_{\text{IC}}(\Lambda'_{t+1},\Lambda'_t)\}-2\max\{\Lambda_{\Sigma,4}\} \nonumber \\
\overset{\eqref{eq:lambda_sigma2_max}}&{>}
 d_{\min}(\Lambda_\text{A})(2+2^{\sum_{i=1}^{\bar{l'}-l'}\alpha_{i+l'}+m_{i-1+l'}}-2^{m_{l'}-2})+2^{\sum_{i=1}^{l'}\alpha_i+m_{i-1}}\left(2-2^{\sum_{l=1}^{\bar{l'}-l'-1}(m_{l+l'}+\alpha_{l+l'})}\right) \nonumber \\
&>2^{\sum_{i=1}^{l'}(\alpha_i+m_{i-1})+2}+ 2^{\sum_{i=1}^{\bar{l'}}\alpha_i+m_{i-1}}-2^{\sum_{i=1}^{l'}(\alpha_i+m_i)-2}-2^{\sum_{i=1}^{\bar{l'}-1}(\alpha_i+m_{i-1})-m_{l'}} \nonumber \\
&>2^{\sum_{i=1}^{l'}(\alpha_i+m_{i-1})+2}  >  d_{\min}(\Lambda_\text{A}).
\end{align}

With \eqref{eq:dmin_CA}, \eqref{eq:break_CA_1}-\eqref{eq:inter_distance_sub_1}, \eqref{eq:edge_distance} and Corollary \ref{corollary1}, we arrive at
\begin{align}\label{eq:dmin_aligned_3}
&d_{\min}(\Lambda_\text{A}+\Lambda_{\Sigma,4})   = d_{\min}\left(\bigcup_{t=1}^{2^{m_{l'}-1}-1}\left(\Lambda'_t+\Lambda_{\Sigma,4}\right)\right) \nonumber \\
&= \min\left\{\min\limits_{t}\{d_{\text{IC}}(\Lambda'_{t+1}+\Lambda_{\Sigma,4},\Lambda'_{t}+\Lambda_{\Sigma,4})\}, \min\limits_{t}\{d_{\min}(\Lambda'_t+\Lambda_{\Sigma,4})\},\min\limits_{t}\{d_{\min}(\Lambda'_t)\}\right\} \nonumber \\
&= d_{\min}(\Lambda_\text{A})= 2^{\sum_{i=1}^{l'}\alpha_i+m_{i-1}+\max\{\beta_{kk},\beta_{k\bar{k}}\}}.
\end{align}
The upper bound for the maximum of $\Lambda_\text{A}+\Lambda_{\Sigma,4}$ is obtain with direct calculation
\begin{align}\label{eq:dmin_aligned_3_maxmin}
\max\{\Lambda_\text{A}+\Lambda_{\Sigma,4}\}<2^{\sum_{i=1}^{l'}\alpha_i+m_{i-1}}\left(2^{\sum_{l=1}^{\bar{l'}-l'}(m_{l+l'}+\alpha_{l+l'})-1}-1\right).
\end{align}

For the rest of the proof, one can follow the same line of approach in Appendix \ref{appD} by using Lemma \ref{prop:dmin2} together with \eqref{eq:lambda_sigma3a}-\eqref{eq:lambda_sigma3b} and \eqref{eq:dmin_aligned_3}-\eqref{eq:dmin_aligned_3_maxmin} to show that the conditions \eqref{prop:dmin3_con1} and \eqref{prop:dmin3_con2b} hold for $\Lambda_{\Sigma,3}$ and $(\Lambda_\text{A}+\Lambda_{\Sigma,4})$ as well as $(\Lambda_\text{A}+\Lambda_{\Sigma,4})$ and $\Lambda_{\Sigma,5}$. Then, one can use Lemma \ref{prop:dmin3} to obtain that
\begin{align}
d_{\min}(\Lambda_{\Sigma}) = d_{\min}(\Lambda_{\Sigma,3}+(\Lambda_\text{A}+\Lambda_{\Sigma,4})+\Lambda_{\Sigma,5}) =d_{\min}(\Lambda_{\Sigma,3}) = 2^{\alpha_1+\beta_1}\rho_1 \geq 1.
\end{align}
The cardinality of $\Lambda_{\Sigma}$ is the same as \eqref{eq:lambda_sigma_card}.

\section{Analysis of Scenario $4c)$}\label{app:F}
We analyze $d_{\min}(\Lambda_{\Sigma})$ for each user separately. As illustrated in \eqref{demonstrate1} from the D-IC, the superimposed constellation at receiver 1 can be written in the form of \eqref{eq:align_dmin_4b} with the replacement $(\msf{V}_{l'},\msf{V}_{\bar{l'}}) = (\msf{F}_{1,5},\msf{F}_{2,3})$ and $(\msf{V}_{l'+1},\msf{V}_{\bar{l''}}) = (\msf{F}_{2,5},\msf{F}_{1,8})$ with $\bar{l''} \in [1:l'-1]$ are the pairs of signals whose corresponding submatrices in the D-IC occupy the same subset of rows in matrix $[\boldsymbol{A}_1\boldsymbol{G}_1 \; \boldsymbol{B}_1\boldsymbol{G}_2]$. Hence, $\msf{F}_{2,5}$ is a part of $\sum_{l=l'+1}^{\bar{l'}-1}P_l\rho_l\msf{V}_l$ and $\msf{F}_{1,8}$ is a part of $ \sum_{l=1}^{l'-1}P_l\rho_l\msf{V}_l$.

First, it is obvious that the lower bound on the minimum distance and the upper bound on the maximum value of $\sum_{l=1}^{l'-1}P_l\rho_l\msf{V}_l$ are the same as those of $\Lambda_{\Sigma,3}$ from scenario \emph{4b)} in \eqref{eq:lambda_sigma3a} and \eqref{eq:lambda_sigma3b}, respectively. As for $\msf{F}_{2,5}$, we note that its power coefficient is $P_{l'+1} = (2^{\sum_{i=1}^{\bar{l''}}\alpha_i+m_{i-1}}+2^{\sum_{i=1}^{l'+1}\alpha_i+m_{i-1}})2^{\beta_{l'+1}}$. Then, it can be shown that
\begin{align}
d_{\min}\left(\sum_{l=l'+1}^{\bar{l'}-1}P_l\rho_l\Lambda_l\right) \overset{\eqref{eq:218}}&{>} d_{\min}\left(\sum_{l=l'+1}^{\bar{l'}-1}2^{\sum_{i=1}^{l}(\alpha_i+m_{i-1})+\beta_{l}}\rho_l\Lambda_l\right) \nonumber \\
\overset{\eqref{eq:lambda_sigma2}}&{=} 2^{\sum_{i=1}^{l'+1}(\alpha_i+m_{i-1})+\beta_{l'+1}}\rho_{l'+1}.
\end{align}
Since the following holds due to the ``$-2$'' guard bits,
\begin{align}\label{eq:max_Pl1}
\max\{P_{l'+1}\rho_{l'+1}\Lambda_{l'+1}\} &= (2^{\sum_{i=1}^{\bar{l''}}\alpha_i+m_{i-1}}+2^{\sum_{i=1}^{l'+1}\alpha_i+m_{i-1}})2^{\beta_{l'+1}}\rho_{l'+1}\frac{2^{m_{l'+1}-2}-1}{2} \nonumber \\
&<2^{\sum_{i=1}^{l'+1}\alpha_i+m_{i-1}}(2^{m_{l'+1}-1}-1),
\end{align}
then following \eqref{eq:typeI_max} together with the fact that \eqref{eq:upper_bound_lambda_l} holds for $l\in [l'+1:\bar{l'}-1]$, we have
\begin{align}\label{eq:max_Pl2}
\max\left\{\sum_{l=l'+1}^{\bar{l'}-1}P_l\rho_l\Lambda_l\right\}<2^{\sum_{i=1}^{l'}\alpha_i+m_{i-1}}\left(2^{\sum_{l=1}^{\bar{l'}-l'-1}(m_{l+l'}+\alpha_{l+l'})-1}-1\right).
\end{align}
One can see that the lower bound on the minimum distance and the upper bound on the maximum value of $\sum_{l=l'+1}^{\bar{l'}-1}P_l\rho_l\Lambda_l$ are the same as those of $\Lambda_{\Sigma,4}$ from scenario \emph{4b)} in \eqref{eq:lambda_sigma2} and \eqref{eq:lambda_sigma2_max}, respectively. From here, the analysis of scenario $4c)$ is exactly the same as that of scenario $4b)$ and thus Proposition \ref{prop:type2dmin2} applies to $\Lambda_{\Sigma}$ for user 1.

Similarly, for receiver 2 as illustrated in \eqref{eq:type2_example2} from the D-IC, the superimposed constellation can be written in the form of \eqref{eq:align_dmin_4b}, where the differences are $(\msf{V}_{l'},\msf{V}_{\bar{l'}}) = (\msf{F}_{2,5},\msf{F}_{1,1})$ and $(\msf{V}_{l'-1},\msf{V}_{\bar{l''}}) = (\msf{F}_{1,5},\msf{F}_{2,9})$ with $\bar{l''} \in [1:l'-1]$. Since both $\msf{F}_{1,5}$ and $\msf{F}_{2,9}$ are parts of $ \sum_{l=1}^{l'-1}P_l\rho_l\msf{V}_l$, then the lower bound on the minimum distance of $\sum_{l=1}^{l'-1}P_l\rho_l\msf{V}_l$ is the same as that of $\Lambda_{\Sigma}$ in \eqref{eq:align_dmin_1a} from scenario \emph{4a)} as shown in Proposition \ref{prop:type2dmin1}. This means that lower bound on the minimum distance of $\sum_{l=1}^{l'-1}P_l\rho_l\msf{V}_l$ is also the same as that of $\Lambda_{\Sigma,3}$ in \eqref{lambda_3aaa} from scenario \emph{4b)} in \eqref{eq:lambda_sigma3a}. In addition, using the similar arguments as in \eqref{eq:max_Pl1}-\eqref{eq:max_Pl2}, it can be easily shown that the upper bound on $\max\{\sum_{l=1}^{l'-1}P_l\rho_l\msf{V}_l\}$ is the same as \eqref{eq:lambda_sigma3b}. From here, the analysis of scenario $4c)$ is exactly the same as that of scenario $4b)$ and therefore Proposition \ref{prop:type2dmin2} holds for $\Lambda_{\Sigma}$ for user 2.


\bibliographystyle{IEEEtran}
\bibliography{MinQiu}

\end{document}